\documentclass[english,aps,twocolumn,showpacs,showkeys,preprintnumbers,prd,floatfix,nofootinbib,superscriptaddress,notitlepage,hidelinks]{revtex4-2}

\usepackage{bm}
\usepackage{bbm}
\usepackage{graphicx}
\usepackage{nicefrac}
\usepackage{amsmath}
\usepackage{amsfonts}
\usepackage[colorlinks = true,linkcolor = blue, urlcolor  = blue,citecolor = red,anchorcolor = blue]{hyperref}
\usepackage[normalem]{ulem} 
\usepackage{subfig}
\usepackage{xcolor}
\usepackage{enumitem}
\usepackage{multirow}
\usepackage{array}
\usepackage{lipsum}
\usepackage[capitalise]{cleveref}
\crefname{section}{Sec.}{Secs.}
\Crefname{section}{Section}{Sections}
\crefname{subsection}{Sec.}{Secs.}
\Crefname{subsection}{Section}{Sections}
\usepackage{float}
\usepackage[labelfont=bf]{caption}
\usepackage{subfig}
\usepackage{comment}
\usepackage{subcaption}
\def\ncteqHQ{{\tt \textbf{nCTEQ15HQ }}}
\def\apfel{{\tt \textbf{APFEL++ }}}
\def\wz{$W^\pm\!/Z$ }

\definecolor{purple}{RGB}{128,0,128}

\let\oldcite\cite
\renewcommand{\cite}[1]{\mbox{\oldcite{#1}}}

\def\Pbonly{{\tt Pb-only}}
\def\multinuc{{\tt multi-nuclei}}
\newcommand{\cumchi}{cumulative-$\chi^2$}

\begin{document}

\preprint{IFJPAN-IV-2026-3}
\preprint{MS-TP-26-05}
\preprint{SMU-PHY-25-06}
\preprint{JLAB-THY-26-465}
\newcommand{\orcid}[1]{\,\href{https://orcid.org/#1}{\includegraphics[width=9pt]{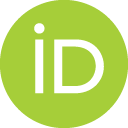}}\,}
\newcommand{\orcidTJ}{0000-0002-1334-7607} 
\newcommand{\orcidMK}{0000-0002-4665-3088} 
\newcommand{\orcidKK}{0000-0003-1412-447X} 
\newcommand{\orcidAK}{0000-0002-4090-0084} 
\newcommand{\orcidND}{0000-0003-0962-631X} 
\newcommand{\orcidPR}{0000-0002-8570-5506} 
\def\smu{\affiliation{Department of Physics, Southern Methodist University,
    Dallas, TX 75275-0175, U.S.A.}}
\def\jlab{\affiliation{Theory Center, Jefferson Lab, Newport News, VA 23606, U.S.A.}}
\def\krakow{\affiliation{Institute of Nuclear Physics Polish Academy of Sciences, PL-31342 Krakow, Poland}}
\def\muenster{\affiliation{Institut f{\"u}r Theoretische Physik, Westf{\"a}lische Wilhelms-Universit{\"a}t M{\"u}nster,             \\Wilhelm-Klemm-Stra{\ss}e 9, D-48149 M{\"u}nster, Germany}}
\title{Determination of Nuclear PDFs using Markov Chain Monte Carlo Methods}
\author{N.~Derakhshanian\orcid{\orcidND}} 
\email{nasim.derakhshanian@ifj.edu.pl}
\krakow{}
\author{P.~Risse\orcid{\orcidPR}}
\email{rissep@jlab.org}
\smu{}
\jlab{}
\author{T.~Je\v{z}o\orcid{\orcidTJ}}
\email{tomas.jezo@uni-muenster.de}
\muenster{}
\author{M.~Klasen\orcid{\orcidMK}}
\muenster{}
\author{K.~Kova\v{r}\'{i}k\orcid{\orcidKK}} 
\muenster{}
\author{A.~Kusina\orcid{\orcidAK}} 
\email{aleksander.kusina@ifj.edu.pl}
\krakow{}

\begin{abstract}
Global QCD analyses of nuclear parton distribution functions (nPDFs) have traditionally relied on the Hessian method for uncertainty estimation. However, the inherent Gaussian approximation and reliance on local curvature often prove insufficient for nPDF fits, which are frequently characterized by limited data constraints and non-Gaussian likelihoods. In this paper, we present the first nPDF determination based on Markov Chain Monte Carlo (MCMC) techniques, implemented within the nCTEQ framework using an adaptive Metropolis–Hastings algorithm. The MCMC approach enables a direct mapping of the posterior distribution and reveals a highly non-trivial parameter-space structure, including multiple modes and pronounced non-Gaussian behavior, particularly for the valence PDFs. We perform the first single-nucleus global analysis of lead PDFs using exclusively lead data and compare it to a multi-nuclei fit employing a standard analytic $A$ dependence. The inclusion of lighter nuclei reduces quark uncertainties and modifies the shape of the lead PDFs, while leaving the gluon distribution largely unaffected. A complementary Hessian analysis exposes systematic limitations of the Gaussian approximation. Our results demonstrate that MCMC methods provide a more reliable framework for uncertainty quantification in nPDF determinations.
\end{abstract}

\date{\today}
\maketitle
\tableofcontents{}

\section{Introduction}
\label{sec:intro}
Nuclear Parton Distribution Functions (nPDFs) are a fundamental ingredient in precision QCD phenomenology involving nuclei. They extend the concept of proton PDFs to bound nucleons by incorporating characteristic nuclear effects such as shadowing, antishadowing, the EMC effect, and Fermi motion \cite{Klasen:2023uqj,Ethier:2020way,Arleo:2025oos}. Reliable nPDFs are essential for interpreting proton–nucleus and heavy-ion measurements at the Large Hadron Collider (LHC), neutrino–nucleus scattering, and future electron–ion collider data.
Over the past decades, several global QCD analyses have produced modern nPDF sets, including those by the nCTEQ~\cite{Kovarik:2015cma,Duwentaster:2022kpv}, EPPS~\cite{Eskola:2021nhw}, nNNPDF~\cite{AbdulKhalek:2022fyi} collaborations, and others~\cite{Helenius:2021tof,Khanpour:2020zyu,deFlorian:2011fp,Hirai:2007sx}. These determinations differ in their choice of $x$ parametrization, treatment of $A$ dependence, dataset selection, and methodology for uncertainty estimation. Despite the progress, a statistically rigorous quantification of nPDF uncertainties remains a significant challenge due to the limited constraining power of nuclear data and the presence of dataset tensions. 
In addition, the presence of both protons and neutrons within the nucleus introduces further correlations into the parameter space, as following the isospin symmetry, the densities for the proton's up-quark are identified with the neutron's down-quark densities and vice versa.
As a result, the likelihood in nPDF fits is typically non-Gaussian and may contain multiple competing minima.

Most global nPDF analyses employ the Hessian method \cite{Stump:2001gu, Pump014011} for uncertainty estimation. This approach assumes an expansion of the $\chi^2$ function around a single minimum up to quadratic terms, corresponding to a multivariate Gaussian posterior. Such a local approximation cannot capture skewed distributions, nonlinear correlations, or multimodal structures. Moreover, in the presence of tensions among data sets, a phenomenological tolerance \mbox{$\Delta \chi^2>1$} must be introduced to enlarge the uncertainties, weakening the direct probabilistic interpretation of the confidence intervals. An alternative strategy, adopted in the nNNPDF framework, is the Monte Carlo replica method \cite{Giele:1998gw, AbdulKhalek:2019mzd}. In this approach, pseudo-data replicas are generated and fitted independently, and the resulting ensemble of PDFs is interpreted statistically. Although this method avoids an explicit quadratic approximation, it does not directly sample the posterior distribution of the fit parameters. It was shown that in situations where theoretical predictions depend nonlinearly on parameters and likelihoods are non-Gaussian, the Monte Carlo replica method does not faithfully reproduce the true parameter probability density~\cite{Costantini:2024wby}.

A statistically consistent treatment requires direct sampling of the full posterior distribution. Markov Chain Monte Carlo (MCMC) methods \cite{Hogg:2017akh, brooks2011handbook} provide precisely this capability. By sampling parameters according to Bayes’ theorem, MCMC methods naturally accommodate nonlinear correlations, asymmetric tails, and multiple minima without relying on Gaussian approximations or externally imposed tolerance parameters. This provides uncertainty estimates with a well-defined probabilistic interpretation. However the complexity of MCMC makes obtaining a successful nPDF fit much more challenging.

The application of MCMC methods to PDF determination has its origins in proton PDF analyses, beginning with the pioneering work of Giele \textit{et al.}~\cite{Giele:2001mr}, who first employed these techniques to infer both PDFs and detector systematics within a unified Bayesian framework. A~subsequent proof-of-principle study by Gbedo \textit{et al.}~\cite{Gbedo:2017eyp}, demonstrated the capability of MCMC sampling to extract PDFs from DIS data. More recently, efforts have concentrated on improving algorithmic efficiency and benchmarking uncertainty estimation strategies. Hunt-Smith \textit{et al.}~\cite{Hunt-Smith:2022ugn, Hunt-Smith:2023ccp} carried out systematic comparisons of MCMC, Hessian, and nested sampling approaches in controlled settings, while Capel \textit{et al.}~\cite{Capel:2024qkm} developed a dedicated MCMC framework based on forward modeling, predicting bin-level event counts directly. The recently introduced \texttt{Colibri} framework~\cite{Costantini:2025agd, Costantini:2025wxp} further advances Bayesian treatments by employing orthogonal functional decompositions to linearize PDF parameterizations and facilitate statistically consistent inference. Finally, in our recent work~\cite{Risse:2025qlo} we have performed a full MCMC proton PDF analysis including both DIS and hadronic data.

Building on these developments,
particularly our proton PDF determination~\cite{Risse:2025qlo},
we extend the MCMC framework to nuclear PDF analyses and present the first comprehensive MCMC-based global analysis of nPDFs using real-world experimental data.
To disentangle methodological effects from modeling assumptions related to the nuclear mass-number dependence and to balance the physical coverage and computational efficiency, we perform two complementary analyses. 
The primary study, \Pbonly{} fit, is restricted to data involving only the lead (Pb, \mbox{$A = 208$}) nucleus. Such a setup allows for a focused investigation of the parameter-space structure within a single nuclear system.
However, the restriction to Pb data removes important constraints from the parameter space, amplifying the aforementioned non-Gaussianity. We also compare in detail the differences occurring between the fully Bayesian MCMC determination of uncertainties with the usual Hessian uncertainty determination.
The secondary analysis, called \multinuc{} fit, introduces the constraints from other nuclei through the $A$ dependence.
It incorporates measurements from a range of nuclear targets and, apart from the $A$ dependence, is carried out within the same framework. As in the primary study, we compare the MCMC- and Hessian-based uncertainties. 
Furthermore, the comparison of the \Pbonly{} and \multinuc{} analyses enables an assessment of the assumed $A$ dependence of nuclear modifications and the stability of the extracted lead PDFs under expanded data input.

In summary, this work establishes a fully Bayesian methodology for nPDF determination and provides a more rigorous and detailed characterization of nuclear PDF uncertainties compared to traditional methods. Finally, it provides first insights into the long-standing questions about constraints and biases that are introduced through the $A$ dependence.
%
\section{Formalism: Nuclear PDFs and MCMC}
\label{sec:formalism}
In this section, we establish the essential theoretical and computational formalism for this work. First we briefly describe the nCTEQ framework for global analyses of nuclear PDFs. We then introduce the Bayesian statistical approach that guides our uncertainty determination, followed by a detailed description of the MCMC algorithm employed to sample the posterior distribution of nPDF parameters.

\subsection{Nuclear PDF Framework }
In global QCD analyses, nPDFs are extracted by fitting theoretical predictions to experimental data from lepton–nucleus and proton–nucleus collisions adopting the collinear factorization ansatz. Within the nCTEQ framework which we use here, the PDF of a nucleus with charge $Z$ and mass number $A$, denoted $f_{i}^A{(x,Q)}$, is expressed in terms of effective bound-proton and bound-neutron PDFs as
\begin{equation}
    f_{i}^A(x,Q)= \frac{Z}{A}\, f_i^{p/A}(x,Q) + \frac{A-Z}{A}\, f_i^{n/A}(x,Q)\,,
    \label{eq.1}
\end{equation}
where isospin symmetry is assumed to relate the bound-neutron PDFs, $f_i^{n/A}$, to the bound-proton PDFs, $f_i^{p/A}$, by interchanging up and down quark distributions. The $x$-dependence of the bound-proton PDFs is parametrized at a low input scale, here $Q_0=m_c=1.3\,\mathrm{GeV}$. The specific functional form is chosen based on the underlying proton PDF set, which is used as a baseline for the nPDFs. In this analysis, similarly to the upcoming nCTEQ nPDF release~\cite{ncteq25} and to our recent proton MCMC analysis~\cite{Risse:2025qlo}, we use the CJ15 PDF set~\cite{Accardi:2016qay} as the proton baseline. For each parton flavor 
\[
i = \{u_v,\bar{u}+\bar{d},g,s+\bar{s}\}\,,
\]
the bound-proton PDFs at the input scale are parameterized as
\begin{equation}
x f_i^{p/A}(x,Q_0) =
c_0^i, x^{c_1^i} (1-x)^{c_2^i}
\left(1 + c_3^i \sqrt{x} + c_4^i x\right)\,,
\label{eq.2}
\end{equation}
In addition, the $d$-valence, $d_v$, distribution incorporates an extra mixture of $u$-valence, $u_v$, through:
\begin{equation}
\begin{split}
    xd_v^{p/A}(x,Q_0) &= c^{i}_0 \Big[
      x^{c^{i}_1} (1-x)^{c^{i}_2} (1+c^{i}_3\sqrt{x}+c^{i}_4x) \\
      &\qquad\qquad+ c^{i}_6x^{c^{i}_7} \,xu_v^{p/A}(x,Q_0)
      \Big]\,,\\
      i&=d_v\,.
\label{eq.3}
\end{split}   
\end{equation}
The asymmetry between light sea quarks is parameterized through the ratio:
\begin{equation}
\begin{split}
  \frac{\bar{d}}{\bar{u}}(x,Q_0) &= c_0^{i}x^{c_1^{i}}(1-x)^{c_2^{i}} + 1 + c_3^{i}x(1-x)^{c_4^{i}},\\
  i&=\bar{d}/\bar{u}\,.
  \end{split}
\label{eq.4}
\end{equation}
Furthermore, we assume there is no strange quark asymmetry in the nucleus $s(x) - \bar{s}(x) = 0$. To account for nuclear modifications, the parameters entering the bound-proton PDFs are made explicitly dependent on the nuclear mass number $A$. For each parameter $c^j$, this dependence is modeled as
\begin{equation}
c_j(A) = p_j + a_j\ln A + b_j\ln^2 A\,, \qquad j = 1, \dots, 5\,.
\label{eq.5}
\end{equation}
This adjustment ensures that the free-proton limit is recovered for $A=1$. The coefficients $p_j$ define the fixed proton baseline, while the parameters $a_j$ and $b_j$ encode nuclear effects and are constrained by nuclear data. This parametrization is applied consistently in both the \Pbonly{} and \multinuc{} analyses. In the \Pbonly{} fit, where data from a single nuclear target are used, the $\ln A$ dependence effectively reduces to a constant evaluated at $A=208$. Expressing nuclear modifications through Eq.~\eqref{eq.5} guarantees that the fitted coefficients $a_j$ preserve a clear interpretation as nuclear-modification parameters and are directly comparable between the \Pbonly{} and \multinuc{} fits.

To determine the best-fit nPDF parameters, global analyses minimize a figure-of-merit function, typically the $\chi^2$, which quantifies the agreement between theory predictions and experimental data. The most common method for quantifying parameter uncertainties in these fits is the Hessian method~\cite{Pumplin:2001ct,Pumplin:2000vx}. This approach starts by identifying the best-fit parameters $\mathbf{a}_{0}$, which minimize the $\chi^2$ function, denoted as $\chi^2_0$. Around this minimum, the method assumes that $\chi^2$ can be approximated by a second-order Taylor expansion:
\begin{align}
    \chi^2 \approx \chi^2_0 + \sum_{i,j} H_{ij}\,y_i\,y_j, \quad y_i \equiv a_i - a_{i,0}\,,
     \label{eq.6}
\end{align}
where $H_{ij}$ is the Hessian matrix, defined as the second derivatives of $\chi^2$:
\begin{align}
    H_{ij} = \frac12 \left(\frac{\partial^2 \chi^2}{\partial y_i \partial y_j}\right)_{a_i=a_{i,0}}\,.
     \label{eq.7}
\end{align}
This quadratic expansion corresponds to a Gaussian approximation of the parameter likelihood near the minimum. Furthermore, the method assumes a linear response of observables within the uncertainty region. Diagonalizing the Hessian matrix defines the principal directions of uncertainty, and the resulting eigenvectors and eigenvalues describe the shape and orientation of the error ellipsoid in parameter space. Uncertainties are typically expressed via PDF error sets, $S_{k}^{\pm}$, constructed along these eigenvectors. Each set corresponds to a fixed deviation in $\chi^2$, known as the Hessian tolerance $\Delta \chi^2$. In the ideal case of Gaussian errors and fully consistent data, the value of $\Delta \chi^2$ yielding 68\% confidence intervals follows from Wilks' theorem~\cite{Wilks:1938dza}. However, global QCD fits often involve partially incompatible or correlated datasets, requiring a phenomenologically motivated tolerance, which weakens the direct statistical interpretation.
The uncertainty on any PDF-dependent observable $\mathcal{O}$ can be propagated using the following asymmetric formula:
\begin{align}
\delta_k^+ &= \max \{ 0, \mathcal{O}(f_k^+) - \mathcal{O}(f_0),  \mathcal{O}(f_k^-)-  \mathcal{O}(f_0)\}\,, \nonumber\\
\delta_k^- &= \max \{ 0, \mathcal{O}(f_0) - \mathcal{O}(f_k^+), \mathcal{O}(f_0) - \mathcal{O}(f_k^-) \}\,, \nonumber\\
\Delta^\pm &= \sqrt{ \sum_k \big[ \delta_k^\pm \big]^2 }\,.
\label{eq.9}
\end{align}
where $\mathcal{O}(f_k^{\pm})$ are observable values calculated using the PDF error sets corresponding to the positive and negative variations along the $k$-th eigenvector of the Hessian matrix and $\mathcal{O}(f_0)$ represents the central value of the observable. 

While the Hessian method is widely used for estimating uncertainties in global fits, it has notable limitations, including its reliance on a Gaussian approximation, sensitivity to the empirically chosen tolerance $\Delta \chi^2$, and potential failure to capture the global minimum in complex parameter spaces.

\subsection{Bayesian Formulation and Likelihood}\label{sec:bayes}
Within the Bayesian framework, the determination of nuclear PDF parameters is formulated as an inference problem, in which the posterior probability distribution of the parameter vector $\bm{a}$ is constrained by experimental data $\bm{D}$. The parameter set $\bm{a} = \{a_k\}$ represents the coefficients describing the $A$-dependent modifications of the nuclear PDFs for each parton flavor as defined in Eq.~(\ref{eq.5}), where the index $k$ runs over the dimensions of the parameter vector. Using Bayes’ theorem, the posterior distribution is given by
\begin{equation}
p(\bm{a}|\bm{D}) \propto l(\bm{D}|\bm{a})\,p(\bm{a})\,,
\label{eq.10}
\end{equation}
where $l(\bm{D}|\bm{a})$ is the likelihood function and $p(\bm{a})$ denotes the prior distribution. In this work, flat priors are assumed unless stated otherwise, such that the posterior is directly proportional to the likelihood. The likelihood is constructed from the standard $\chi^2$ function used in global nPDF analyses,
\begin{equation}
\chi^2(\boldsymbol{a}) =
\left(\bm{D} - \bm{T}(\bm{a})\right)^{\mathrm{T}}
\bm{C}^{-1}(\bm{a})
\left(\bm{D} - \bm{T}(\bm{a})\right)\,.
\label{eq.11}
\end{equation}
Here, $\bm{T}(\bm{a})$ denotes the theoretical predictions for the observables, computed by evaluating the PDFs defined by $\bm{a}$, evolving them via the DGLAP equations, and convolving them with the relevant partonic cross sections. The covariance matrix $\bm{C}$ encodes all experimental uncertainties and follows the nCTEQ formulation from~\cite{Muzakka:2022wey}:
\begin{equation}
C_{T,ij}(\bm{a}) = \sigma_i^2 \delta_{ij} + \sum_\alpha \bar{\sigma}_{i\alpha} \bar{\sigma}_{j\alpha} + \sigma_{\text{norm}}^2 T_i(\bm{a}) T_j(\bm{a})\,,
\label{eq.12}
\end{equation}
where $\sigma_i$ are uncorrelated statistical uncertainties, $\bar{\sigma}_{i\alpha}$ denote correlated systematic uncertainties associated with source $\alpha$, and $\sigma_{\text{norm}}$ represents a normalization uncertainty incorporated via the T-method \cite{Muzakka:2022wey,Ball:2009qv}. The final term introduces a coherent shift proportional to the theory predictions, ensuring a consistent treatment of overall normalization effects across the dataset. The likelihood is then defined as
\begin{equation}
l(\bm{D}|\bm{a}) \propto
\exp\!\left(-\frac{1}{2}\chi^2(\boldsymbol{a})\right)\,,
\label{eq.13}
\end{equation}
which serves as the target distribution for the Markov Chain Monte Carlo sampling. This formulation is identical to that employed in our recent proton PDF analysis~\cite{Risse:2025qlo} and provides a statistically well-defined basis for sampling the full posterior distribution of the nuclear PDF parameters.

\subsection{MCMC Algorithm}\label{sec:algorithm}
In the Bayesian framework, the traditional Hessian approach corresponds to a Gaussian approximation of the posterior distribution and therefore provides only a local representation of parameter uncertainties. Markov Chain Monte Carlo methods offer a more general approach by directly sampling the full posterior distribution without such approximations. By constructing a chain that converges to the target distribution, MCMC enables the estimation of posterior densities and associated uncertainties in high-dimensional parameter spaces. The Metropolis–Hastings (MH) algorithm \cite{metropolis1953equation, Hastings:1970aa} generates a Markov chain that converges to a desired target distribution. The chain evolves in the parameter space, where each iteration corresponds to a state represented by a $k$-dimensional vector \mbox{$\bm{X}_{t} =[ X_{t}^1, X_{t}^2, \cdots, X_{t}^k]^T\in \mathbb{R}^k$}, where $k$ denotes the number of fitted nuclear parameters.\footnote{Here, a ``state'' refers to a specific configuration of parameters in the parameter space at a given iteration, while a ``step'' denotes the transition between successive samples in the chain.} In the present analysis, $\bm{X}$ corresponds to the full set of nPDF fit parameters, $\bm{X} \equiv \bm{a} = \{ a_k \}$. Starting from an initial state $\bm{X}_0$, a candidate state $\tilde{\bm{X}}_{t+1}$ is proposed from a transition (or proposal) distribution $q(\tilde{\bm{X}}_{t+1} \mid \bm{X}_t)$, typically chosen as a multivariate Gaussian centered on the current state,
\begin{equation*}
\tilde{\bm{X}}_{t+1} \sim q(\tilde{\bm{X}}_{t+1} \mid \bm{X}_t) = \mathcal{N}(\bm{X}_t, C_0)\,,
\end{equation*}
where $C_0$ is the fixed covariance matrix. The proposed candidate is accepted with probability
\begin{equation}
\alpha = \min \left( 1, \frac{\pi(\tilde{\bm{X}}_{t+1}) \, q(\bm{X}_t \mid \tilde{\bm{X}}_{t+1})}{\pi(\bm{X}_t) \, q(\tilde{\bm{X}}_{t+1} \mid \bm{X}_t)} \right)\,,
\end{equation}
where $\pi(\bm{X})$ denotes the target probability density, corresponding to the unnormalized posterior defined in Eq.~(\ref{eq.10}).\footnote{More precisely, the MH transition kernel is constructed such that the target posterior $\pi(\bm{X})$ is an invariant distribution of the chain. Under standard regularity conditions (e.g.\ irreducibility and aperiodicity), this invariant distribution is unique and therefore constitutes the stationary distribution.} If accepted, the chain moves to $\bm{X}_{t+1} = \tilde{\bm{X}}_{t+1}$; otherwise, $\bm{X}_{t+1} = \bm{X}_t$.\footnote{Unlike standard Monte Carlo algorithms, where rejected points are discarded, the MH algorithm retains rejected proposals as repetitions of the current state, ensuring the chain maintains the correct stationary distribution \cite{robert2005monte}.}

While the standard MH algorithm is effective for moderately correlated parameter spaces, its efficiency can degrade in high-dimensional problems with strong parameter correlations, motivating the adaptive extensions. The Adaptive Metropolis–Hastings (aMH) algorithm \cite{Haario2001amh} addresses this limitation by updating the proposal covariance matrix during sampling, allowing the proposal distribution to gradually approximate the structure of the target posterior. The algorithm is initialized as a standard MH sampler with a fixed covariance matrix $C_0$, After a pre-run of $N_{\text{pre}}$ iterations, the proposal covariance is updated recursively using the accumulated chain history. At iteration $t$, the self-learned  covariance matrix, $\overline C_{t+1}$, is computed as
\begin{equation}
    \overline C_{t+1} = \frac{1}{t} \left( S_{t+1} - (t+1) \bm{\mu}_{t+1} \bm{\mu}_{t+1}^T \right)\,,
    \label{eq.15}
\end{equation}
where $S_{t+1} = S_t + \bm{X}_{t+1}\bm{X}_{t+1}^{T}$ is the updated cross-product matrix, and the mean vector is updated recursively as $\bm{\mu}_{t+1} = \bm{\mu}_{t} + \frac{1}{t+1} (\bm{X}_{t+1}-\bm{\mu}_{t})$. This formulation avoids recomputing the covariance from all previous samples and captures the evolving parameter correlations up to iteration $t+1$. The proposal distribution in the aMH algorithm is then defined as a mixture of adaptive and fixed components:
\begin{equation}
\begin{split}
\tilde{\bm{X}}_{t+1} &\sim q(\tilde{\bm{X}}_{t+1} \mid \bm{X}_t) \\
                          &= (1 - \beta)\, \mathcal{N}\left(\bm{X}_t, s_d \overline C_t\right) + \beta\, \mathcal{N}\left(\bm{X}_t, C_0\right)\,,
\end{split}
\label{eq.16}
\end{equation}
where $s_d = (2.4)^2/d$ is the optimal scaling factor for Gaussian targets and Gaussian proposals~\cite{Gelman1996, Gelman1997}, and $d$ denotes the dimension. The mixing coefficient \mbox{$\beta \in (0, 1)$} balances adaptation and stability, ensuring that the proposal remains well-defined even if $\overline C_t$ collapses to zero due to the presence of the fixed covariance $C_0$. The validity of the aMH algorithm relies on two key conditions. First, each proposal distribution must preserve the target distribution, which is satisfied here since both terms in Eq.~\eqref{eq.16} are valid random-walk kernels with the posterior as an invariant distribution. Second, the effect of adaptation must diminish as \mbox{$t \rightarrow \infty$}. The $1/t$ scaling in Eq.~\eqref{eq.15} enforces this diminishing-adaptation condition, thereby preserving the ergodicity of the chain. To reduce biases from early samples and improve long-term mixing, the self-learned covariance matrix is periodically reset to the fixed non-adaptive covariance $C_0$. This Reset-to-Mean strategy prevents the chain from becoming overly influenced by its initial states and stabilizes the self-learned covariance matrix $\overline C_t$. As in the standard MH algorithm, rejected proposals are retained as repeated states. In this context, the multiplicity of a state, defined as the number of consecutive iterations in which the chain remains at that state, serves as a useful diagnostic. Large multiplicities indicate regions where the sampler struggles to move efficiently, often indicating issues with adaptation or proposal tuning. During posterior analysis, multiplicity is also used for weighted estimation, where states with higher multiplicity contribute proportionally more to posterior averages, ensuring accurate representation of the target distribution.

The aMH algorithm updates the $\chi^2$-based covariance matrix recursively at each iteration, substantially reducing the computational cost compared to recalculating full empirical covariances and making it efficient for high-dimensional parameter spaces. Unlike more advanced MCMC methods such as Hybrid/Hamiltonian Monte Carlo \cite{Duane:1987de}, which require multiple $\chi^2$ evaluations per iteration, aMH performs only a single evaluation per step, maintaining low computational overhead. Its effectiveness, however, depends on the structure of the $\chi^2$ surface. aMH performs best when the target distribution exhibits a single, well-defined global minimum, but it may struggle in multimodal posteriors where transitions between distant modes are rare~\cite{Albert:2024zsh}. 
However, as we will see in Sec.~\ref{sec:res}, in the current analysis aMH still allows to obtain reliable results even in the situation when multiple minima are present.

\begin{figure}[b]
    \centering
    \includegraphics[width=\linewidth]{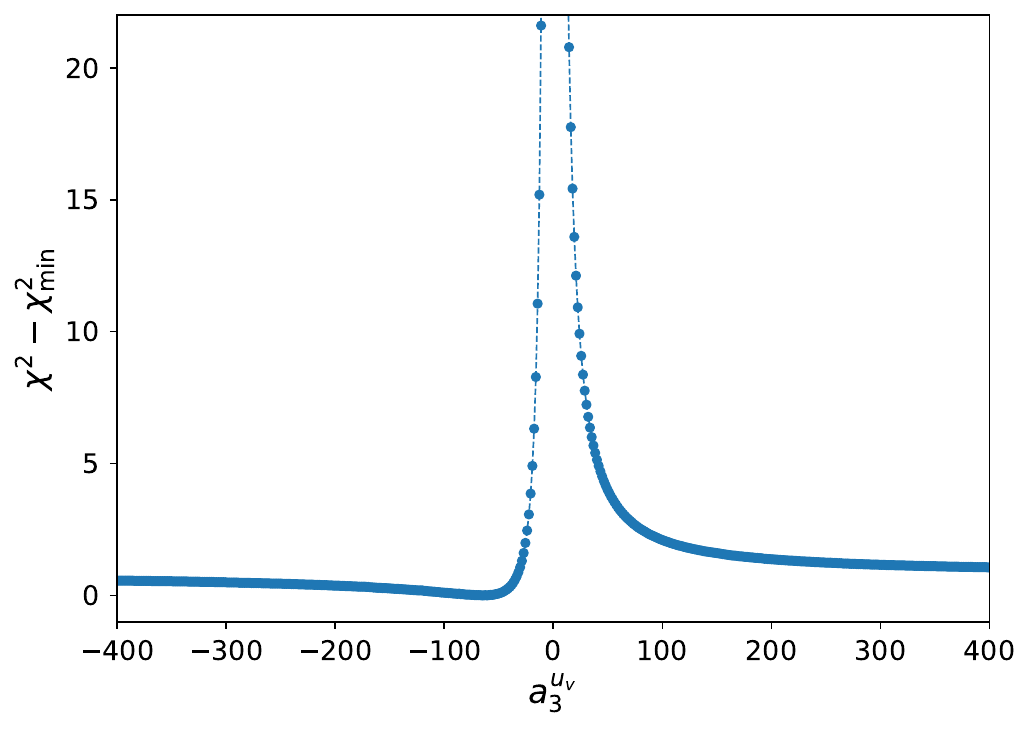}
    \caption{Scan of the $\chi^2$ profile as a function of the $a_3^{u_v}$ parameter for the \Pbonly{} fit. The $a_3^{u_v}$ parameter is varied while all other parameters are kept fixed at the MCMC global minimum. A pronounced peak at $a_3^{u_v}=0$ indicates this value is strongly disfavored, whereas the flat behavior elsewhere reflects the limited sensitivity of the fit to this parameter.}
    \label{fig:chi2Scan}
\end{figure}
In the Bayesian framework, the prior distribution encodes assumptions about model parameters before incorporating experimental data. In the present analysis, most free nPDF parameters are assigned non-informative flat priors, allowing the experimental data to determine the structure of the posterior distribution through the likelihood function. We intentionally avoid introducing additional prior information on the fitted parameters so that the results are driven as directly as possible by the data. At the same time, certain constraints are implicitly imposed through parameters that are not fitted, for example by enforcing momentum and number sum rules or by fixing specific baseline quantities. An exception arises in the \texttt{Pb-only} fit for the parameter $a_3^{u_v}$. During MCMC sampling, this parameter exhibits large drifts in value, reflecting the limited sensitivity of the available data in the intermediate-$x$ region. A dedicated $\chi^2$ scan, see Fig.~\ref{fig:chi2Scan}, shows a nearly flat profile over a broad interval, indicating that the likelihood does not meaningfully constrain its value. Additionally, the scan reveals a localized barrier near $a_3^{u_v}=0$ characterized by a sharp\footnote{The $\chi^2$ rises to values of several ten-thousands.} increase in $\chi^2$; this demonstrates that while the surrounding parameter space is nearly degenerate, specific values remain strongly disfavored by the data. A common approach for such weakly constrained parameters is to fix them to a specific value. While this eliminates the flat direction and, with it, the associated non-Gaussian correlations, it also reduces the available parameter space and constrains the flexibility of the fit. In contrast to other methods such as the Hessian, the MCMC framework is well suited to handle non-Gaussian structures in the posterior, allowing such correlations to be explored explicitly. For this reason, rather than fixing $a_3^{u_v}$, we modify the target distribution by introducing a bounded uniform prior, $a_3^{u_v} \sim U(-300, 300)$, which regularizes the flat direction without artificially restricting the parameter space. This choice stabilizes the sampling while preserving the ability of the data to determine the posterior wherever meaningful constraints are present.
%
\section{Methodology }
\label{sec:method}
Building on the Bayesian framework introduced in the previous section, this part outlines the methodology used to perform the nuclear PDF fits. We begin by describing the setup of the global analysis, including the experimental datasets and their corresponding theoretical predictions, followed by the configuration of the MCMC sampling, the diagnostics used to assess chain convergence, and the procedures for estimating uncertainties from the posterior samples.
\subsection{Nuclear Fit Setup} \label{subsec:3.1}
We perform two complementary global analyses. The \Pbonly{} analysis, as the primary fit, includes data exclusively from lead nuclei ($A=208$) and is used for the full MCMC study. A secondary \multinuc{} analysis incorporates data from a wide range of nuclear targets. The complete list of datasets for the \multinuc{} analysis is provided in Appendix~\ref{appen:A}, while the \Pbonly{} datasets are discussed below. The \Pbonly{} fit includes three categories of experimental data: vector boson production, heavy-quark production, and neutrino deep inelastic scattering (DIS). Each of these processes probes distinct aspects of the partonic structure of lead nuclei and they collectively provide complementary constraints on quark and gluon distributions.
Among the \Pbonly{} datasets, $W^{\pm}$ and $Z$ vector boson production in proton–lead collisions at the LHC provide one of the most direct and precise probes of nuclear parton distributions. Since the electroweak bosons interact only weakly, their production cross sections are largely unaffected by final-state effects, offering a clean constraint on the quark and antiquark distributions, and through higher-order QCD corrections sensitivity to the gluon density. For more detailed information about vector boson production in proton-lead collisions at the LHC and their impacts on nuclear PDF fits, see \cite{Kusina:2020lyz,Kusina:2016fxy,Kusina:2012vh}. In this study, we include the \wz data sets exclusively from proton-lead collision measurements, from LHC Runs I and II \cite{Aad:2015gta, AtlasWpPb, Khachatryan:2015pzs, Khachatryan:2015hha, CMS:2019leu, Aaij:2014pvu, ALICE:2016rzo}, as summarized in Table~\ref{tab:WZdata} in App.~\ref{appen:A}. Theoretical predictions are calculated at next-to-leading order (NLO) in QCD using MCFM 6.8~\cite{Campbell:2015qma}, interfaced with APPLgrid~\cite{Carli:2010rw} to enable fast numerical evaluation during the MCMC sampling. Since these data lie in a fully perturbative region and only rapidity spectra are considered, resummation effects are negligible, and no additional kinematic cuts are applied.

Heavy-quark production in proton–lead (pPb) collisions provides complementary sensitivity to the nuclear gluon distribution, particularly in the small-$x$ region ($x< 0.01$). We include measurements of open heavy-flavor mesons (e.g., $D^0$) and heavy quarkonia ($J/\Psi$, $\Upsilon$) production from ALICE, LHCb, ATLAS, and CMS \cite{ALICE:2013snh, LHCb:2013gmv, ALICE:2014xjz, ALICE:2014cgk, ALICE:2014ict, ALICE:2015sru, ATLAS:2015mpz, ALICE:2016yta, ATLAS:2017prf, CMS:2017exb, LHCb:2017yua, LHCb:2017ygo, CMS:2018gbb, ALICE:2019fhe, ALICE:2020vjy}, summarized in Table~\ref{tab:HQdata} in App.~\ref{appen:A}.
To ensure computational efficiency, we use a data-driven Crystal Ball parameterization that assumes gluon dominance and is calibrated to proton–proton reference data~\cite{Kusina:2017gkz,Kusina:2020dki,Duwentaster:2021ioo}. This fast approach accurately reproduces~\cite{Duwentaster:2021ioo,ncteq25} QCD-based results from GM-VFNS for open heavy-flavor mesons~\cite{Kniehl:2004fy,Kniehl:2005mk} and NRQCD for quarkonia~\cite{Butenschoen:2010rq,Bodwin:1994jh}. Further details of the theoretical treatment and validation procedure can be found in the dedicated \ncteqHQ heavy-flavor analysis~\cite{Duwentaster:2021ioo}. In our analysis, we apply the following kinematic selections:
\begin{equation*}
  P_T > 3.0 \, \text{GeV}\ , \qquad
  -4.1<y<4.1\ .
\end{equation*}

The charged-current (CC) neutrino DIS data on lead from the CHORUS experiment~\cite{Onengut:2005kv} provide essential flavor sensitivity allowing to distinguish between the light quarks. They are summarized in Table~\ref{tab:nudata}. The corresponding theory predictions are computed using the massive SACOT-$\chi^2$ scheme~\cite{Aivazis:1993kh, Aivazis:1993pi} at NLO, implemented via the \apfel library~\cite{Bertone:2017gds, Risse:2025smp}.
To ensure the validity of the used perturbative calculations the following kinematic cuts are applied:
\begin{equation*}
  Q^2 > 4.0 \, \text{GeV}, \qquad
  W^2 > 12.25 \, \text{GeV}\,.
\end{equation*}

The nuclear PDFs are parameterized following the functional form defined in~\cref{eq.1,eq.2,eq.3,eq.4,eq.5}. In this study, we fit 10 $a_j$ parameters: three ($a_1$, $a_2$ and $a_3$) for the $u$- and $d$-valence distributions, and two ($a_1$ and $a_2$) for the $\bar{d}+\bar{u}$ and gluon distributions. For reasons of limited constraining power of nuclear data and in order to keep computational complexity manageable, we choose to fit 10 parameters.
The $p_j$ proton parameters are fixed to the CJ values~\cite{Accardi:2016qay}, while the remaining nuclear parameters ($b_j$ and the remaining $a_j$) are set to zero and fixed. 
Although the \Pbonly{} fit does not exploit the $A$ dependence of the parameterization, we chose to use the same functional form for both \Pbonly{} and \multinuc{} fits. This allows for easy comparison of the two fits and maintains the same interpretation of the parameters.

\subsection{Sampling Setup} \label{subsec:3.2}
To sample the posterior distribution of the nuclear PDFs, we employ an MCMC approach using the Adaptive Metropolis–Hastings algorithm described in \cref{sec:algorithm}. Multiple independent chains were generated in parallel allowing faster accumulation of sufficient statistics and ensuring robust exploration of the parameter space while enabling convergence diagnostics.
Each MCMC chain is initialized by perturbing its parameters within $\pm 20\%$ of the Hessian best-fit value obtained from a preliminary minimization. This ensures that sampling begins in a physically meaningful region while still allowing sufficient freedom for unbiased posterior exploration. Independent random seeds are assigned to all chains to maintain statistical independence. The initial proposal distribution is defined by a fixed diagonal covariance matrix, $C_0=\operatorname{diag}(0.001 \times \mathbf{1})$, where $\mathbf{1}$ is a vector of ones with dimensionality equal to the number of fit parameters. For the first $N_\text{pre}=5000$ iterations, the proposal covariance remains fixed. After this pre-adaptation phase, the aMH algorithm updates the proposal covariance dynamically using the evolving sample history. To minimize potential biases from early samples, the adaptive covariance is reset at 10\,000 and 20\,000 steps. The adaptation scaling factor is set to $\beta=0.1$, providing a gradual and ergodic adjustment of the proposal distribution.

For the \Pbonly{} fit, the sampling procedure is computationally efficient, producing about 20\,000 MCMC iterations per day on a single CPU core. In contrast, extending the analysis to multiple nuclei substantially increases computational cost. Each additional nucleus requires initialization of a dedicated high-speed lookup grid encoding its specific nuclear cross section, which makes every $\chi^2$ evaluation slower and reduces the overall sampling rate. Consequently, the \multinuc{} fit generates only about 10\,000 samples per week. Given this limitation, the \multinuc{} analysis is treated as complementary and exploratory, while the \Pbonly{} fit remains the central result of this work and is given more attention.

\subsection{MCMC Diagnostics and Processing}
After generating the MCMC chains with the aMH algorithm, we perform diagnostic tests to verify that the samples accurately represent the target posterior distribution. These diagnostics assess stationarity, convergence, and sampling efficiency, and ensure that the resulting ensemble provides reliable estimates of the nuclear PDF parameters and their uncertainties. The main components of this procedure are burn-in removal, convergence assessment, autocorrelation analysis, and thinning.

\subsubsection{Thermalization and Convergence}
Markov chains require an initial set of iterations to approach the stationary distribution that represents the target posterior. This initial segment, referred to as the \emph{burn-in} or \emph{thermalization} phase, is influenced by the starting values and must be discarded to avoid bias in the final analysis. \emph{Convergence} denotes the stage at which the chain produces samples representative of the stationary distribution. In practice, convergence cannot be theoretically proven for a finite chain; the diagnostics therefore provide heuristic but informative criteria. If convergence is not achieved, the chain may retain memory of its initialization, mix poorly, or remain confined to local modes, resulting in biased parameter estimates. We assess convergence using several complementary diagnostics. Trace plots are inspected to identify long-term drifts, poor mixing, or mode trapping. Cumulative mean plots monitor the stability of running averages and provide an indication of stationarity. Stability of cumulative means alone is insufficient to establish convergence and must be interpreted together with trace behavior and additional diagnostics. Since no single test is definitive, convergence is evaluated through a combination of these criteria. For further discussion of MCMC convergence diagnostics, see Refs. \cite{Hogg_2018, brooks2011handbook}.

\subsubsection{Autocorrelation and Thinning}
\emph{Autocorrelation} in MCMC quantifies the statistical dependence between successive samples of a Markov chain. Because the chain is constructed sequentially, neighboring samples are correlated, reducing the effective amount of independent information and impacting statistical precision.
For independent samples, the Monte Carlo standard error is
\begin{equation}
    \sigma_{\text{uncor}} = \frac{S}{\sqrt{N}}\,,
\end{equation}
where $N$ is the total number of independent samples and
\begin{equation}
    S^2 =  \frac{1}{N-1} \sum_{i=1}^{N} (\bm{X}_i - \bar{\bm{X}})^2\,.
    \label{eq.17}
\end{equation}
is the sample variance with mean $\bar{\bm{X}}$. In MCMC, however, correlations must be explicitly accounted for. The autocorrelation function (ACF),
\begin{equation}
\rho(k) = \frac{\Gamma(k)}{\Gamma(0)}, \qquad
\Gamma(k) = \text{Cov}(\bm{X}_i, \bm{X}_{i+k})\,,
\end{equation}
measures the correlation between samples separated by lag $k$, with $\Gamma(0)=\text{Var}(\bm{X}_i)$. The decay rate of $\rho(k)$ characterizes the mixing efficiency of the chain.
The integrated autocorrelation time,
\begin{equation}
    \tau_{\text{int}} \approx \frac{1}{2} + \sum_{k=1}^\infty \rho(k)\,,
    \label{eq.19}
\end{equation}
represents the average number of steps required to produce one effectively independent sample. The effective sample size and corresponding standard error are
\begin{equation} 
N_{\text{eff}} = \frac{N}{2 \tau_{\text{int}}},\qquad \sigma_{\text{cor}} = \frac{S}{\sqrt{N_{\text{eff}}}}= S \sqrt{\frac{2 \tau_{\text{int}}}{N}}\,.
\end{equation}
In practice, estimating $\tau_{\text{int}}$ is non-trivial due to the infinite sum in Eq.~(\ref{eq.19}). We employ the $\Gamma$-method~\cite{Wolff:2003sm, Risse:2025qlo}, which determines an optimal cutoff $W_{\text{opt}}$ using an automatic windowing procedure that truncates the sum when correlations become negligible:
\begin{equation}
    \tau_{\text{int}} \approx \frac{1}{2} + \sum_{k=1}^{W_{\text{opt}}} \rho(k)\,.
\end{equation}

While the $\Gamma$-method provides reliable estimates of autocorrelation and $\tau_{\text{int}}$, we further apply a \emph{thinning} procedure \cite{link2012thinning, riabiz2022optimal} to obtain approximately independent samples. In this approach, every $\eta$-th point of the Markov chain is kept while the rest are discarded. The thinning rate $\eta$ is chosen such that the estimated $\tau_{\text{int}}$ approaches the uncorrelated limit $\tau_{\text{int}}=0.5$ (corresponding to $\rho(k) \to 0$ for all $k \geq 1$ according to Eq.~(\ref{eq.19})), ensuring that the remaining samples are effectively independent. Thinning is employed for two main reasons. Firstly, it reduces autocorrelation to a level where standard Monte Carlo error estimation applies, simplifying statistical analysis. Secondly, by reducing the number of samples while preserving their statistical representativeness, thinning ensures that the resulting PDFs are both computationally practical and user-friendly. Such reduced sample of nPDFs can then be stored in the form of LHAPDF files~\cite{Buckley:2014ana} that can be easily used in practical applications.

\subsection{Error Estimations}
\label{subsec:thin}
Accurate error estimation is a crucial step in the analysis of Markov chains, as it quantifies the uncertainties associated with estimated parameters and allows their propagation to observables. Since the chains have been thinned, the resulting samples can be treated as approximately uncorrelated, allowing the use of standard Monte Carlo error estimation techniques that assume sample independence. In this study, we consider three complementary methods to determine central values and confidence intervals (CIs) of the posterior distributions. These methods are formulated for a general confidence level $\alpha$. However, in this analysis, we will mostly adopt $\alpha = 68\%$ ($1 \sigma$).%
\footnote{Note that in our proton PDF MCMC study \cite{Risse:2025qlo}, we used the $90\%$ ($1.64 \sigma$) interval per default.}
Uncertainties are evaluated both for the fitted nuclear PDF parameters $\{a_k\}$ and for derived quantities such as the nuclear PDFs $f_i(x,Q^2)$ or any PDF-dependent observable. The three methods are:
\begin{itemize}
    \item The {\em Monte Carlo Standard Error} (MCSE) provides symmetric confidence intervals based on the sample mean and standard deviation. For a given observable $\mathcal{O}$, the central value is computed as
    \begin{equation}
        \mu_\mathcal{O} = \frac{1}{N} \sum_{n=1}^{N} \mathcal{O}^{(n)},
    \end{equation}
    where $\mathcal{O}^{(n)}$ denotes the value of $\mathcal{O}$ in the $n$-th MCMC sample and $N$ is the total number of samples. The sample standard deviation, $S_{\mathcal{O}}$, is given by Eq. (\ref{eq.17}) and the corresponding 68\% confidence interval is
    \begin{equation}
        \textbf{CI}_{68\%} = [\mu_{\mathcal{O}} - S_{\mathcal{O}},\,\, \mu_{\mathcal{O}} + S_{\mathcal{O}}].
    \end{equation}
    \item The {\em percentile} method, which is used e.g.~in~the nNNPDF3.0 PDF analysis~\cite{AbdulKhalek:2022fyi}, yields asymmetric $\alpha$\% confidence intervals. It is particularly suitable for skewed or non-Gaussian distributions. The central value of an observable $\mathcal{O}$ is defined as the median (50th percentile) of its sampled distribution,\footnote{In the proton PDF study using the MCMC approach \cite{Risse:2025qlo}, the central value in the percentile method was instead defined as the minimum-$\chi^2$ configuration.} and the confidence bands are given by the $(100-\alpha)/2$ and $(100+\alpha)/2$ percentiles. For 68\% CI
    \begin{equation}
        \textbf{CI}_{68\%} = [\mathcal{O}_{16th}, \mathcal{O}_{84th}].
    \end{equation}
    \item The {\em cumulative-$\chi^2$} method incorporates the goodness-of-fit metric $\chi^2$ directly into the uncertainty estimation~\cite{Putze:2008ps,Risse:2025qlo}. The central value corresponds to the minimum-$\chi^2$ sample, $\chi^2_{\text{min}}$.
    The 68\% confidence interval is determined by the range of samples that fall within the 68\% percentile of $\chi^2$ distribution, satisfying
    \begin{equation}
        \textbf{CI}_{68\%} = \{ x| \chi^2(x) - \chi^2_{\text{min}} < \Delta \chi^2_{68\%} \},
    \end{equation}
    where $\Delta \chi^2_{68\%}$ is determined from the 68\% cumulative probability of the $\chi^2$ distribution, i.e., it retains the 68\% of samples with the lowest $\chi^2$ values.
\end{itemize}
The percentile and \cumchi{} methods define conceptually different 68\% credible regions, corresponding to local and global constraints in parameter space. The percentile method determines the uncertainty of each observable (e.g., a PDF value at fixed $x$) directly from the marginal posterior distribution at that point, independent of the global goodness-of-fit of individual samples. In contrast, the \cumchi{} method enforces a global constraint by retaining only parameter configurations within the 68\% probability volume of the $\chi^2$ distribution. As a result, the \cumchi{} approach yields uncertainty bands that are globally consistent with the fit quality, whereas the percentile method provides pointwise intervals reflecting the local posterior spread.
%
\section{Results and Discussion}
\label{sec:res}
\begin{figure*}[ht!]
\centering
\includegraphics[width=\textwidth]{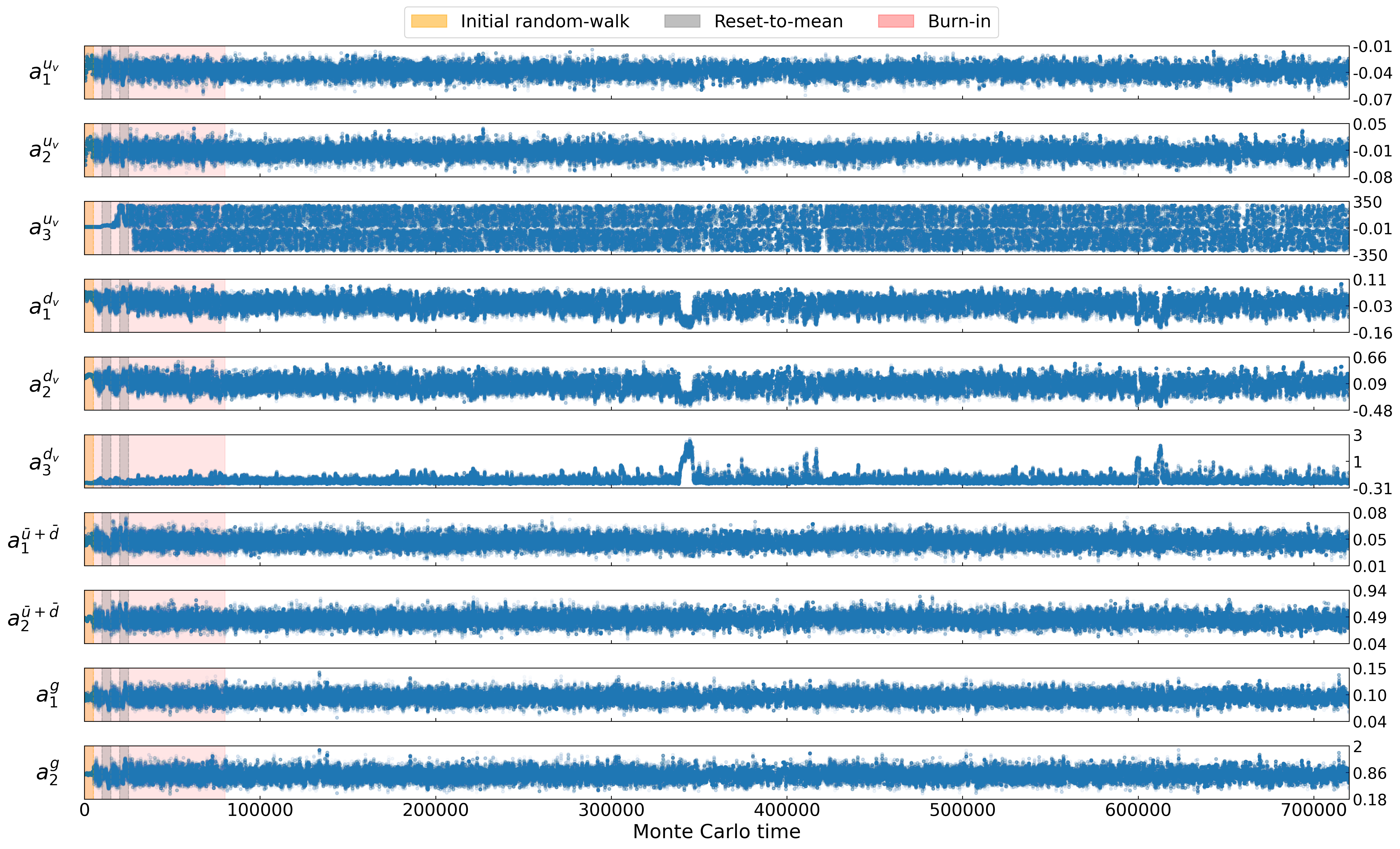}
\caption{The time series of the ten nPDF parameter values for the \Pbonly{} analysis, shown for an example chain. The red-shaded region marks the thermalization (burn-in) phase, approximately the first 80\,000 samples, which are removed for further analysis. The yellow region indicates the part of the chain where the standard random walk MH algorithm with a fixed covariance matrix is used, and the gray areas show the Monte Carlo times where the reset-to-mean mechanism was used.}
\label{fig:timeseries} 
\end{figure*}

\begin{figure*}[ht!]
\centering
\includegraphics[width=\textwidth]{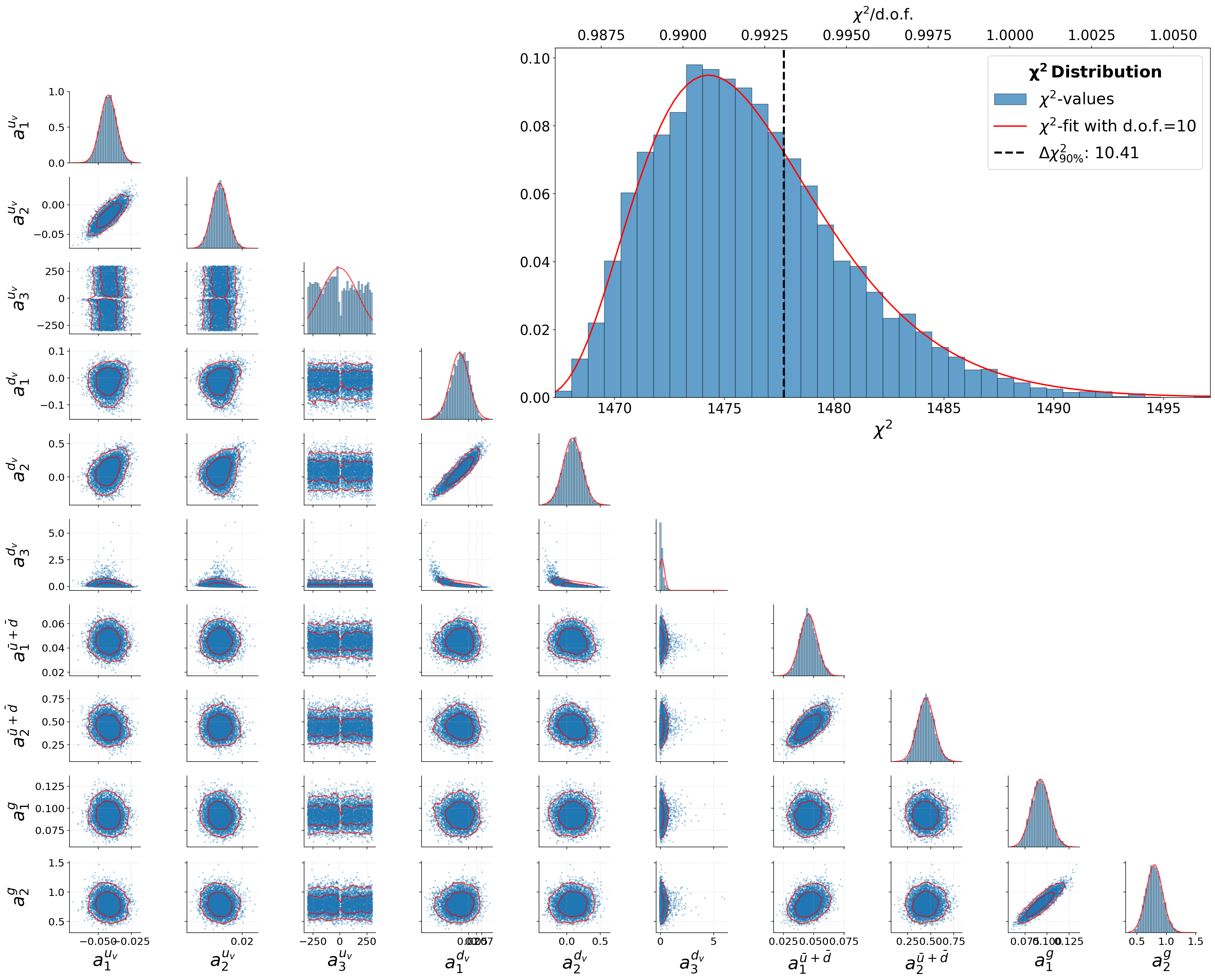}
\caption{Pairwise plot of the \Pbonly{} posterior distributions, showing one-dimensional marginals (diagonal panels) and two-dimensional projections of pairwise correlations (off-diagonal panels). Estimates of credible regions are indicated by 68\% and 95\% contours. The top-right panel displays the $\chi^2$ distribution, which is fitted by the theoretical $\chi^2$ distribution (red line).}
\label{fig:pairwise} 
\end{figure*}

This section presents our results with emphasis on the MCMC analysis within the \Pbonly{} setup. We begin with a brief summary of the sampling performance to demonstrate the stability of the algorithm. We then discuss the posterior distributions of the fitted parameters and the resulting nuclear PDFs and their uncertainties.
Lastly we also present selected results of the \multinuc{} analysis and compare them with those of the \Pbonly{} fit.

\subsection{Sampling Performance }
Several independent Markov chains were generated for the \Pbonly{} setup using the Adaptive Metropolis–Hastings algorithm described in Sec.~\ref{sec:algorithm}. After applying standard convergence diagnostics (details are provided in Appendix~\ref{appen:B}), 11 chains were retained that had consistent stationary behavior and satisfied all convergence criteria. An example chain is presented in Fig.~\ref{fig:timeseries} as a time series, plotting parameter values as a function of the Monte Carlo time. Chains that failed the convergence tests typically showed large $\chi^2$ values (well above the Hessian minimum) and very low acceptance rates, leading to long sequences of repeated states and poor exploration of the parameter space.
For each converged chain, the initial thermalization (burn-in) phase was removed; its length varied slightly between chains but typically ranged from 60\,000 to 80\,000 samples. Following burn-in removal, the converged chains collectively contained approximately 3.8 million samples (including accepted and rejected states). To remove autocorrelations and ensure statistical independence, an appropriate thinning factor $\eta$ was applied to each chain. The optimal thinning rate, chosen to yield an integrated autocorrelation time of $\tau_\text{int}=0.5$ for all parameters, varied between $\eta=400$ and $\eta=600$. The resulting uncorrelated samples were then merged into a single comprehensive ensemble containing $N_{tot}= 7069$ statistically independent points.

\subsection{Posterior of the Combined Chain}
After applying burn-in removal and thinning, all 11 converged \Pbonly{} chains were merged into a unified posterior ensemble representing the full Bayesian distribution of the fitted parameters. This combined dataset forms the statistical foundation for all subsequent analyses, including posterior estimation and uncertainty quantification of the nuclear PDFs. To visualize the structure of this posterior, Fig.~\ref{fig:pairwise} presents a corner (pairwise) plot for all ten fitted nPDF parameters. 
The diagonal panels show the one-dimensional marginal distributions of PDF parameters, and the off-diagonal panels display the two-dimensional marginal distributions showing correlations between the parameter pairs. Additionally, to make the plots easier to read, in the diagonal plots we include a fitted Gaussian distribution in red, and in the two-dimensional correlation plots we provide contours indicating estimates of the 68\% and 95\% probability density regions. Finally, in the upper part of Fig.~\ref{fig:pairwise} we display the histogram of $\chi^2$ distribution together with the fitted analytic $\chi^2$ function, again in red.

We start by examining the $\chi^2$ distribution. We can see that the obtained histogram closely follows the analytic curve (in red) given by the $\chi^2$ probability density function with 10 degrees of freedom (corresponding to the 10 fitted parameters), validating the statistical consistency of the model and the experimental uncertainties. The 68\% confidence interval for the joint uncertainty of the ten parameters is computed based on the samples and yields \mbox{$\Delta \chi^2_{68\%}=10.41$}, which is slightly smaller than the value resulting from the analytic $\chi^2$ cumulative distribution function 11.50. This computed value is adopted as the tolerance criterion in the corresponding Hessian fit (used later for comparisons), ensuring a consistent treatment of uncertainties between the MCMC and Hessian approaches.

In the next step we discuss the marginal distributions obtained for the nPDF parameters.
Focusing on the sea-quark and gluon parameters ($a_1^{\bar{u}+\bar{d}}$, $a_2^{\bar{u}+\bar{d}}$ and $a_1^g$, $a_2^g$) first we find the one- and two-dimensional distributions to be very close to a Gaussian. Apart from the positive correlation (the tilt of the ellipses) between the $a_1$ and $a_2$ parameters for each flavor respectively ($a_1^{\bar{u}+\bar{d}}$ vs.\,$a_2^{\bar{u}+\bar{d}}$ and $a_1^g$ vs.\,$a_2^g$), no significant correlation to the remaining parameters is found. Importantly there is no correlation of the gluon and sea parameters to the parameters of the valence PDFs. Thus we expect the PDF uncertainties for the sea quarks and the gluon to be entirely Gaussian distributed. This result is not unexpected since most of the constraints from the data are on the gluon (c.f.~\cref{tab:WZdata,tab:HQdata}) and the sea quarks (c.f.~\cref{tab:nudata}) with only a limited sensitivity on the valence quarks. Stronger constraints usually yield posterior distributions close to a Gaussian.
And in fact, the distributions for the less constrained valence quarks show significant deviations from it, especially the $a_3$ parameters.  
Already from the one-dimensional marginal distributions, we can see that $a_3^{d_v}$ yields a long tail towards larger values and $a_3^{u_v}$ seems to follow a flat distribution, except for values close to zero, where no samples are found. The absence of samples with $a_3^{u_v}$ in the immediate vicinity of zero hints at the fact that the combined chain samples a minimum with a very complicated shape or even more than one minimum, a subject matter we explore in \cref{sec:Multiple_minima}.

\begin{figure*}[htb!]
\centering
\includegraphics[width=0.49\textwidth]{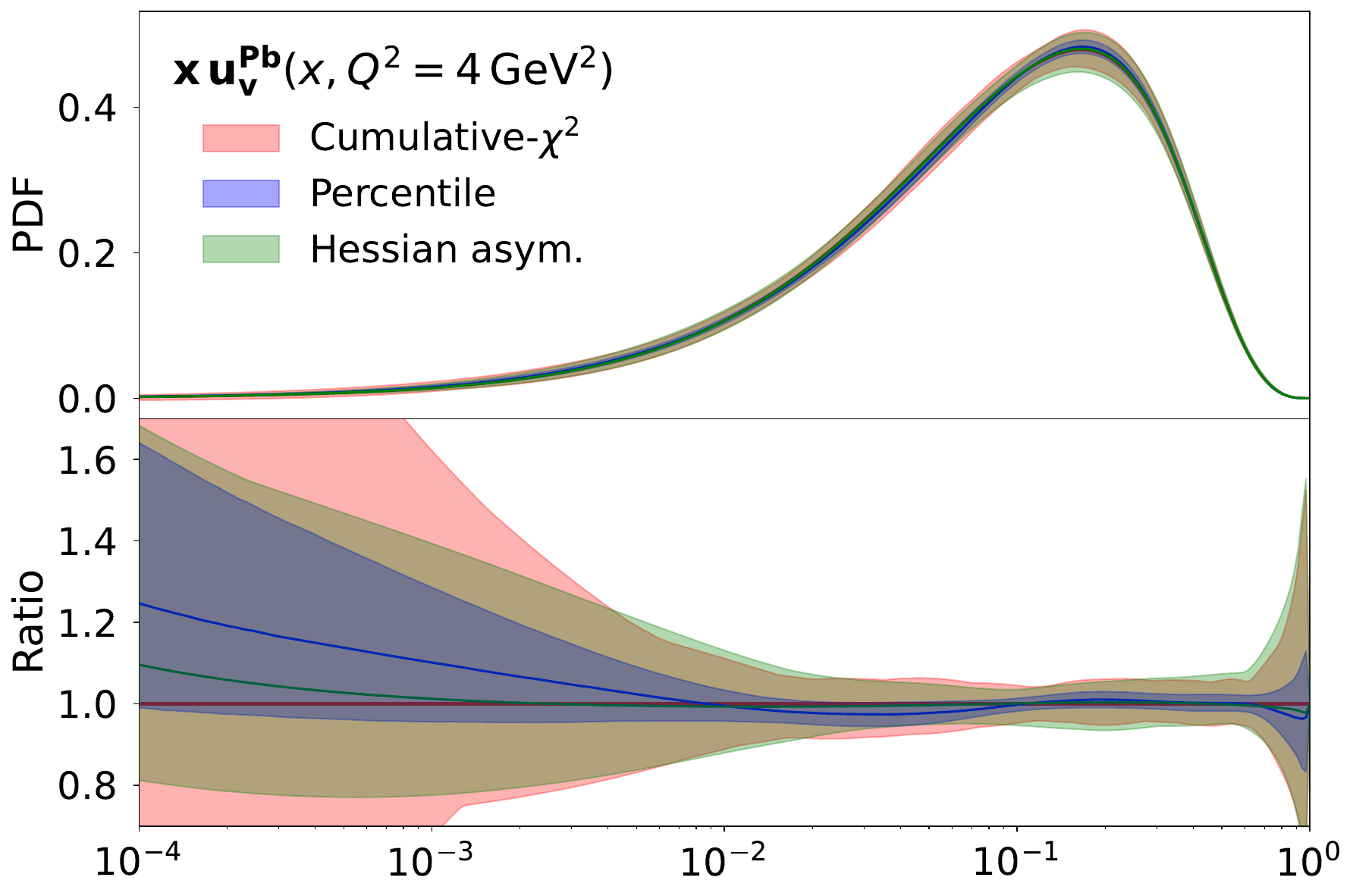}
\includegraphics[width=0.49\textwidth]{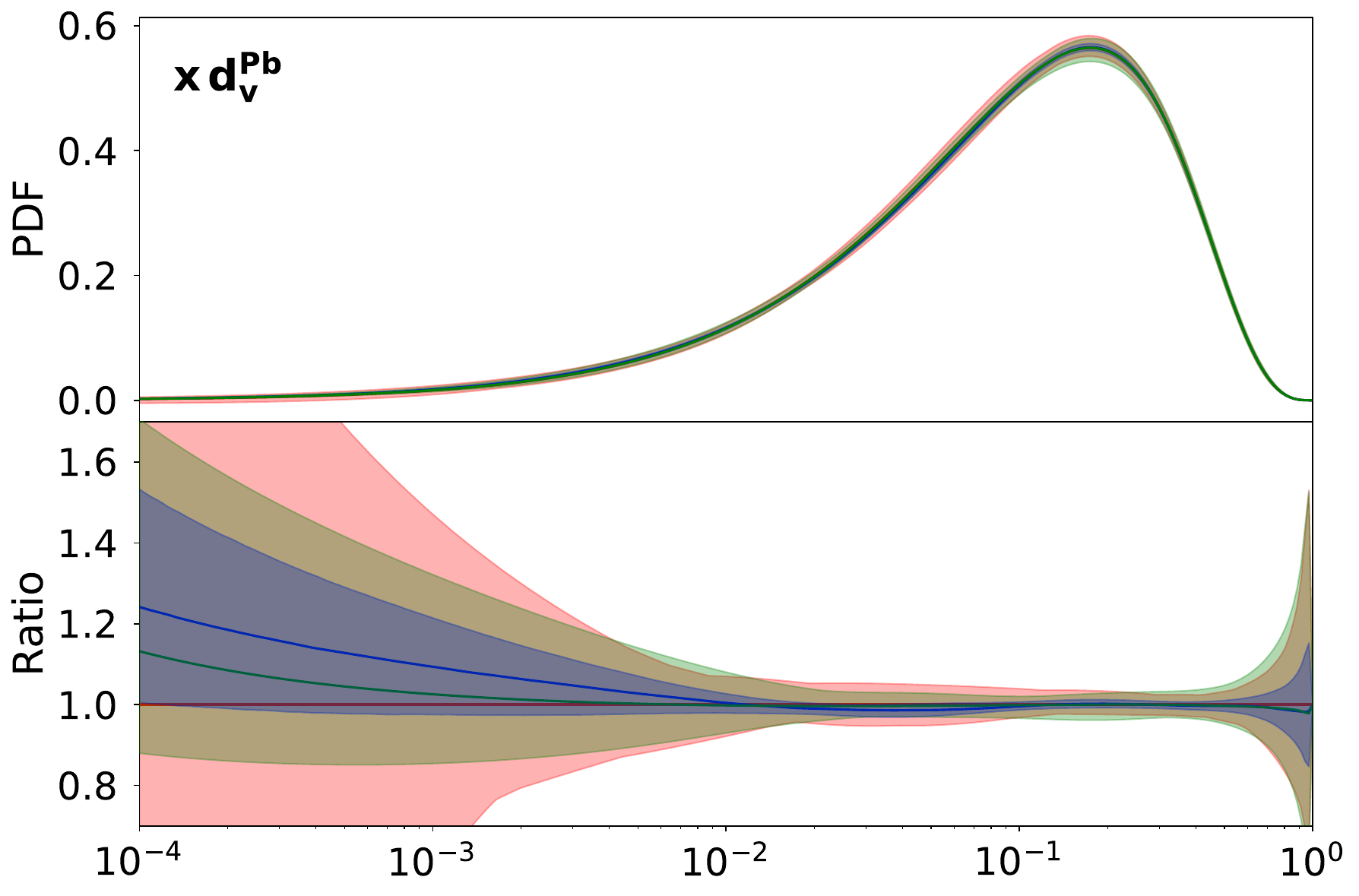}
\includegraphics[width=0.49\textwidth]{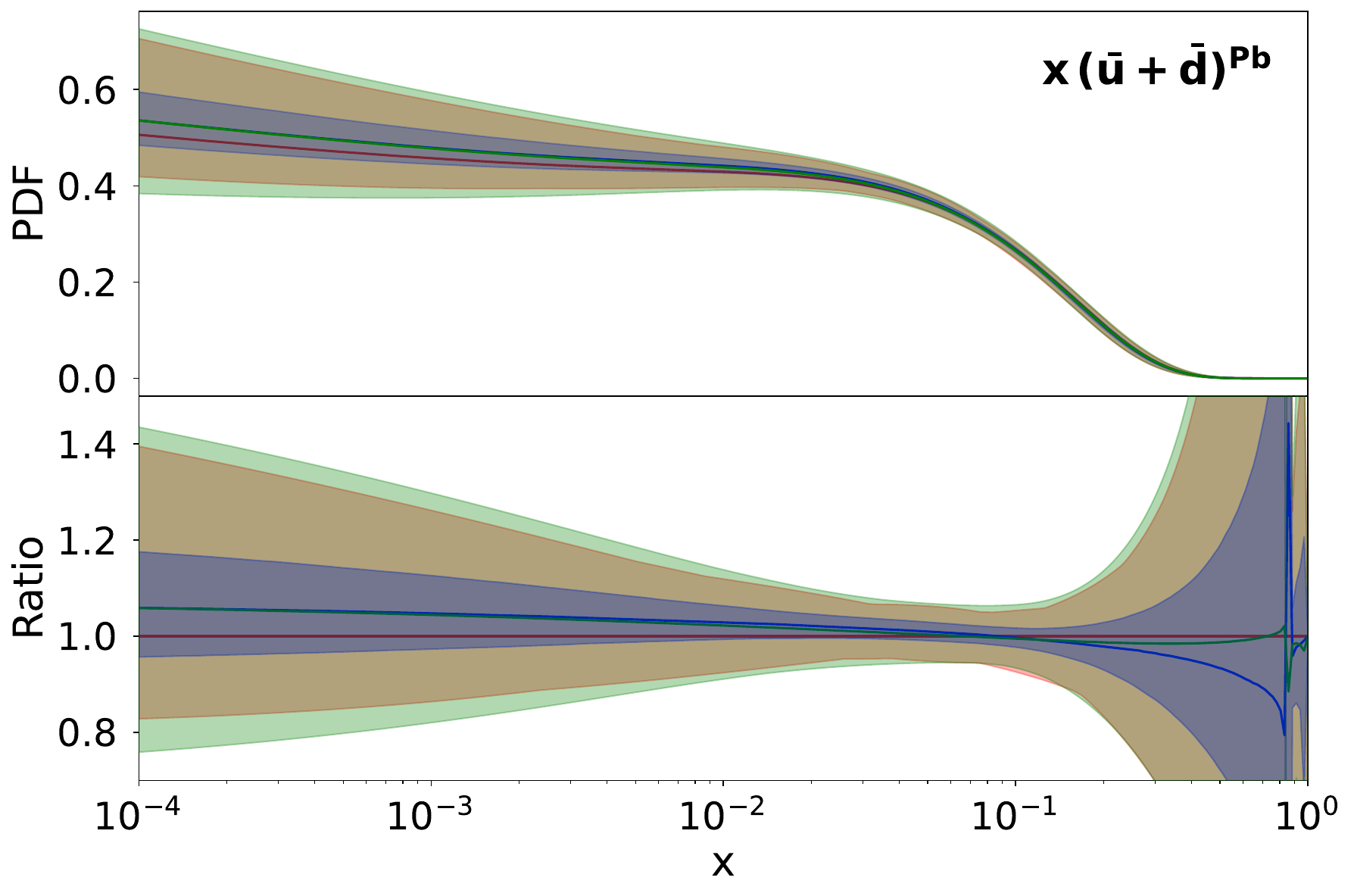}
\includegraphics[width=0.49\textwidth]{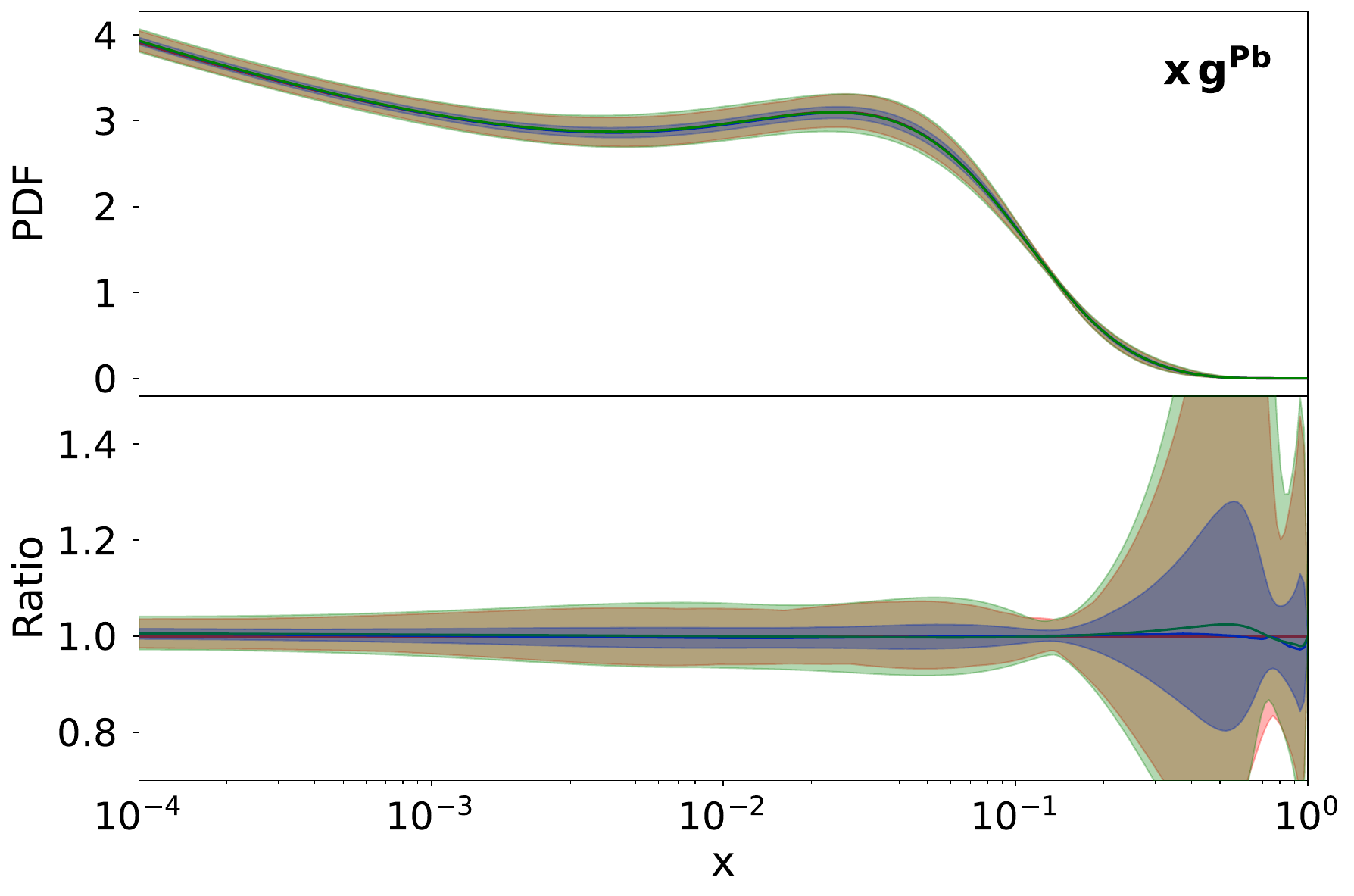}
\caption{Comparison of lead PDFs uncertainty estimations for $u_v$, $d_v$, $\bar{u}+\bar{d}$, and $g$ distributions for the \Pbonly{} analysis. The upper panels show the absolute nPDFs, while the lower panels display the ratios with respect to the best-fit values obtained from the \cumchi{} method. The colored bands correspond to different uncertainty estimation methods: red for the MCMC \cumchi{}, blue for the MCMC percentile, and green for the asymmetric Hessian.}
\label{fig:PDFcompPb-MCMChess} 
\end{figure*}

Finally, we note that all four of the respective pairwise correlations between the first two up- and down-valence parameters have a distinct triangular shape. This is a feature noticed also in our earlier nuclear PDF extractions, although it was not explicitly discussed. The correlations between up- and down-valence PDFs are stronger compared to the free proton (or neutron) simply because the lead nucleus contains both, and the imposed isospin symmetry relates the up PDF of the (bound) proton with the down PDF of the (bound) neutron ultimately blurring the difference between the up and down PDFs.

To summarize, we find that the combined chain samples a vastly complex $\chi^2$ landscape, yielding insight into the correlations and interplay between the parameters that no other diagnostic tool so far has been able to provide. The second non-trivial result is that the sea-quark and gluon distributions are unaffected by this and show simple Gaussian behavior.

\subsection{Nuclear PDFs and Uncertainty Bands}

Using the combined posterior ensemble, the nuclear PDFs for Pb are evaluated together with their corresponding uncertainty bands. As discussed in Sec.~\ref{subsec:thin}, we considered three methods for computing their uncertainties based on the generated samples; however, we display results only for two of them, the percentile and the cumulative-$\chi^2$. We leave out the MCSE which is extended by the percentile method to be applicable for situations like ours where the distributions are skewed.

In Fig.~\ref{fig:PDFcompPb-MCMChess} we display the lead PDFs with their uncertainty bands obtained using the percentile and \cumchi{} methods as well as the corresponding results obtained using the Hessian method.
First, we compare the two results from MCMC analysis.
As expected, the ``central'' values are different because the \cumchi{} method uses the sample with the lowest $\chi^2$ for this purpose, whereas in the percentile method it is given by the median (50th percentile). Second, we can see that the error bands are quite different both in size and shape. This is also expected and was already discussed in Sec.~\ref{subsec:thin} and our earlier work~\cite{Risse:2025qlo}. In what follows, we will concentrate on the more conservative estimates given by the \cumchi{} method which we can match to the Hessian approach by using the same $\Delta \chi^2_{68\%}$ tolerance~\cite{Risse:2025qlo}.\footnote{Note that for the Hessian method, a tolerance of \mbox{$\Delta \chi^2=1$} corresponds to a 68\% CI based on the distribution of a given one-dimensional observable, thereby matching the interpretation of the MCSE or the percentile method. If we base the $\Delta \chi^2$ tolerance on the distribution of the $\chi^2$ values, we can match the interpretation of the Hessian uncertainties to the interpretation of the \cumchi{} method. In a statistically idealistic scenario this is equivalent to applying Wilks' theorem~\cite{Wilks:1938dza} to define the Hessian tolerance.}

The \cumchi{} and the Hessian methods yield rather consistent central values across all parton flavors and $x$ regions with only small deviations at low $x$ for valence and sea quark distributions. Both methods use their ``best fit'' estimate as their central PDF, but since it is unlikely to exactly land in the minimum in a ten dimensional random walk, the central value of the MCMC method is expected to be close to the Hessian but not an exact match.
However, noticeable differences appear in the size and shape of the uncertainty bands, reflecting the distinct statistical assumptions underlying each method.

It should be noted that the size of the Hessian uncertainty bands is determined using the tolerance criterion of $\Delta \chi^2_{68\%}=10.41$, extracted from the MCMC $\chi^2$ distribution (see Fig.~\ref{fig:pairwise}). This ensures a consistent definition of the 68\% confidence level between the two approaches, which is manifest when comparing the uncertainties for the gluon distribution. In this case the uncertainties obtained from the two methods are very consistent throughout the entire $x$ range, since the PDF parameters exhibit Gaussian behavior.
In the case of light sea quark distributions ($\bar{u}+\bar{d}$), for which the parameter distributions exhibit only small deviations from Gaussianity, the Hessian error bands slightly overshoot the uncertainties, making them also more symmetric than the more realistic MCMC result.
Finally, for the valence-quark distributions, $u_v$ and $d_v$, the MCMC posterior exhibits clear deviations from Gaussian behavior. 
We find that the \cumchi{} method produces broader uncertainty bands. Specifically at low $x$ we observe a large asymmetry in the MCMC errors, and the uncertainty band is much larger than the one obtained from the Hessian method. The only way to capture the uncertainty in this region within the Hessian approach is to increase the tolerance, which has the undesired secondary effect of also inflating the uncertainty band for all other $x$ ranges and PDF flavors. Thus, providing a conservative estimate in one region would yield an overestimate in the others. To conclude, we can see that the MCMC estimate provides more accurate PDF uncertainties.

\subsection{Multiple Minima}\label{sec:Multiple_minima}

\begin{figure*}[ht]
\centering
\subfloat[Single chain: alternative regions in $d_v$ parameters ($N_{\text{blue}}/N_{\text{red}}=37.02$ and $L_{\text{blue}}/L_{\text{red}}= 42.57$, computed after removing the thermalization region and thinning)]{
\label{fig:timeseries_min_dv_oneChain}
\includegraphics[width=\textwidth]{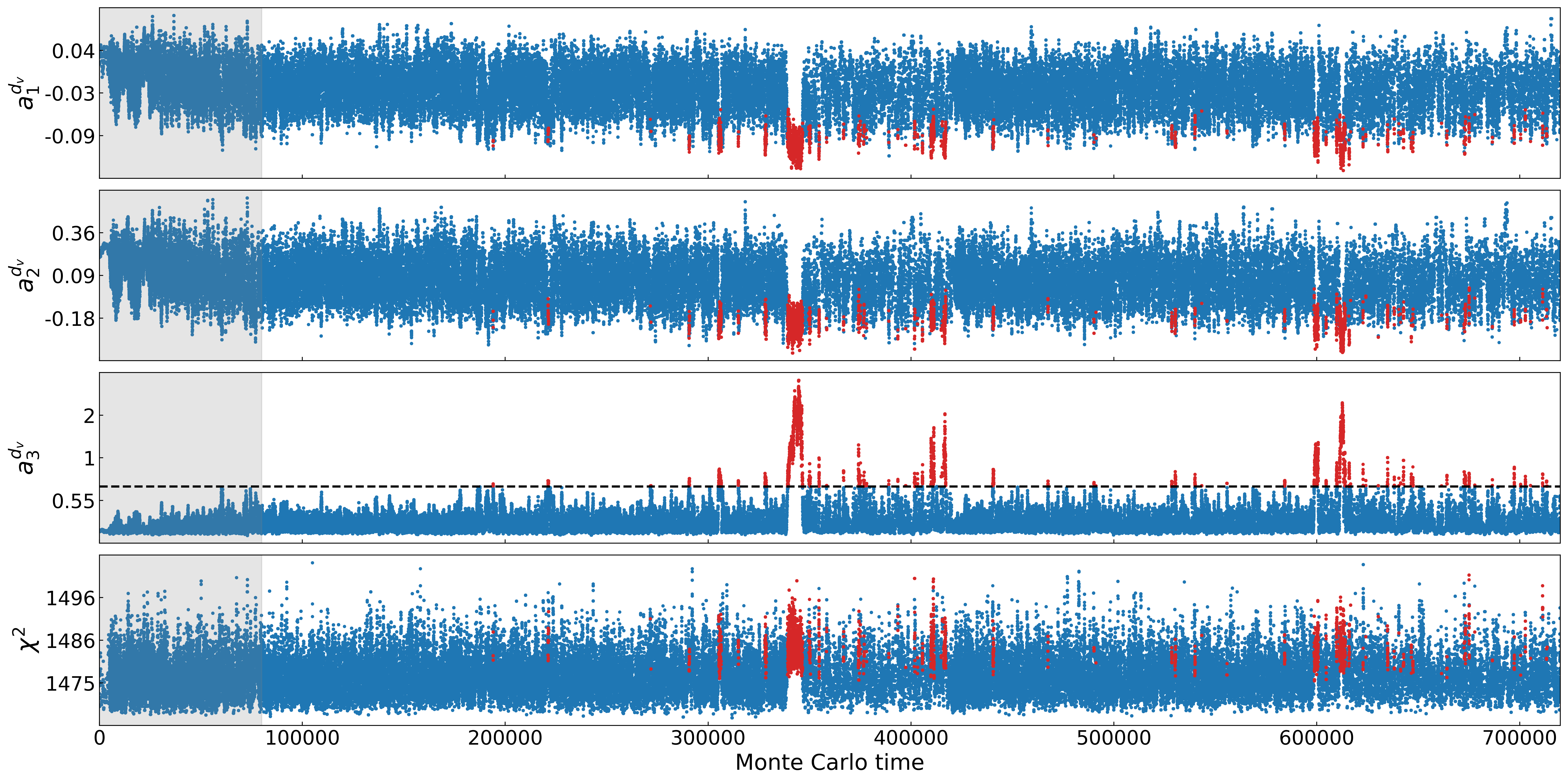}}\\
\subfloat[Combined chain: alternative regions in $d_v$ parameters]{
\label{fig:timeseries_min_dv_fullChain}
\includegraphics[width=\textwidth]{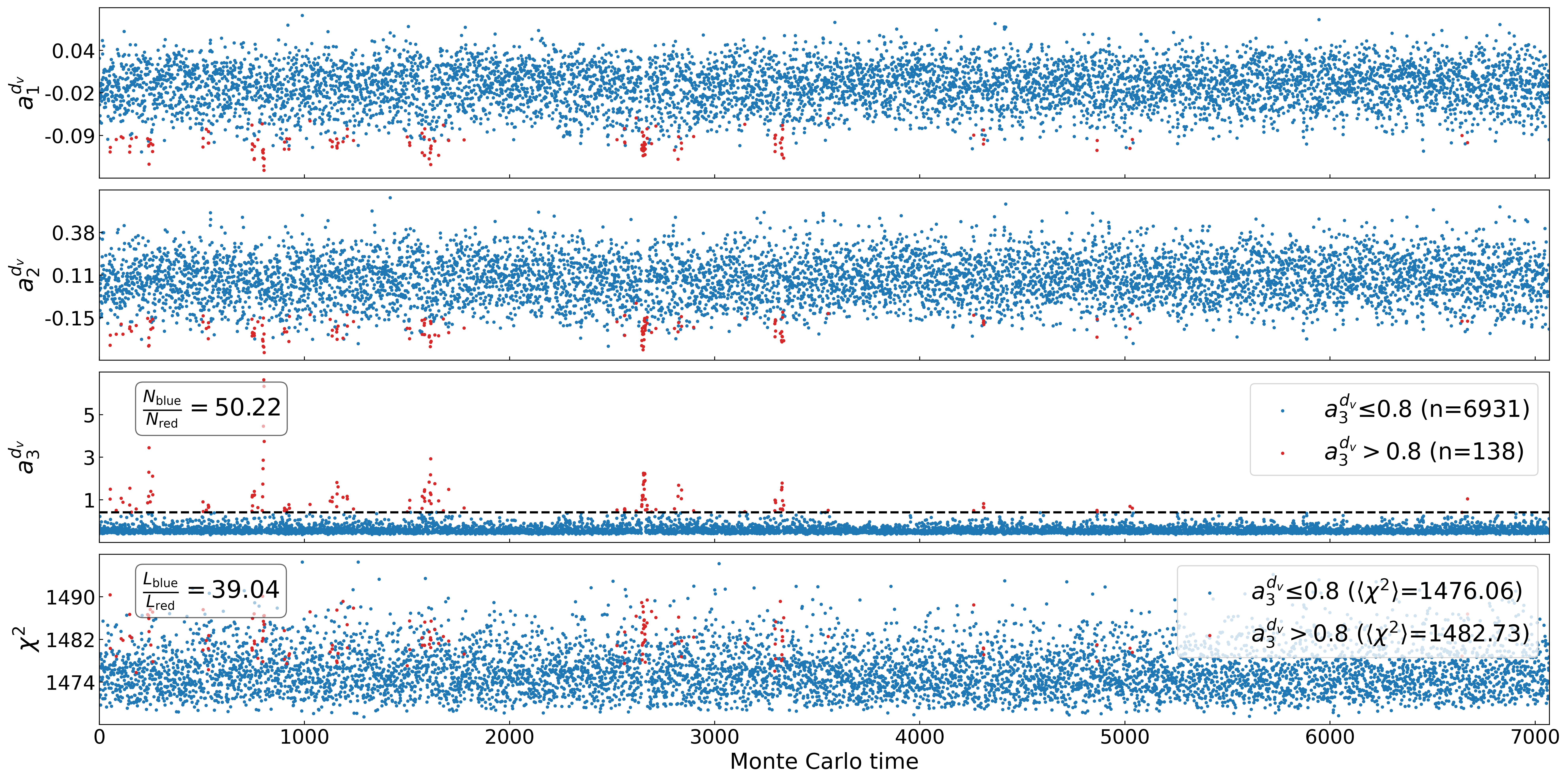}}
\caption{To investigate the presence of local minima in the MCMC exploration, we apply a threshold to the $a_3^{d_v}$ parameter, which exhibits a particularly prominent structure. Samples with $a_3^{d_v}>0.8$ are shown in red, while those with $a_3^{d_v} \le 0.8$ are shown in blue. We compute the mean $\chi^2$ and the corresponding likelihood $L \propto \exp{ (- \frac{1}{2} \chi^2)} $ separately for both regions. The upper plot (a) shows one of the analyzed chains before thinning, and the lower plot (b) shows the same for the final (thinned and combined) chain.}
\label{fig:timeseries_min_dv}
\end{figure*}

\begin{figure*}[hpbt!]
\centering
\includegraphics[width=\textwidth]{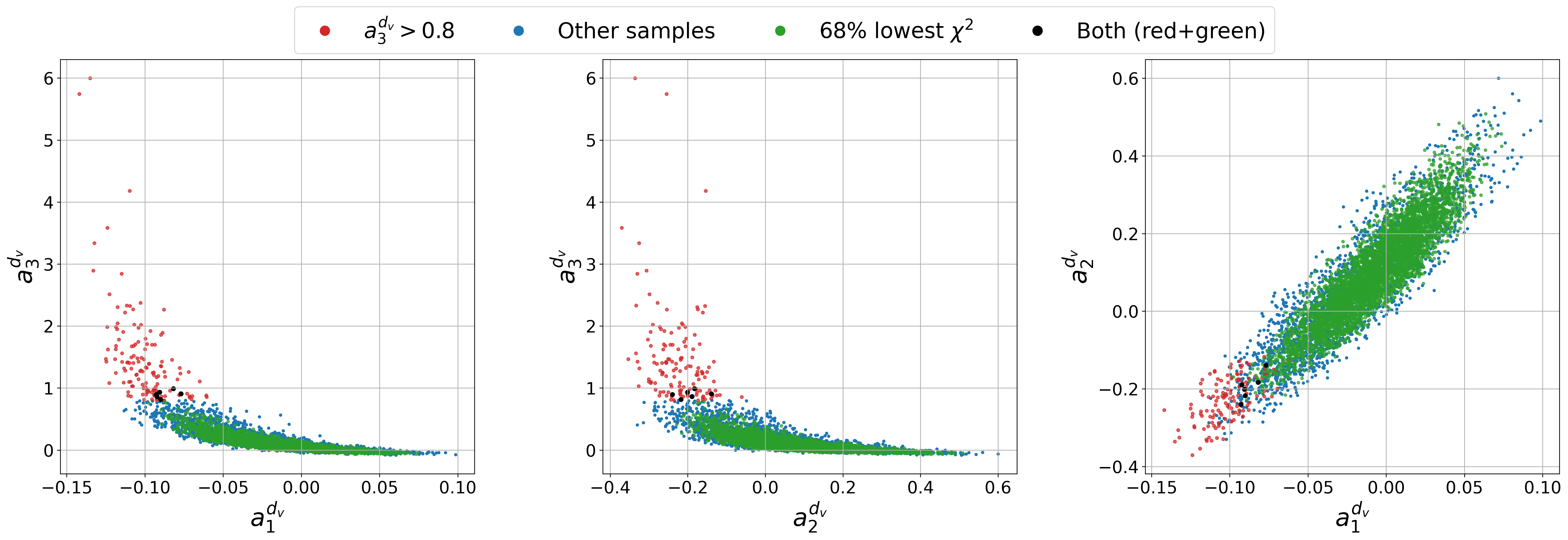}
\includegraphics[width=\textwidth]{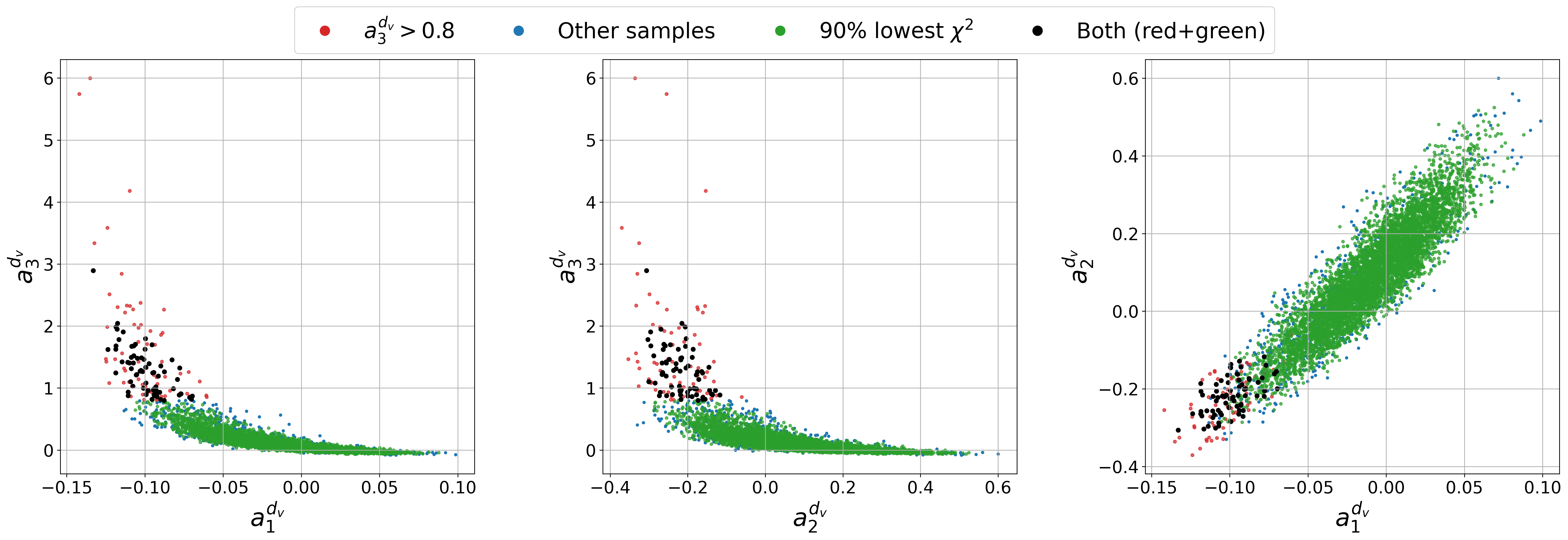}
\caption{Pairwise scatter plots of the $d_v$ parameters for two confidence levels: top: 68\% CL, bottom: 90\% CL. Samples are split into two sets using a threshold in the $a_3^{d_v}$ parameter, $v=0.8$. In red we show points with $a_3^{d_v} > v$, in green points with the lowest 68\% (90\%) in $\chi^2$, and in black points where both selections overlap. The remaining points are plotted in blue.}
\label{fig:dv68CLvs90CL}
\end{figure*}

Returning to Fig.~\ref{fig:timeseries}, it exhibits further interesting structures for the valence parameters. If we specifically look at the $d_v$ parameters we can see that for certain Monte Carlo times, e.g. $\sim350\,000$, the chain switches states and jumps to a different location in the parameter space with a substantially larger value of $a_3^{d_v}$ ($a_3^{d_v}\gtrsim0.8$) and smaller values of $a_1^{d_v}$ and $a_2^{d_v}$, and spends appreciable time exploring that region before returning. The other parameters exhibit no such structures. 
This feature is not characteristic for the particular chain displayed but can also be found in other chains, and consequently also in the combined sample that is used for the final analysis. Such behavior hints at the existence of another minimum different from the main one.

\begin{figure*}[ht]
\centering
\subfloat{
\includegraphics[width=0.49\textwidth]{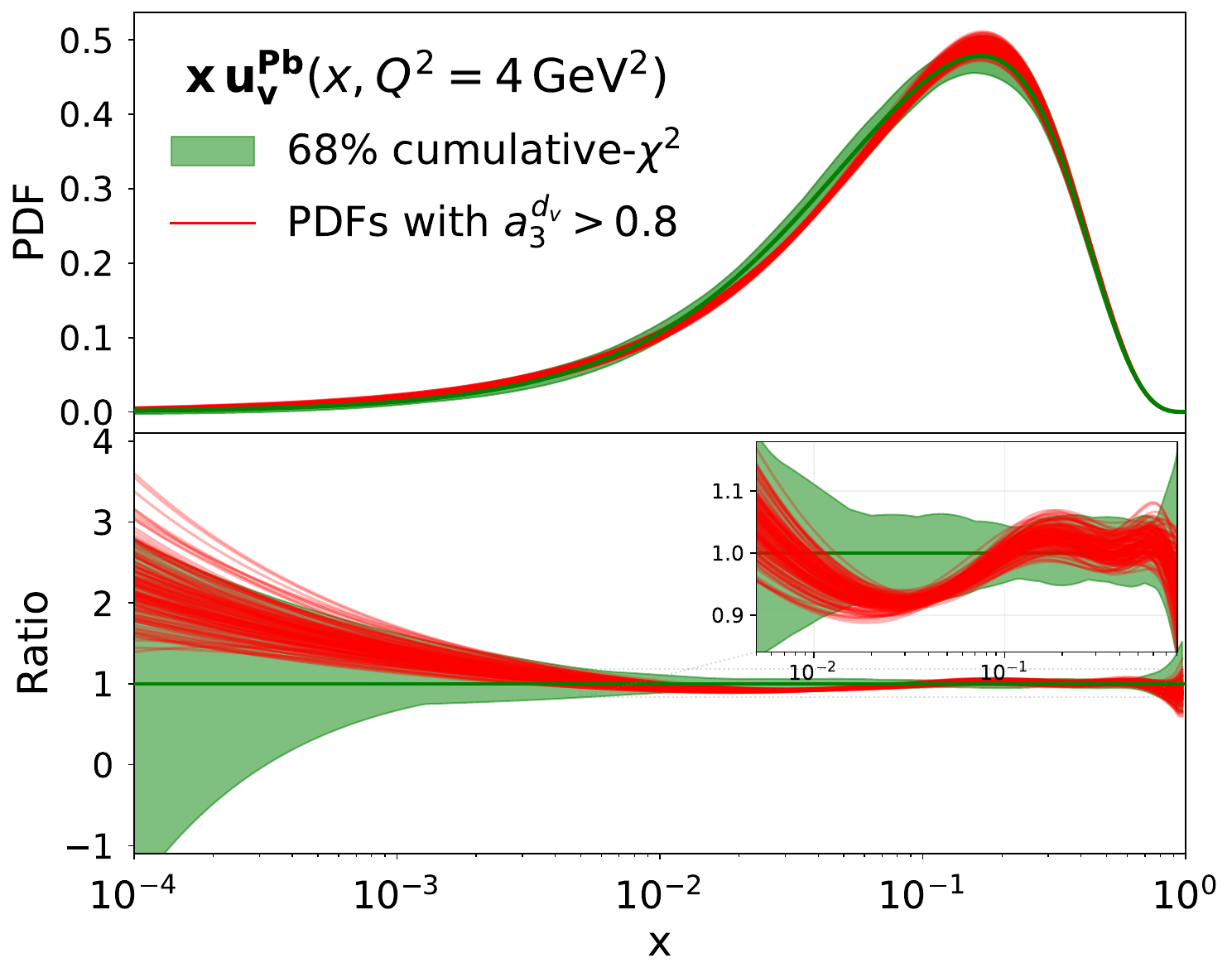}
\includegraphics[width=0.49\textwidth]{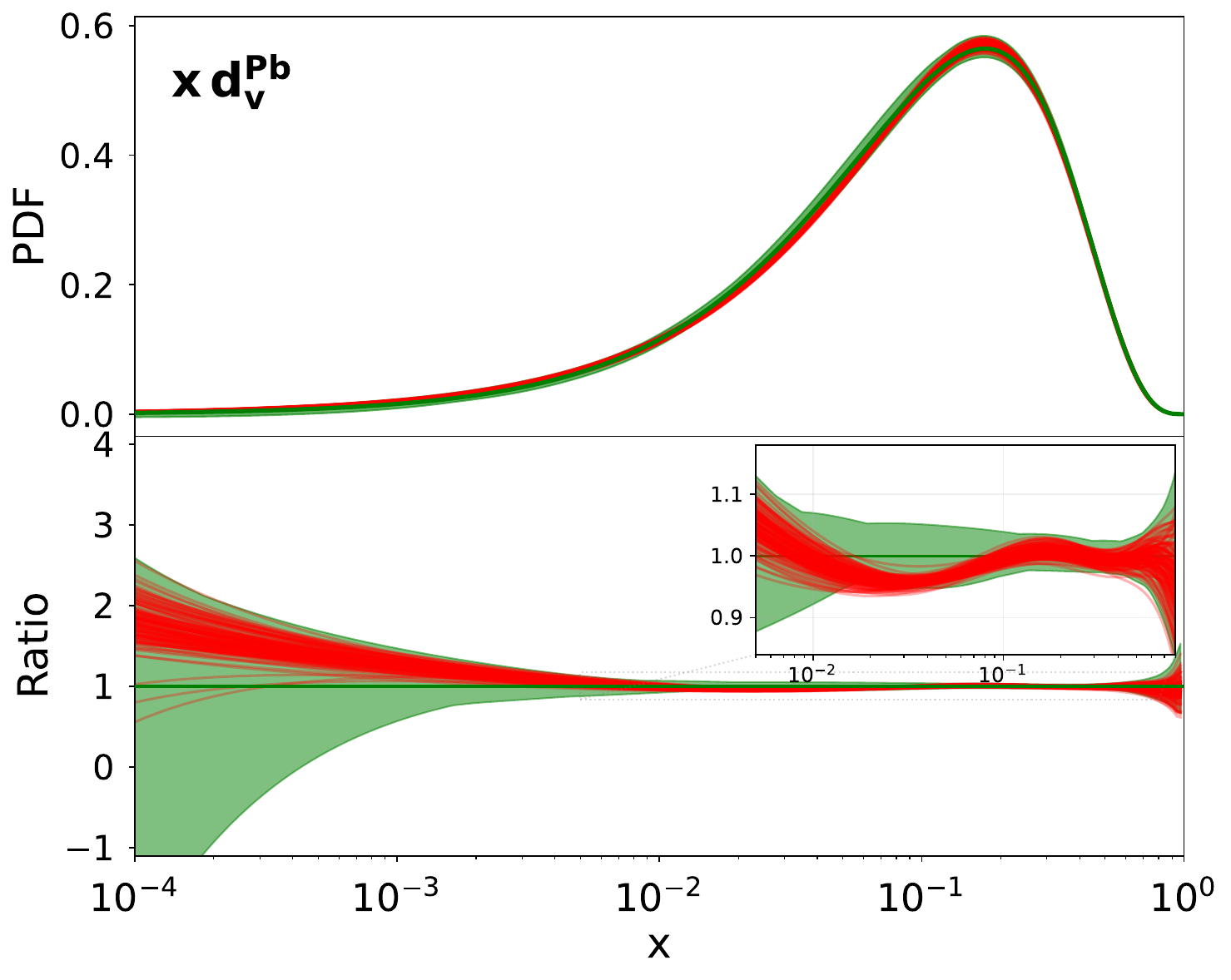}}\\
\addtocounter{subfigure}{-1}
\subfloat[Full Pb PDFs]{
\label{fig:PDFsaltMinPb_fullNuc}
\includegraphics[width=0.49\textwidth]{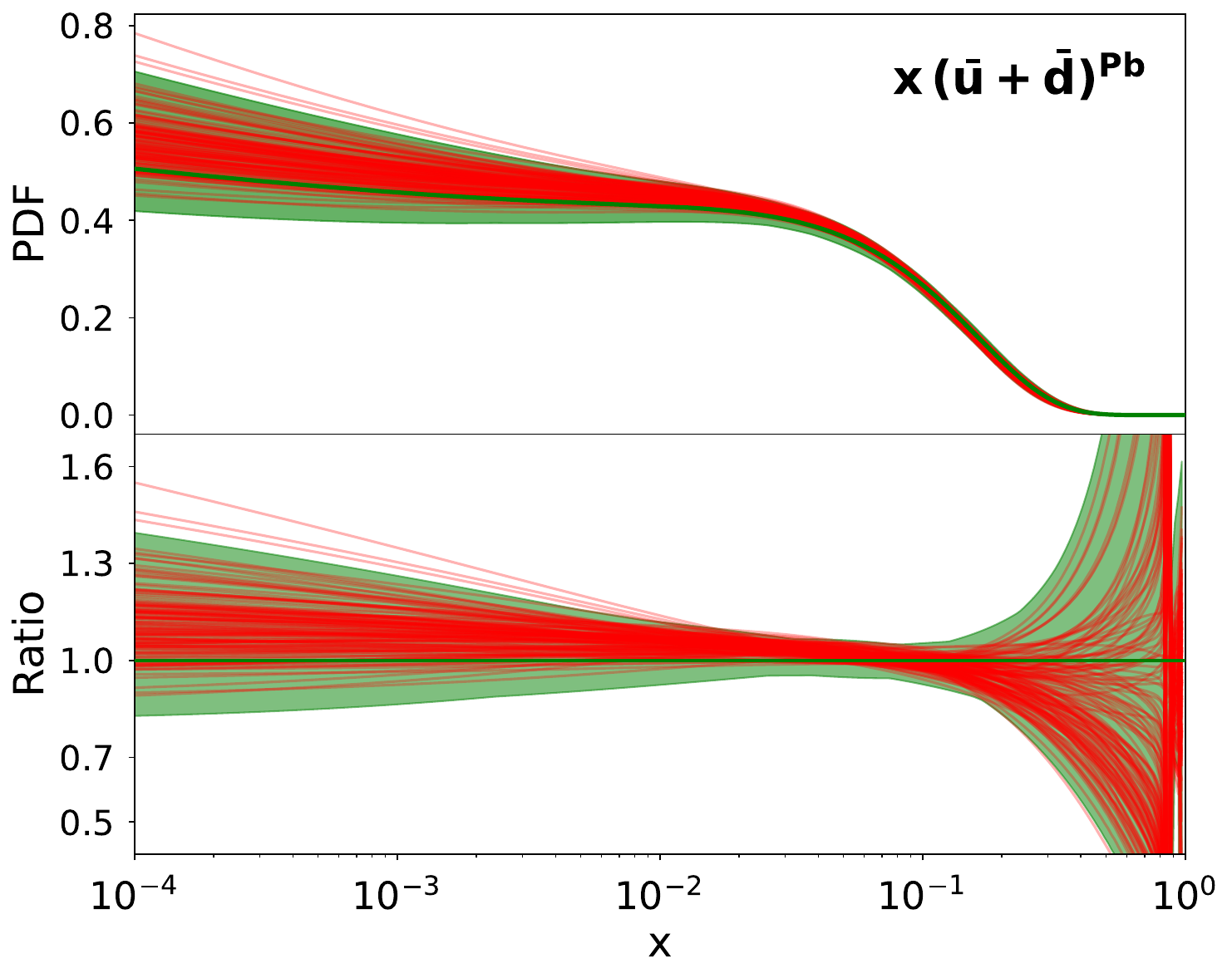}
\includegraphics[width=0.49\textwidth]{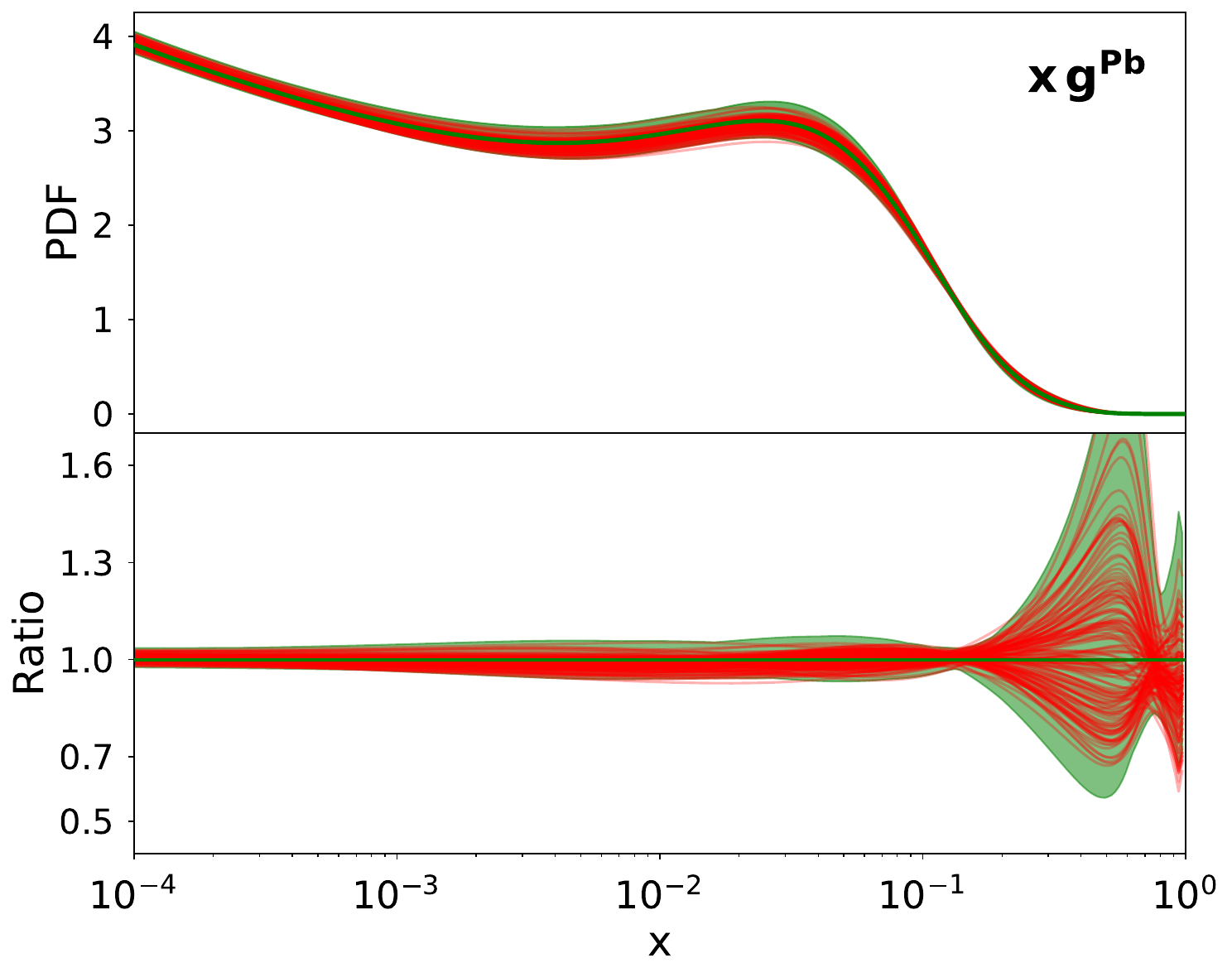}}\\
\subfloat[Bound-proton Pb PDFs]{
\label{fig:PDFsaltMinPb_bound}
\includegraphics[width=0.49\textwidth]{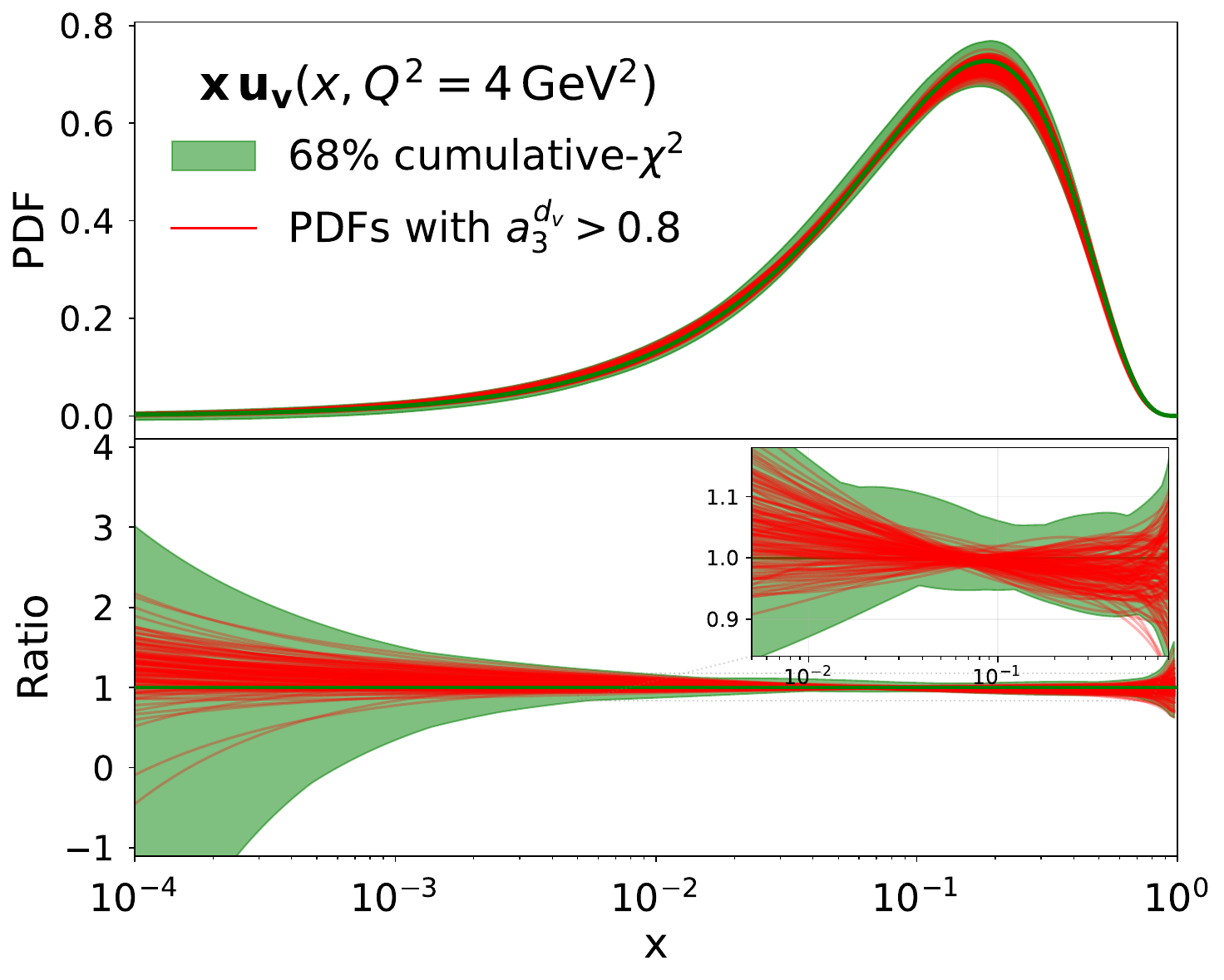}
\includegraphics[width=0.49\textwidth]{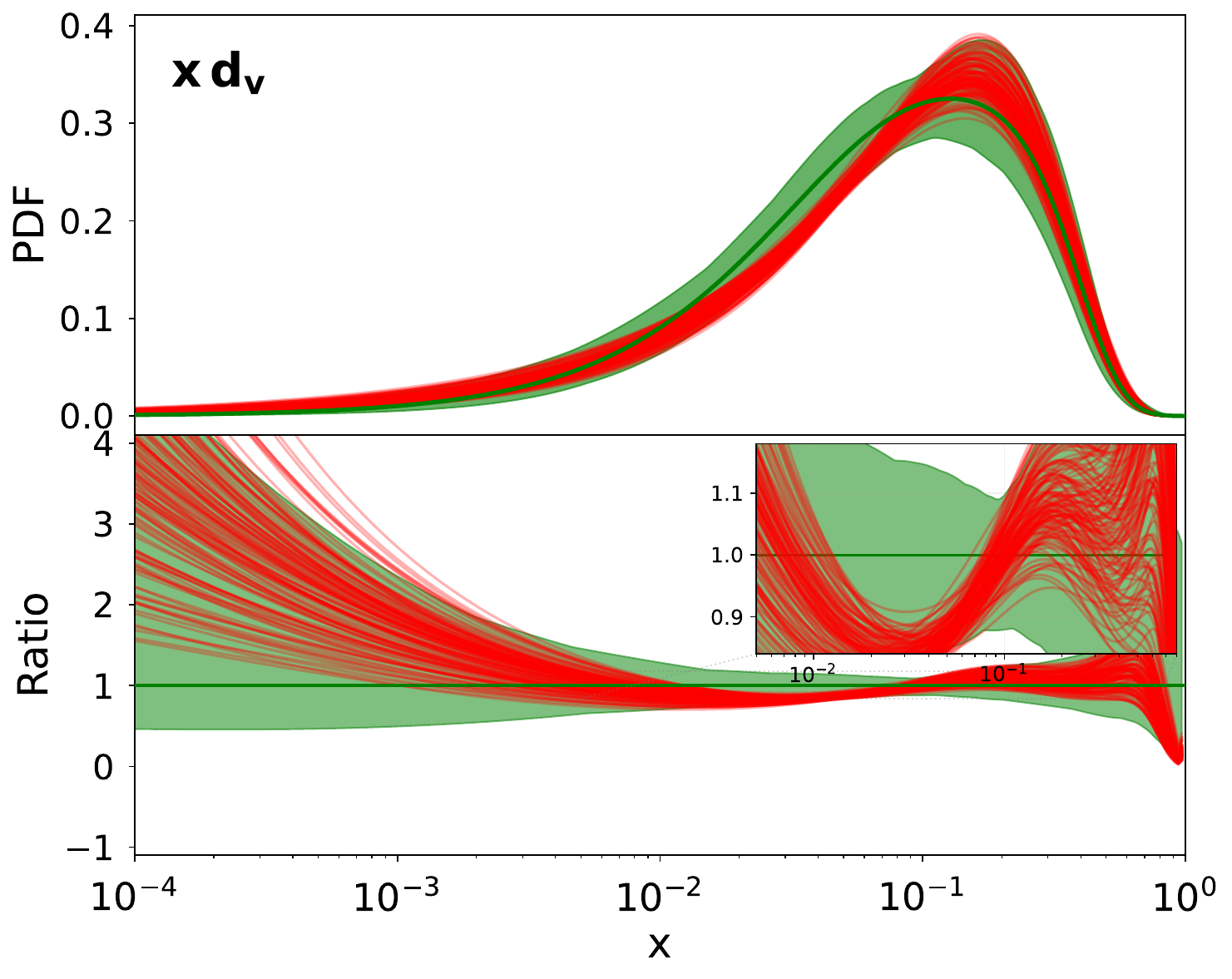}}
\caption{In green we show Pb PDFs with \cumchi{} uncertainties computed at 68\% CL level, and in red we plot individual PDF replicas corresponding to the alternative minimum defined by criterion of $a_3^{d_v}>v=0.8$. Red curves correspond to the red points in Fig.~\ref{fig:timeseries_min_dv_fullChain}. All ratios are plotted with respect to the \cumchi{} central value. 
Distributions in figure (a) are for full Pb PDFs, as defined by \cref{eq.1}, and figure (b) shows the bound proton distributions corresponding directly to~\cref{eq.2}.}
\label{fig:PDFsaltMinPb}
\end{figure*}

To scrutinize this further, in Fig.~\ref{fig:timeseries_min_dv}, we show the time series for the $d_v$ parameters including also the corresponding $\chi^2$ values for the chain from Fig.~\ref{fig:timeseries} and for the final combined and thinned sample.\footnote{Due to the thinning only a very small number of samples from the alternative parameter configuration survive in the combined chain.}
We split the samples into two sets, one with $a_3^{d_v} \le v$ colored in blue and the other with $a_3^{d_v} > v$ colored in red, where the value of $v=0.8$ was chosen to separate the visibly distinct regions in the time series.
The second set, which we treat as a potential second minimum, has a larger $\chi^2$ value but can nevertheless contribute to the final uncertainty estimate when using the \cumchi{} method, depending on the difference in likelihoods between the two sets.\footnote{Note that for the MCSE or the percentile method these samples would enter the uncertainty estimation regardless of their  $\chi^2$-value as the uncertainty is calculated solely based on the density of samples in the PDF space. See \cref{fig:PDFsaltMinPb} for the location of the PDF samples of the second set (shown in red).} To illustrate this, in Fig.~\ref{fig:dv68CLvs90CL}, we show correlations between the $d_v$ parameters based on the final combined sample, color-coding the individual points. In green we indicate points that enter 68\%\,CL (upper panel) and 90\%\,CL (lower panel) uncertainty estimated using the \cumchi{} method. In red we show points from the second set, and in black are points that fulfill both criteria, meaning they are from the second set but they also enter the 68\%\,CL (90\%\,CL) uncertainty estimate. The remaining samples are plotted in blue.
We can see that already at 68\%\,CL the fraction of the samples with alternative parameter configuration contributes to the uncertainty estimate, and its contribution increases when extending to higher confidence levels.

To rule out a possible degeneracy in the parameter space we verify that the parameters in the second set do not combine to the same PDF shape as the main minimum. For this purpose, in \cref{fig:PDFsaltMinPb}, we plot the 68\%\,CL uncertainties and overlay them with the individual PDF replicas corresponding to the second set (red points in \cref{fig:dv68CLvs90CL}). This clearly demonstrates a difference in the shape of the valence PDFs originating from the second set (increase in the low-$x$ region and depletion around \mbox{$x\sim10^{-2}$}). At the same time the shape of the gluon PDF is unchanged, which is reassuring since we observed the absence of any correlations between the $d_v$ parameters corresponding to the potential alternative minimum and the gluon parameters earlier.
Note that in \cref{fig:PDFsaltMinPb_fullNuc} we display the full nuclear Pb PDFs, and additionally in \cref{fig:PDFsaltMinPb_bound}, we show the corresponding bound proton PDFs. The former is the quantity of interest in a nuclear PDF fit, but the corresponding bound proton PDFs are more directly related to the parameters we are extracting. This is clear when investigating the valence distributions in both figures. We observe a very distinct shape change of the bound-proton $d_v$ distribution between the two sets of samples, occupying different regions of the $d_v$ parameter space, which is not present for the bound $u_v$ distribution. In case of full lead PDFs of \cref{fig:PDFsaltMinPb_fullNuc}, because of the averaging of $u$ and $d$ distributions, this effect is less pronounced but it occurs in both valence distributions.

\begin{figure*}[t]
\centering
\includegraphics[width=0.49\textwidth]{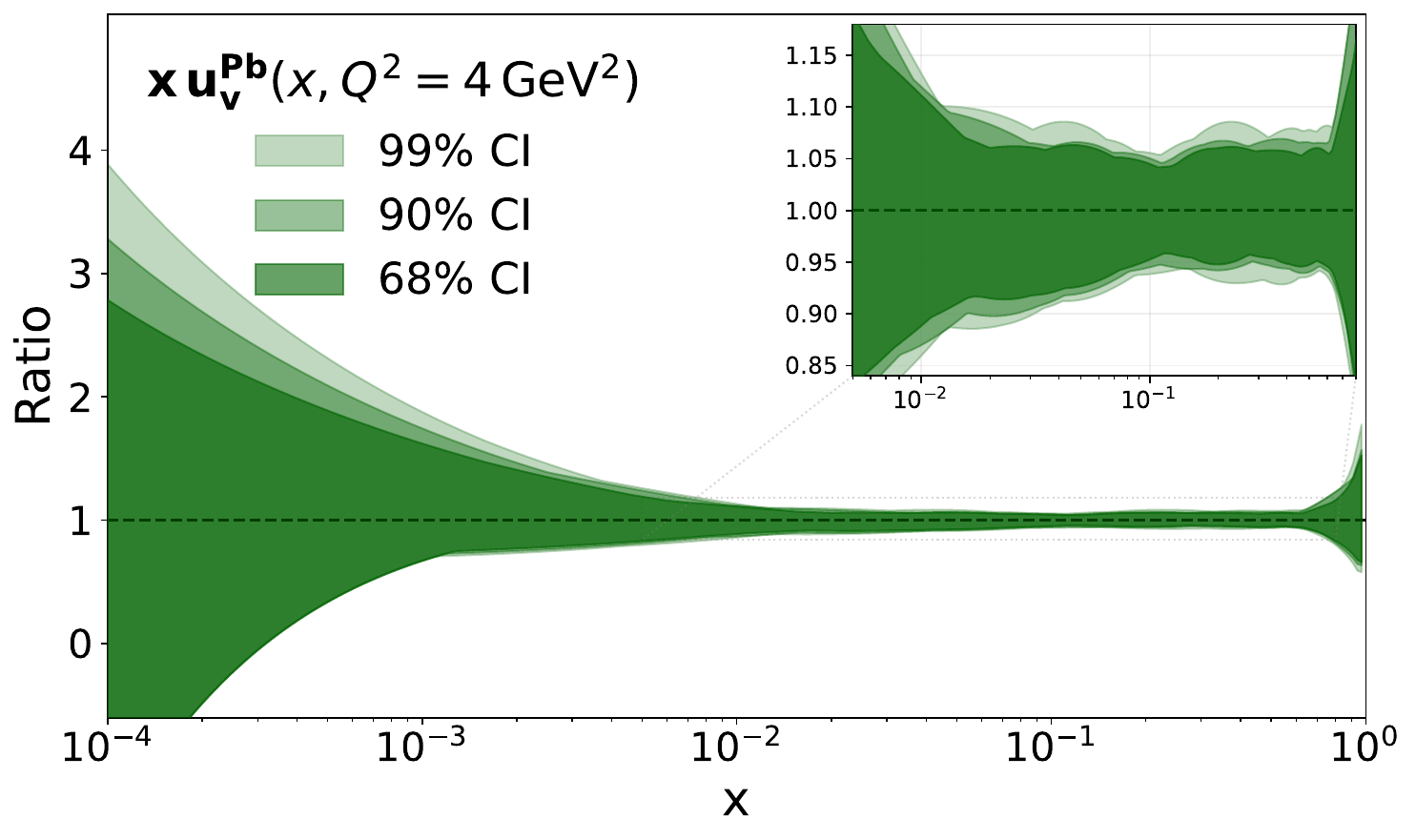}
\includegraphics[width=0.49\textwidth]{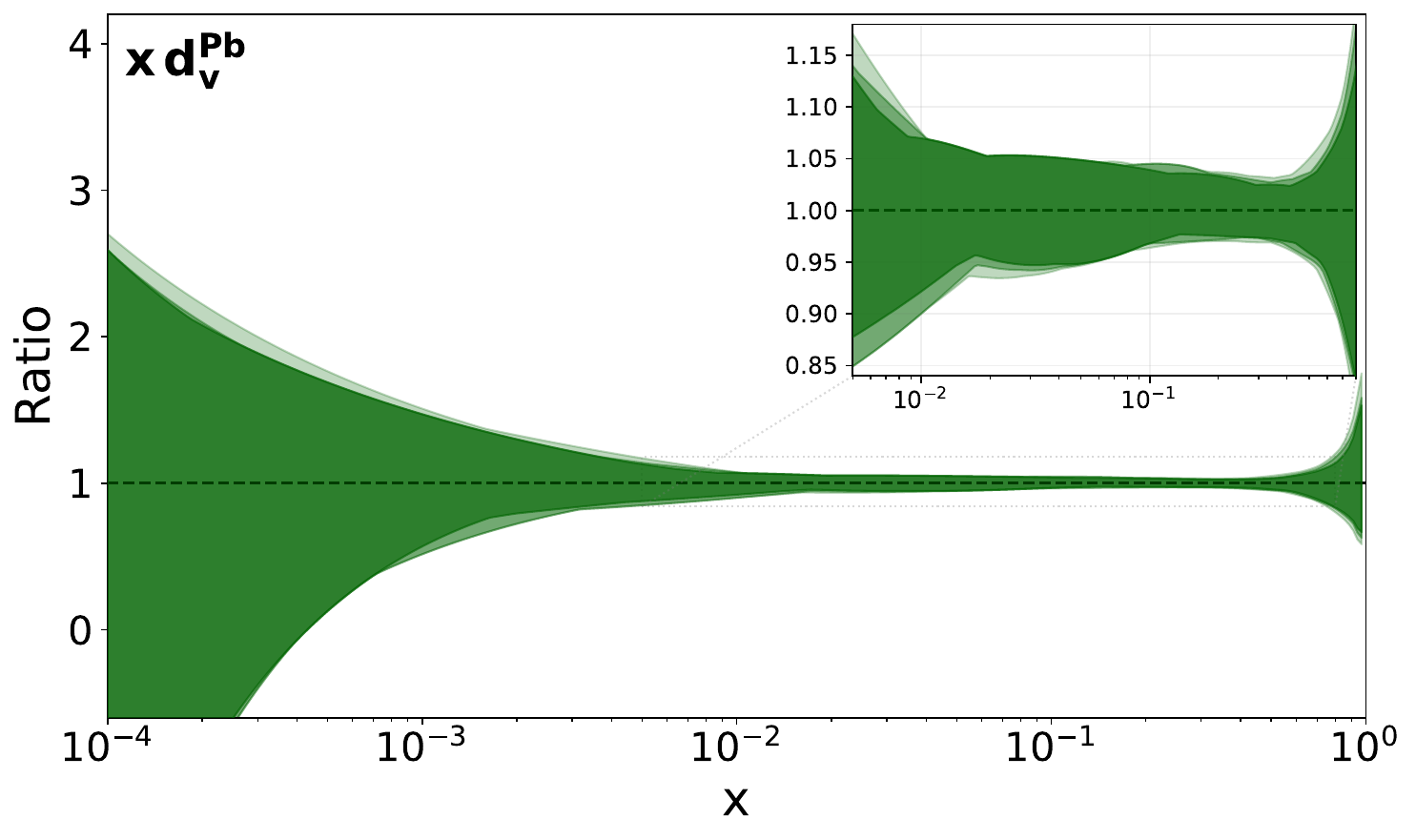}
\includegraphics[width=0.49\textwidth]{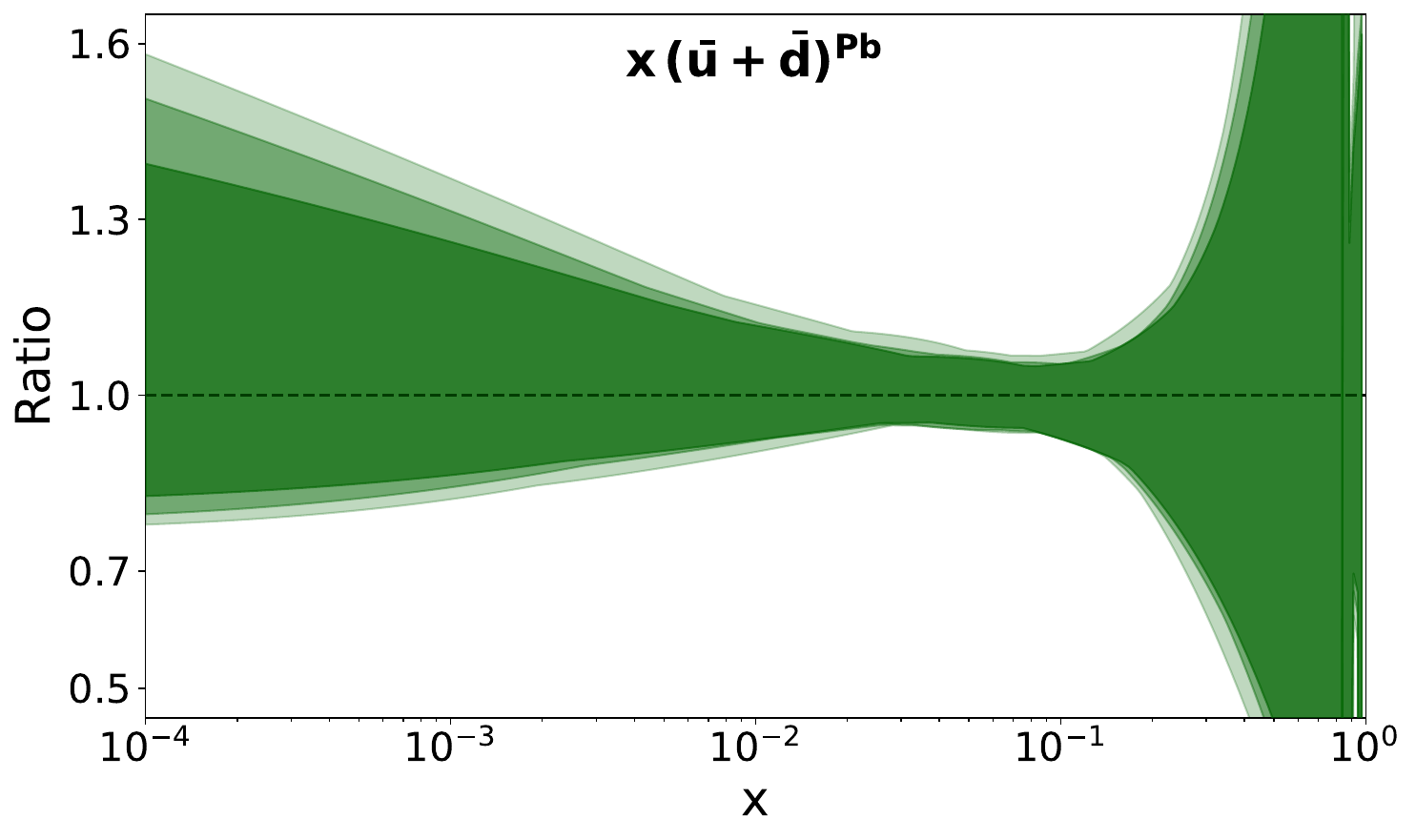}
\includegraphics[width=0.49\textwidth]{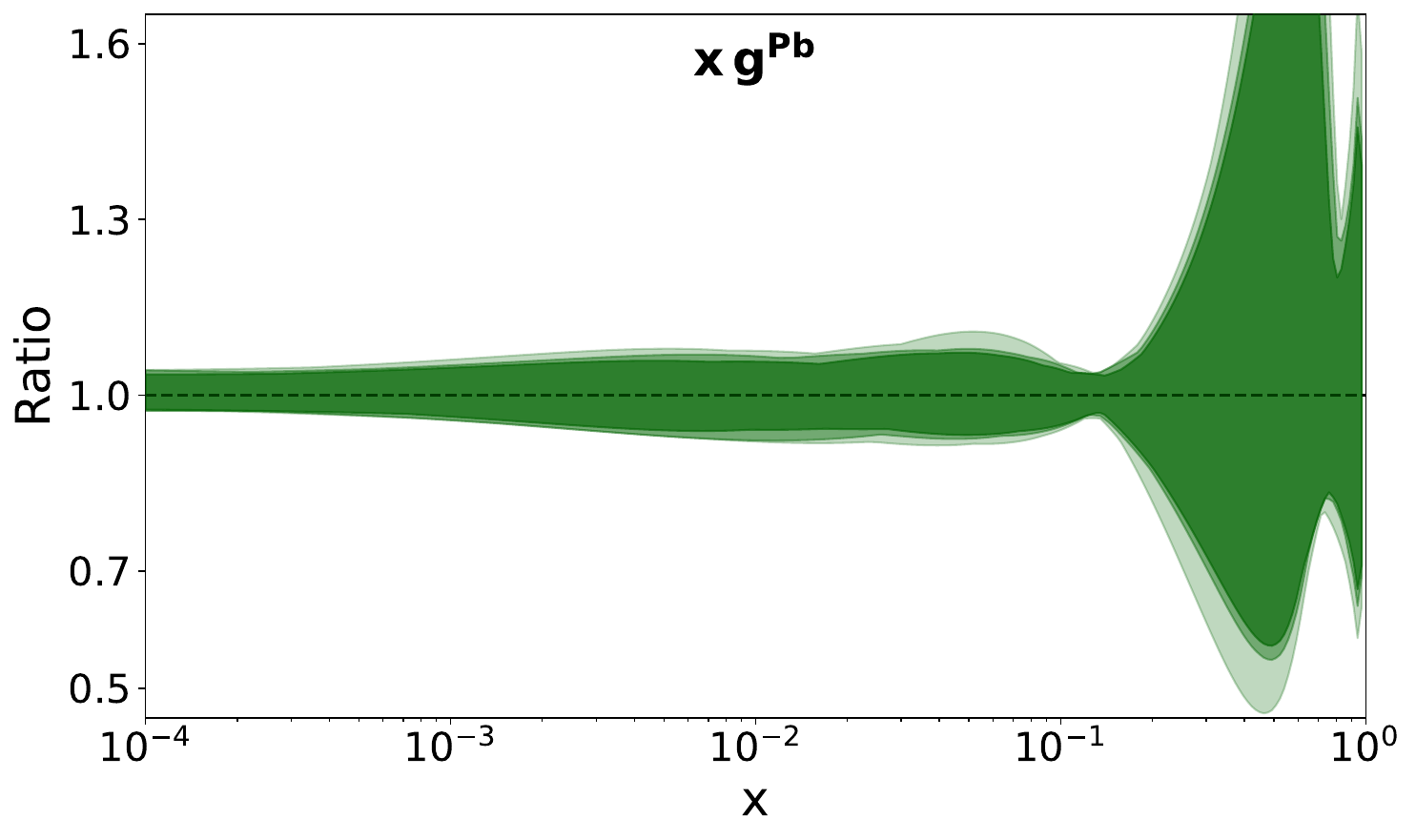}
\caption{Comparison of the full Pb PDFs uncertainty estimations from the \cumchi{} method at different confidence levels: \{68\%, 90\%, 99\%\} CL. Displayed are ratios of PDFs of different flavors to the corresponding central values.}
\label{fig:PbdiffCL} 
\end{figure*}
This behavior can be further illustrated by looking at the PDF uncertainty estimates performed at different confidence levels. In Fig.~\ref{fig:PbdiffCL}, we show them for 68\%\,CL, 90\%\,CL, and 99\%\,CL. When increasing the confidence level, we are including more and more points from the supposed second minimum and it impacts the error bands in an asymmetric way -- the upper band of $u_v$ uncertainty increases, whereas the lower one remains unchanged (similarly for $\bar{u}+\bar{d}$).\footnote{Note that this localized change in the shape of the uncertainties is also in contrast with the Hessian method where a larger CL is achieved by increasing the $\Delta\chi^2$ tolerance. Then, the uncertainties increase uniformly with the square root of the tolerance.}
As already pointed out earlier, the fact that in \cref{fig:PDFsaltMinPb_fullNuc,fig:PbdiffCL} we observe changes for both $u_v$ and $d_v$ PDFs, even though we are investigating here the alternative region in the $d_v$ parameters, is caused by the fact that we are plotting the full lead PDFs (see Eq.~\eqref{eq.1}) in which valence distributions mix.

\begin{figure*}[t]
\centering
\includegraphics[width=\textwidth]{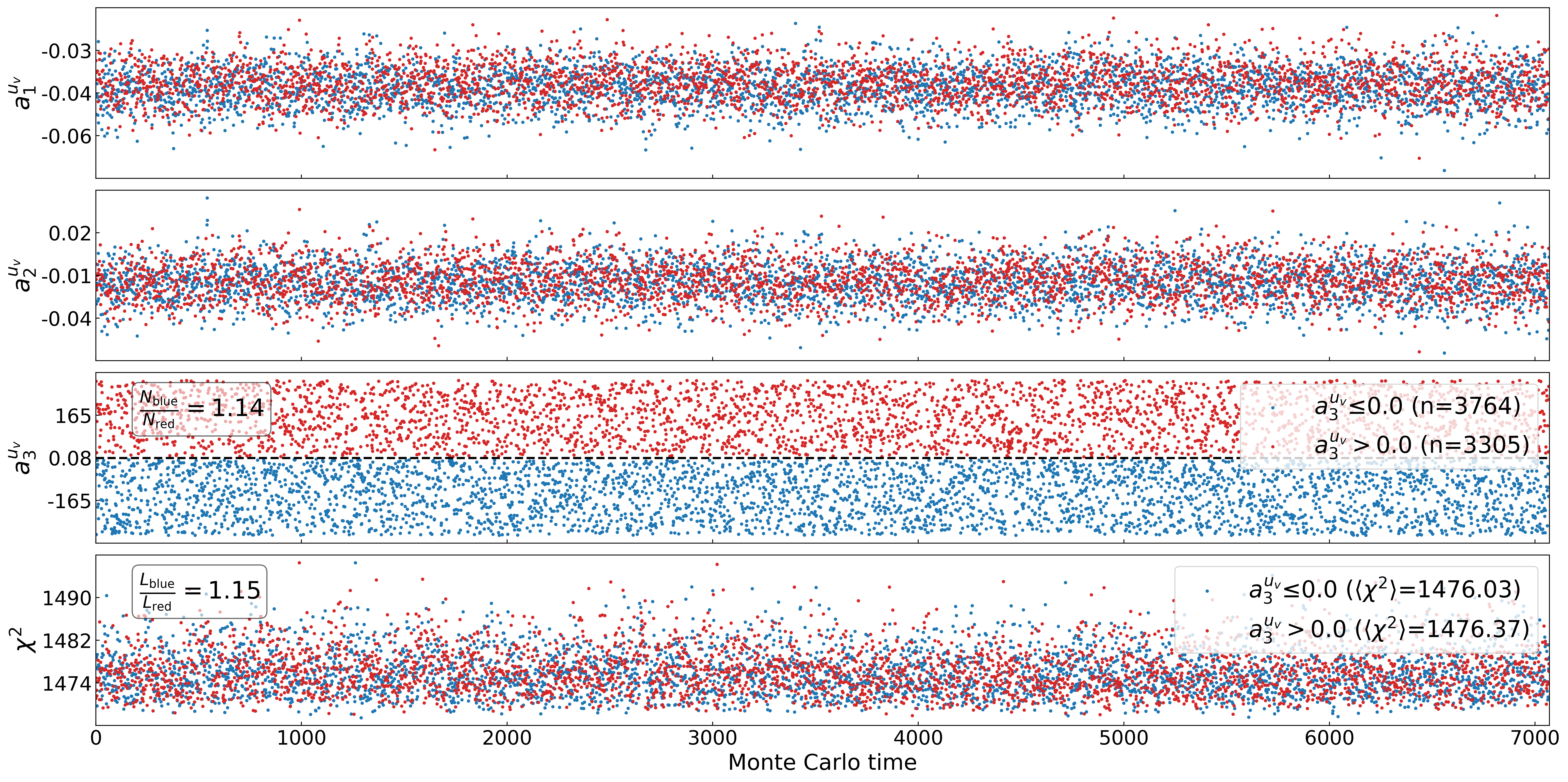}
\caption{Time series of the $u_v$ parameters of the \Pbonly{} analysis plotted together with the corresponding $\chi^2$ values. The points drawn in red correspond to $a_3^{u_v} > 0$, while the points in blue correspond to $a_3^{u_v} < 0$.
}
\label{fig:timeseries_min_uv} 
\end{figure*}

We now take a step back and look again at the chain in Fig.~\ref{fig:timeseries} and the correlations in Fig.~\ref{fig:pairwise}.
As already observed earlier, there is a continuous split between the negative and positive values of $a_3^{u_v}$ in every subplot, thereby indicating that the samples cover two regions separated by a barrier. 
Additionally, in the vicinity of zero, the $a_3^{u_v}$ parameter exhibits non-negligible correlations to $a_1^{u_v}$, where $a_1^{u_v}$ prefers lower values if $a_3^{u_v}$ is negative and larger values if $a_3^{u_v}$ is positive. This correlation diminishes when the absolute value of $a_3^{u_v}$ becomes large.
This is further demonstrated in Fig.~\ref{fig:timeseries_min_uv} where time series for the $u_v$ parameters is plotted together with the $\chi^2$ values. In this case we do not observe any significant difference on the level of $\chi^2$ between the two regions, only the split in the value of $a_3^{u_v}$ and exclusion of the region of $a_3^{u_v}\sim0$. 
Such a behavior, with clearly separated regions in $a_3^{u_v}$, suggests that what we observe here is a secondary minimum.
This can be seen in more detail in Fig.~\ref{fig:altMINuvpairwise} in App.~\ref{appen:C}.

A crucial issue concerns the sampling of multiple, possibly disjoint, regions: are they properly sampled by our algorithm? Assuming proper sampling, the number of points drawn from each region/minimum should scale with its likelihood. Consequently, the ratio of the sample counts associated with the two regions will be equal to the ratio of their respective likelihoods. To verify this, we calculated the values of these ratios and display them in \cref{fig:timeseries_min_dv,fig:timeseries_min_uv}.%
\footnote{The likelihood ratio is computed as the ratio of the mean likelihood values in the two regions, $\langle e^{-\chi^2/2}\rangle_{\text{blue}} / \langle e^{-\chi^2/2}\rangle_{\text{red}}$. The sample ratio is obtained by counting the number of samples in each region.} In an ideal case they should be the same, but in practice we expect some deviations.

We first discuss the split in the $a_3^{u_v}$ parameter. In this case we have a very similar number of points with negative and positive values of $a_3^{u_v}$, namely 3764 samples with $a_3^{u_v}<0$ and 3305 with $a_3^{u_v}>0$, which results in the ratio of 1.14. The corresponding ratio of likelihoods of both regions is nearly the same and equals 1.15. This gives us confidence that both regions are properly sampled by our algorithm, giving a robust representation of this behavior.

In the case of the alternative configuration of the $d_v$ parameters, shown in Fig.~\ref{fig:timeseries_min_dv}, the situation is more subtle. Here the alternative parameter configuration has a substantially larger $\chi^2$ value and hence is less likely, which means we expect a much smaller number of samples from this region.
If we now look at the final combined sample after thinning (lower sub-figure), the number of points from the first set is 6931 and from the second set only 138 points, giving a ratio of 50.22. The corresponding ratio of likelihoods is equal to 39.04. We can see a much larger discrepancy between the two ratios as compared to the split in $a_3^{u_v}$ parameter.
However, we need to take into account the fact that the separation cut between the two regions, defined  by choosing a particular value of $a_3^{d_v}$ (here $a_3^{d_v} = 0.8$), is somewhat arbitrary.%
\footnote{We varied the cut at $a_3^{d_v} = 0.8$ and found similar results.} Additionally, the error estimates of the ratios scale (inversely) with the number of samples, which is small for the alternative $d_v$ minimum.
Nevertheless, we believe that with this number of points in the second set, the sampling obtained by the employed aMH algorithm is surprisingly good. Even though we did not tune the sampling algorithm to explore multiple minima, it was able to find supposedly disjoint regions and to sample them representatively. And since the aMH algorithm is known to be a suboptimal choice for multimodal posteriors, one could, of course, choose a fine-tuned algorithm or, e.g., try to generate separate chains for individual minima to explore them even better. This is, however, beyond the scope of our paper.%
\footnote{A discussion of different approaches for sampling multimodal distributions can be found e.g. in~\cite{Hunt-Smith:2023ccp,Craiu:2014}.}

Finally, note that although the structure in $a_3^{d_v}$ parameter visible in \cref{fig:timeseries_min_dv} suggests the existence of a secondary minimum, it does not constitute proof of its presence, especially since there is no clear separation in the pairwise distributions (cf. \cref{fig:pairwise,fig:dv68CLvs90CL}).
Another possible interpretation would be a single minimum with a complicated shape in the three-dimensional subspace given by the $\{a_1^{d_v}, a_2^{d_v}, a_3^{d_v}\}$ parameters. The apparent state switching in the time series would then be a consequence of a poor sampling efficiency due to the change of geometry in the parameter space~\cite{brooks2011handbook}. As the aMH algorithm draws more samples from the region below the $a_3^{d_v} = 0.8$ cut the proposal distribution is optimized to yield adequate sampling efficiency for that region and thereby making traverses through the shape less likely.
The situation is more clear for the $u_v$ parameters, where we observe a barrier in $a_3^{u_v}$, giving us a stronger indication that it is a secondary minimum.

What is more important is that regardless of the nature of the observed structures in the likelihood, the employed MCMC approach allowed us to sample from it appropriately, ultimately providing a reliable estimate of uncertainties.
It should not go unnoticed that such a complicated landscape in the parameter space can not be reliably explored using other methods, in particular the Hessian method. 
After all, if we only consider the information gained from the Hessian analysis, we find no indication of the presence of these structures. 
More importantly, as a consequence, only a subset of the samples identified by the MCMC approach is included within the Hessian uncertainty estimate. This therefore compromises the reliability of the Hessian method in presence of such complex likelihoods.

\subsection{Multi-Nuclei PDF Fit}
\label{subsec:comp}
In this section we concentrate on the complementary analysis, the \multinuc{} analysis, that we have performed using data taken on several different nuclear targets to extract nPDFs for a range of nuclei. 
We note that this is a standard approach for a nuclear PDF analysis employed in all current extractions~\cite{Eskola:2021nhw,Duwentaster:2021ioo,AbdulKhalek:2022fyi,Helenius:2021tof,Khanpour:2020zyu}. Here, we use it as an additional study, which allows us to obtain further insights into the impact of the analytic $A$-dependence (see \cref{eq.5}) on nPDFs and investigate the indirect impact of data from different nuclei on lead PDF uncertainties.
Additional details about this analysis are provided in App.~\ref{app:multnuc}.

The \multinuc{} analysis is performed with the MCMC method using an analogous setup as for the \Pbonly{} fit.\footnote{One difference which is worth noting is the need for a smaller value of the fixed covariance matrix, $C_0=\operatorname{diag}(\sigma_0 \times \mathbf{1})$ in the proposal distribution. In \multinuc{} analysis we use $\sigma_0=10^{-5}$ instead of $10^{-3}$ used in the \Pbonly{} fit.}
Several independent MCMC chains were generated for this setup, of which 7 passed all convergence diagnostics. After removing the thermalization segments, these chains collectively accumulated approximately $N_{\text{tot}}=873\,000$ samples. Following the thinning procedure, the total number of statistically independent samples was reduced to $N_{\text{tot}}=1\,515$. Although smaller than for the \Pbonly{} case, this sample size is still large enough to perform a meaningful analysis and extract reliable PDF uncertainties.
For the sake of space, we do not report here the time series or the marginal distributions of the resulting samples, as we observed similar features as in the \Pbonly{} case.
Instead, we concentrate on the comparison of the resulting lead PDFs.

For the following discussion, note that the only difference in the two setups is the indirect constraints from the light nuclei on the lead PDFs through the imposed $A$ dependence. The questions we want to focus on are: ``How do the additional constraints manifest in the results?'' and ``Are the findings consistent between using MCMC and the Hessian method?''.

\begin{figure*}[!htbp]
\centering
\includegraphics[width=\textwidth]{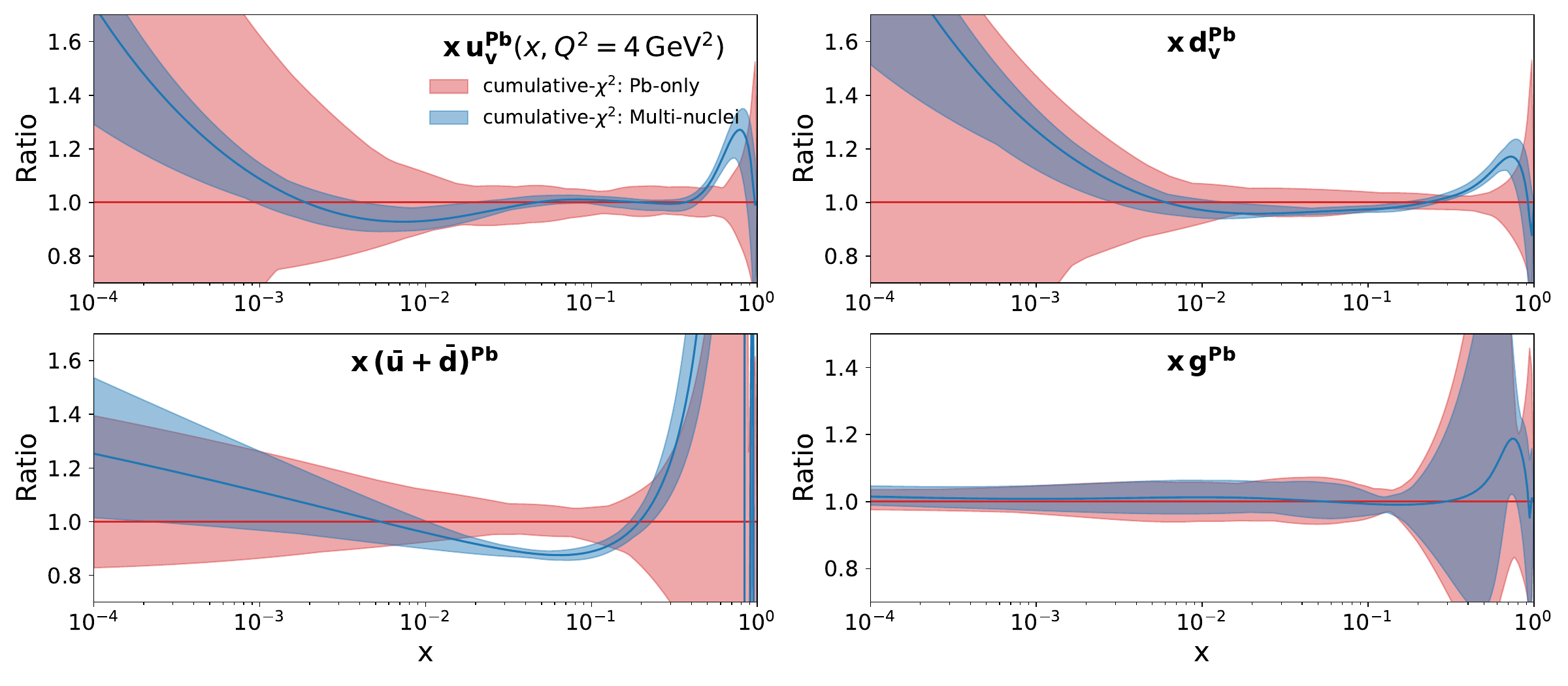}
\caption{Comparison of PDFs from the \Pbonly{} and \multinuc{} MCMC analyses plotted as a ratio to the best-fit PDF from the \Pbonly{} analysis.}
\label{fig:PbvsNucMCMC} 
\end{figure*}

\begin{figure*}[!htbp]
\centering
\includegraphics[width=\textwidth]{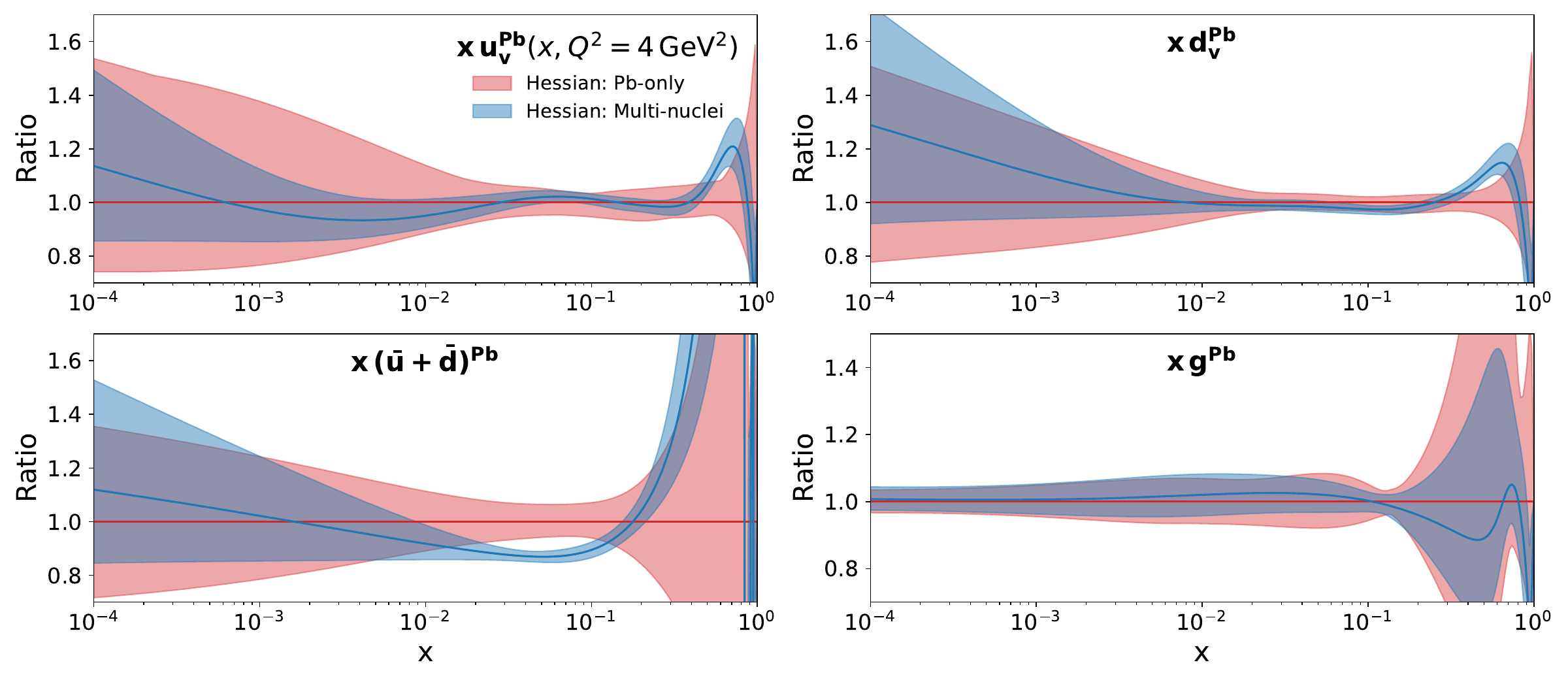}
\caption{Comparison of PDFs from the \Pbonly{} and \multinuc{} Hessian fits plotted as a ratio to the central PDF from the \Pbonly{} Hessian fit.}
\label{fig:PbvsNucHess} 
\end{figure*}

In Fig.~\ref{fig:PbvsNucMCMC} we compare the lead PDFs obtained from the \Pbonly{} and the \multinuc{} analyses with errors computed using the \cumchi{} method. Specifically, we plot the ratio of the lead PDFs with respect to the ``central'' PDF (sample with lowest $\chi^2$) from the \Pbonly{} analysis. In general the extracted lead PDFs from the two analyses agree within their error bands, the exceptions being the $\bar{u}+\bar{d}$ at mid-$x$ and in the extrapolation regions at extremely large $x$-values. However, there are also a number of differences between them which we discuss now.

First, we see a considerable reduction of uncertainties for the valence quarks in the \multinuc{} fit for almost the entire $x$-range shown. This can be understood from the fact that the \multinuc{} analysis includes many data points from neutral-current DIS (see \cref{tab:NCDIS1,tab:NCDIS2}) for a broad range of nuclei, which provide constraints for the valence distributions. Hence, we can reproduce the known fact that they can impact lead PDFs indirectly through the assumed $A$ dependence and result in reduced uncertainties. In the case of the light-sea quarks we also observe reduced uncertainties, but here the reduction happens mostly in the mid- and large-$x$ region ($x\gtrsim0.01$). This can be partially associated with the already mentioned DIS data and further with the fixed-target Drell-Yan data listed in~\cref{tab:DY}.
On the other hand, the gluon PDF and its uncertainty are basically identical between the two analyses. 
This confirms the reverse side of the previous argument, i.e.~it shows that since almost all data constraints on the gluon distribution originate from the data taken on lead (especially the LHC data), the lighter nuclei do not impose further constraints.

In addition to the changes in the size of the uncertainty bands, we also observe changes in the shape of quark PDFs. Specifically, both valence distributions as well as the sea-quark distributions obtained from the \multinuc{} fit are enhanced at low-$x$. Furthermore, especially in case of the $\bar{u}+\bar{d}$ distribution and to a lesser extent for the valence distributions, we observe a suppression when going to larger $x$-values. For the sea-quarks and $d_v$ the affected regions start from $x\sim\{0.01,0.1\}$, whereas in case of $u_v$ PDF it occurs already at lower $x\lesssim0.01$.
Looking at the shape of the ratio of quark distributions in the \multinuc{} and \Pbonly{} fits, we notice that it has a pattern of an inverted nuclear modification (instead of shadowing at low-$x$ we observe enhancement; instead of anti-shadowing we have suppression; in the EMC region we observe enhancement and a drop at extremely large-$x$).
It appears that the lighter nuclei weaken nuclear corrections in lead, leading to such inverted nuclear modification shape. This indicates that introducing a fixed $A$ dependence can have systematic effects on the extracted nPDFs, and we consider this difference as a first approximation of the systematic difference between the two approaches. In retrospect, this is not unexpected, as the use of the $A$ dependence was adopted in the first place because of the need to include more constraints, since data for individual nuclei were not sufficient on their own. Today, with the help of the proton-lead LHC data, it is the first time we can meaningfully extract PDFs for a single nucleus -- lead.
One should note that from the point of view of factorization theorems there is no reason to assume an analytic $A$ dependence~\cite{Ruiz:2023ozv,Qiu:2003cg}. On the other hand, from the phenomenological perspective we observe correlations between different nuclei with similar mass.

Finally, we have also performed a twin Hessian fit for the \multinuc{} analysis. In Fig.~\ref{fig:PbvsNucHess}, we compare its results with the earlier PDFs from the \Pbonly{} Hessian fit. Qualitatively we observe the same patterns as we have found in Fig.~\ref{fig:PbvsNucMCMC} -- reduced uncertainties and a shape change in terms of ``reversed nuclear modification''. However, the details are different and they have the potential to provide insight about the differences between the MCMC and Hessian methods. Specifically, if we look at the low-$x$ valence distributions, we can see the largest deviations between the Hessian and MCMC approaches. This is the region where the valence distributions are the least constrained, especially in the \Pbonly{} analyses. We can see that the Hessian errors in this region are closer in shape and size between the \Pbonly{} and \multinuc{} fits, whereas the uncertainties from MCMC are significantly reduced in the \multinuc{} fit compared to \Pbonly{}.

Another noteworthy feature we observe is the difference between the gluon error bands. As pointed out earlier, the majority of the data constraints for the nuclear gluon PDFs come from the data on lead. As a result, we expect the gluon PDFs extracted from \Pbonly{} and \multinuc{} analyses to be very similar. This is exactly what we find in our MCMC analyses in \cref{fig:PbvsNucMCMC}, where both ``central`` curves and uncertainties are nearly identical (with only small deviations at extremely large values of $x$). However, when we compare with the corresponding plot done with the Hessian fits we see more striking differences. Although the central predictions agree within their uncertainties, the error bands show more variation. Specifically, they agree well only at low-$x$ where most stringent data constraints reside (coming from the pPb heavy-quark data). At higher $x$-values $x\gtrsim0.01$ we observe deviations which are rising when going to larger $x$-values. This behavior points towards limited reliability of the Hessian method in the region where data constraints are sparse or absent.
Of course, such behavior could also be related to the indirect impact of the $A$ dependence. However, the lack of any significant gluon constraints from non-lead data and the fact that MCMC yields much more compatible uncertainties in both the \Pbonly{} and the \multinuc{} analyses points towards potential issues with the Hessian method.
%
\section{Conclusion}
\label{sec:con}
We performed the first-ever analysis of nuclear PDFs using the MCMC methods. We used the adaptive Metropolis-Hastings algorithm to produce a collection of independent Markov chains, which were further processed by removing the initial non-thermalized samples and applying a thinning procedure, thereby producing the final sample of independent points that were used to extract PDFs and their uncertainties.

The analysis concentrated on the determination of lead PDFs and consisted of two parts: first, we performed a \Pbonly{} extraction using only lead data, and second, in the \multinuc{} fit, we extended the dataset by including data from a range of other nuclei, making use of a standard $A$ dependence to introduce additional constraints on the lead PDFs. The main motivation of the analysis was to demonstrate that the parameter space that must be explored in the nPDF determination is highly intricate, warranting the need for a more sophisticated method for the estimation of uncertainties. Further, we  have shown that MCMC methods allow to provide a robust mapping of this space, offering unprecedented insights into the parameter correlations and as a result, yielding more reliable uncertainty estimates.

We have demonstrated that the likelihood landscape and the resulting structure of the minimum is highly non-trivial. It includes multiple modes and complicated structures, leading to non-Gaussian distributions of the fitted parameters, especially in the case of the valence PDFs. This finding raises the question of whether traditional uncertainty estimation methods, such as the Hessian approach, can reliably determine nPDF uncertainties. A complementary analysis using the Hessian method allowed for a detailed comparison and provided insights on the limitations and shortcomings of the Hessian method. Of course, similar considerations are applicable to other problems in which the corresponding parameter space has a complicated structure and features large deviations from Gaussianity, a situation that occurs often for optimization problems where only limited data constraints are available.

Our \Pbonly{} analysis is the first extraction of nuclear PDFs for a single nucleus -- lead. Such an analysis was possible because of the wealth of recent LHC data from proton-lead collisions. We have also performed a \multinuc{} analysis, following a more standard approach to nPDF fits, which included data from a range of nuclei and enabled the extraction of nPDFs for different nuclei.
Comparing both setups allowed us to investigate the impact of the $A$ dependence on nPDFs/lead PDFs and to obtain a first approximation of the systematic effects imposed by the $A$ dependence. We found that for quarks the inclusion of lighter nuclei allows for a reduction of uncertainties and additionally impacts the shape of the lead PDFs by reducing the nuclear effects (see Fig.~\ref{fig:PbvsNucMCMC}). However, these reduced uncertainties are not observed for the gluon PDF, for which the majority of the data constraints already come from the lead data. We conclude from this observation that the imposed $A$ dependence behaves as expected: it provides additional constraints from different nuclei in cases where the information is limited (e.g., for quarks),  while preserving sufficient flexibility not to bias the results when strong data constraints are present for a given nucleus. The comparison between the MCMC and Hessian method extractions for both setups revealed systematic effects inherent to the Hessian analysis that could overshadow a meaningful assessment of the differences between a single-nuclei vs.~a multi-nuclei fit.
\appendix
%
\section{Additional Details about the Single- and Multi-Nuclei Analyses}
\label{appen:A}
This appendix provides additional details about the \Pbonly{} and \multinuc{} analyses, in particular specific information on the used datasets.
Also in \cref{tab:chi2_comparison} we provide $\chi^2$ values for both analyses.
\begin{table}[h]
\centering
\setlength{\tabcolsep}{4pt}
\renewcommand{\arraystretch}{1.25}
\begin{tabular}{lcc}
\hline
\textbf{$\chi^2_{\min}$} & \textbf{Pb-only} & \textbf{Multi-nuclei} \\
\hline
Hessian (per d.o.f.) 
  & 1466.57\; (0.985) 
  & 2056.10\; (1.035) \\
MCMC (per d.o.f.) 
  & 1467.30\; (0.992) 
  & 2051.82\; (1.028) \\
[3pt]
\hline
\end{tabular}
\caption{Comparison of total and reduced $\chi^2$ values (for best fit/central PDFs) using Hessian and MCMC methods.}
\label{tab:chi2_comparison}
\end{table}

\subsection{\Pbonly{} Analysis}
The following tables summarize the datasets used in the \Pbonly{} analysis, including the number of data points before and after applying the kinematic selection cuts. Specifically, \cref{tab:WZdata} lists the $W$ and $Z$ boson production data, \cref{tab:HQdata} lists the heavy-quark data, and \cref{tab:nudata} shows the neutrino DIS data. The $N_{\text{data}}$ column in the tables indicates the number of data points after/before kinematic cuts.

\begin{table}[hb!]
  \centering
  \renewcommand{\arraystretch}{1.2}
  \begin{tabular}{|l l l c c c|}
    \hline
    \multicolumn{6}{|c|}{LHC $p$Pb $W^{\pm}/Z$ data $d\sigma/dy$}\\
    \hline
    \hline
    Exp. & Obser. & $\sqrt{s}$ [TeV] &  \;ID &  Ref. & $N_{\text{data}}$ \\
    \hline
    \multirow{3}{*}[-0.0em]{ATLAS} 
    & $Z$ & \;\;\;5.02 & 6215 & \cite{Aad:2015gta} &  14/14 \\
    & $W^{+}$ & \;\;\;5.02 & 6213 &\cite{AtlasWpPb} & 10/10 \\
    & $W^{-}$ & \;\;\;5.02 & 6211 &\cite{AtlasWpPb} & 10/10 \\
    \cline{1-6}
    \multirow{5}{*}[-0.0em]{CMS\;} 
    &  $Z$  & \;\;\;5.02 & 6235 & \cite{Khachatryan:2015pzs} & 12/12 \\
    &  $W^{+}$ & \;\;\;5.02 & 6233 & \cite{Khachatryan:2015hha} & 10/10 \\
    &  $W^{-}$ & \;\;\;5.02 & 6231 & \cite{Khachatryan:2015hha} & 10/10 \\
    &  $W^{+}$  & \;\;\;8.16 & 6234 & \cite{CMS:2019leu} & 24/24 \\
    &  $W^{-}$  & \;\;\;8.16 & 6232 & \cite{CMS:2019leu} & 24/24 \\
    \cline{1-6}
    \multirow{1}{*}[-0.0em]{LHCb\;}  
    & $Z$  & \;\;\;5.02 & 6275 & \cite{Aaij:2014pvu} & 2/2 \\
    \cline{1-6}
    \multirow{2}{*}[-0.0em]{ALICE\;}  
    & $W^{+}$  & \;\;\;5.02 & 6253 & \cite{ALICE:2016rzo} & 2/2 \\
    & $W^{-}$  & \;\;\;5.02 & 6251 & \cite{ALICE:2016rzo} & 2/2 \\
    \hline 
    \hline 
    \textbf{Total} &  &  & & & \hspace{-0.35cm}\textbf{120/120} \\
    \hline
  \end{tabular}
  \caption{\wz data from the proton-lead collisions at the LHC.}
  \label{tab:WZdata}
\end{table}

\begin{table}[h!]
  \centering
  \renewcommand{\arraystretch}{1.2}
  \begin{tabular}{|l l l c c c|}
    \hline
    \multicolumn{6}{|c|}{LHC heavy flavour production in $p$Pb $d^2\sigma/dydp_t$}\\
    \hline
    \hline
    Exp. & Observable & $\sqrt{s}$ [TeV] & \;ID & Ref. & $N_{\text{data}}$ \\
    \hline 
    \hline
    \multirow{10}{*}[-0.0em]{ALICE} 
    & $\text{prompt}~D^0$ & \;\;\;5.02 & 3101 & \cite{ALICE:2014xjz} & 6/10 \\
    & $\text{incl.}~J/\Psi$ & \;\;\;5.02 & 3103 & \cite{ALICE:2013snh} & 0/12 \\
    & $\text{incl.}~J/\Psi$ & \;\;\;5.02 & 3104 & \cite{ALICE:2015sru} & 10/25 \\
    & $\text{incl.}~J/\Psi$ & \;\;\;8.16 & 3112 & \cite{ALICE:2018mml} & 9/24 \\
    & $\text{incl.}~\Upsilon(1S)$ & \;\;\;8.16 & 3114 & \cite{ALICE:2019qie} & 3/10 \\
    & $\text{prompt}~D^0$ & \;\;\;5.02 & 3122 & \cite{ALICE:2019fhe} & 13/21 \\
    & $\text{prompt}~D^0$ & \;\;\;5.02 & 3123 & \cite{ALICE:2016yta} & 6/11 \\
    & $\text{incl.}~\Psi(2S)$ & \;\;\;8.16 & 3126 & \cite{ALICE:2020vjy} & 3/10 \\
    & $\text{incl.}~\Psi(2S)$ & \;\;\;5.02 & 3127 & \cite{ALICE:2014cgk} & 2/8 \\
    & $\text{incl.}~\Upsilon(1S)$ & \;\;\;8.16 & 3110 & \cite{ALICE:2014ict} & 0/4 \\
    \cline{1-6}
    \multirow{7}{*}[-0.0em]{ATLAS} 
     & $\text{incl.}~\Upsilon(1S)$  & \;\;\;5.02 & 3109 & \cite{ATLAS:2017prf} & 6/8 \\
    & $\text{non-pr.}~J/\Psi$ & \;\;\;5.02 & 3116 & \cite{ATLAS:2017prf} & 8/8 \\
    & $\text{prompt}~J/\Psi$ & \;\;\;5.02 & 3117 & \cite{ATLAS:2017prf} & 8/8 \\
    & $\text{prompt}~J/\Psi$ & \;\;\;5.02 & 3118 & \cite{ATLAS:2015mpz} & 10/10 \\
    & $\text{non-pr.}~J/\Psi$ & \;\;\;5.02 & 3119 & \cite{ATLAS:2015mpz} & 10/10 \\
    & $\text{prompt}~\Psi(2S)$ & \;\;\;5.02 & 3124 & \cite{ATLAS:2017prf} & 8/8 \\
    & $\text{non-pr.}~\Psi(2S)$ & \;\;\;5.02 & 3125 & \cite{ATLAS:2017prf} & 8/8 \\
    \cline{1-6}
    \multirow{3}{*}[-0.0em]{CMS\;} 
    & $\text{prompt}~\Psi(2S)$ & \;\;\;5.02 & 3115 &\cite{CMS:2018gbb} & 17/17 \\
    & $\text{prompt}~J/\Psi$ & \;\;\;5.02 & 3120 &\cite{CMS:2017exb} & 51/53 \\
    & $\text{non-pr.}~J/\Psi$ & \;\;\;5.02 & 3121 &\cite{CMS:2017exb} & 51/53 \\
    \cline{1-6}
    \multirow{5}{*}[-0.0em]{LHCb}
    & $\text{prompt}~D^0$ & \;\;\;5.02 & 3102 & \cite{LHCb:2017yua} & 53/92 \\
    & $\text{prompt}~J/\Psi$ & \;\;\;8.16 & 3105 & \cite{LHCb:2017ygo} & 88/140 \\
    & $\text{non-pr.}~J/\Psi$ & \;\;\;8.16 & 3106 & \cite{LHCb:2017ygo} & 88/140 \\
    & $\text{non-pr.}~J/\Psi$ & \;\;\;5.02 & 3107 & \cite{LHCb:2013gmv} & 25/40 \\
    & $\text{prompt}~J/\Psi$ & \;\;\;5.02 & 3108 & \cite{LHCb:2013gmv} & 25/40 \\
    & $\text{incl.}~\Upsilon(1S)$ & \;\;\;5.02 & 3111 & \cite{LHCb:2014rku} & 0/2 \\
    & $\text{incl.}~\Upsilon(1S)$ & \;\;\;8.16 & 3113 & \cite{LHCb:2018psc} & 36/46 \\
    \hline 
    \hline 
    \textbf{Total} &  &  & & & \hspace{-0.35cm}\textbf{544/717} \\
    \hline
  \end{tabular}
  \caption{The heavy flavor meson and heavy quarkonia data from LHC experiments.}
  \label{tab:HQdata}
\end{table}

\begin{table}
  \centering
  \renewcommand{\arraystretch}{1.2}
  \begin{tabular}{|l l c c c|}
    \hline
    \multicolumn{5}{|c|}{Neutrino DIS data}\\
    \hline
    \hline
    Exp. & Obser. & ID & Ref. & $N_{\text{data}}$ \\
    \hline
    \multirow{2}{*}{Chorus} & $\nu$             & 5946   & \multirow{2}{*}{\cite{Onengut:2005kv}}   & 412/607\\
                            & $\bar{\nu}$       & 5947   &                                         &  412/607\\
    \hline
    \hline 
    \textbf{Total}     &  &   &  & \textbf{824/1214} \\
    \hline 
  \end{tabular}%
  \caption{ Neutrino data sets from the CHORUS
experiment.}
  \label{tab:nudata}
\end{table}

\subsection{\multinuc{} Analysis}
\label{app:multnuc}
The \multinuc\ global fit extends the \Pbonly\ analysis by including experimental data from a broad range of nuclear targets to probe the $A$ dependence of nuclear modifications. In addition to the \Pbonly\ datasets (\wz production, heavy-flavor production, and neutrino DIS on lead), the fit incorporates fixed-target measurements of neutral-current (NC) DIS and Drell–Yan (DY) production. The NC DIS data, taken from the SLAC, NMC, EMC, and BCDMS experiments, cover nuclei ranging from deuterium through iron to lead and are reported as ratios of structure functions or cross sections, $F_2^A/F_2^D$ and $\sigma_{\text{DIS}}^{A}/\sigma_{\text{DIS}}^{D}$. The DY data, from Fermilab FNAL-E772-90 and FNAL-E886-99 experiments, provide ratios of dimuon production cross sections between heavy and light nuclear targets. Together, these datasets extend the kinematic coverage in $x$ and $Q^2$, offering complementary constraints on both valence and sea-quark distributions. The complete list of datasets is given in Tables~\ref{tab:DY}-\ref{tab:NCDIS2}.
\begin{table}
  \centering
  \renewcommand{\arraystretch}{1.2}
  \begin{tabular}{|l l c c c|}
    \hline
    \multicolumn{5}{|c|}{Fixed target DY data $\mathbf{\sigma_{DY}^{pA}/\sigma_{DY}^{pA'}}$}\\
    \hline
    \hline
    Nucleus & Experiment &  \,\,\,ID & Ref. &  $N_{\text{data}}$ \\
    \hline
    $\mathrm{C}/\mathrm{D}$    & FNAL-E772-90 & 5203 & \cite{Alde:1990im}  & 9/9 \\
    \hline 
    $\mathrm{Ca}/\mathrm{D}$ & FNAL-E772-90 & 5204 & \cite{Alde:1990im}  & 9/9 \\
    \hline 
    $\mathrm{Fe}/\mathrm{D}$ & FNAL-E772-90 & 5205 & \cite{Alde:1990im}  & 9/9 \\
    \hline 
    $\mathrm{W}/\mathrm{D}$ & FNAL-E772-90 & 5206 & \cite{Alde:1990im}  & 9/9 \\
    \hline 
    $\mathrm{Fe}/\mathrm{Be}$ & FNAL-E886-99 & 5201 & \cite{Vasilev:1999fa}  & 28/28 \\
    \hline 
    $\mathrm{W}/\mathrm{Be}$ & FNAL-E886-99 & 5202 & \cite{Vasilev:1999fa}  & 28/28 \\
    \hline 
    \hline 
    \textbf{Total} &  &  &   & \textbf{92/92} \\
    \hline 
    \end{tabular}
  \caption{Drell–Yan data as cross-section ratios of heavy to light nuclei.}
  \label{tab:DY}
\end{table}

\begin{table}
  \centering
  \renewcommand{\arraystretch}{1.2}
  \begin{tabular}{|l l c c c |}
    \hline
    \multicolumn{5}{|c|}{NC DIS data $\mathbf{F_{2}^{A}/F_{2}^{A'}}$}\\
    \hline
    \hline
    Nucleus & Experiment &  \,\,\,ID & Ref. &  $N_{\text{data}}$ \\
    \hline 
    $\mathrm{C}/\mathrm{Li}$ & NMC-95,re & 5123 & \cite{Amaudruz:1995tq} & 7/25 \\
    \hline
    $\mathrm{Ca}/\mathrm{Li}$ & NMC-95,re & 5122 & \cite{Amaudruz:1995tq} & 7/25 \\
    \hline
    $\mathrm{Be}/\mathrm{C}$ & NMC-96 & 5112 & \cite{Arneodo:1996rv} & 14/15 \\
    \hline
    $\mathrm{Al}/\mathrm{C}$ & NMC-96 & 5111 & \cite{Arneodo:1996rv} & 14/15 \\
    \hline
    \multirow{2}{*}{$\mathrm{Ca}/\mathrm{C}$} & NMC-95,re & 5120 & \cite{Amaudruz:1995tq} & 7/25 \\
    & NMC-96 & 5119 & \cite{Arneodo:1996rv} & 14/15 \\
    \hline
    $\mathrm{Fe}/\mathrm{C}$ & NMC-96 & 5143 & \cite{Arneodo:1996rv} & 14/15 \\
    \hline
    $\mathrm{Sn}/\mathrm{C}$ & NMC-96 & 5159 & \cite{Arneodo:1996ru} & 111/146 \\
    \hline
    $\mathrm{Pb}/\mathrm{C}$ & NMC-96 & 5116 & \cite{Arneodo:1996rv} & 14/15 \\
    \hline 
    \hline 
    \textbf{Total} &  &  &  & \hspace{-0.15cm}\textbf{202/296}\tabularnewline
    \hline 
\end{tabular}
  \caption{NC DIS data as ratios between two nuclei, $F_{2}^{A}/F_{2}^{A'}$.}
  \label{tab:NCDIS1}
\end{table}

\begin{table}
  \centering
  \renewcommand{\arraystretch}{1.2}
  \begin{tabular}{|l l c c c|}
    \hline
    \multicolumn{5}{|c|}{NC DIS data $\mathbf{F_{2}^{A}/F_{2}^{D}}$ and $\mathbf{\sigma_{\text{DIS}}^{A}/\sigma_{\text{DIS}}^{D}}$}\\
    \hline
    \hline
    Nucleus & Experiment &  \,\,\,ID & Ref. &  $N_{\text{data}}$ \\
    \hline 
    ${}^3\mathrm{He}/\mathrm{D}$ & Hermes & 5156 & \cite{Airapetian:2002fx} & 17/182 \\
    \hline
    \multirow{2}{*}{${}^4\mathrm{He}/\mathrm{D}$} & NMC-95,re & 5124 & \cite{Amaudruz:1995tq} & 12/18 \\
     & SLAC-E139 & 5141 & \cite{Gomez:1993ri} & 3/18 \\
    \hline
    $\mathrm{Li}/\mathrm{D}$ & NMC-95 & 5115 & \cite{Arneodo:1995cs} & 11/24 \\
    \hline
    $\mathrm{Be}/\mathrm{D}$ & SLAC-E139 & 5138 & \cite{Gomez:1993ri} & 3/17 \\
    \hline
    \multirow{6}{*}{$\mathrm{C}/\mathrm{D}$} & FNAL-E665-95 & 5125 & \cite{Adams:1995is} & 3/11 \\
    & SLAC-E139 & 5139 & \cite{Gomez:1993ri} & 2/7 \\
    & EMC-88 & 5107 & \cite{Ashman:1988bf} & 9/9 \\
    & EMC-90 & 5110 & \cite{Arneodo:1989sy} & 0/9 \\
    & NMC-95 & 5113 & \cite{Arneodo:1995cs} & 12/24 \\
    & NMC-95,re & 5114 & \cite{Amaudruz:1995tq} & 12/18 \\
    \hline
    \multirow{2}{*}{$\mathrm{N}/\mathrm{D}$} & Hermes & 5157 & \cite{Airapetian:2002fx} & 19/175 \\
    & BCDMS-85 & 5103 & \cite{Bari:1985ga} & 9/9 \\
    \hline
    \multirow{2}{*}{$\mathrm{Al}/\mathrm{D}$} & SLAC-E049 & 5134 & \cite{Bodek:1983ec} & 0/18 \\
    & SLAC-E139 & 5136 & \cite{Gomez:1993ri} & 3/17 \\
    \hline
    \multirow{4}{*}{$\mathrm{Ca}/\mathrm{D}$} & NMC-95,re & 5121 & \cite{Amaudruz:1995tq} & 12/18 \\
    & FNAL-E665-95 & 5126 & \cite{Adams:1995is} & 3/11 \\
    & SLAC-E139 & 5140 & \cite{Gomez:1993ri} & 2/7 \\
    & EMC-90 & 5109 & \cite{Arneodo:1989sy} & 0/9 \\
    \hline
    \multirow{5}{*}{$\mathrm{Fe}/\mathrm{D}$} & SLAC-E049 & 5131 & \cite{Bodek:1983qn} & 2/14 \\
    & SLAC-E139 & 5132 & \cite{Gomez:1993ri} & 6/23 \\
    & SLAC-E140 & 5133 & \cite{Dasu:1993vk} & 0/10 \\
    & BCDMS-87 & 5101 & \cite{Benvenuti:1987az} & 10/10 \\
    & BCDMS-85 & 5102 & \cite{Bari:1985ga} & 6/6 \\
    \hline
    \multirow{3}{*}{$\mathrm{Cu}/\mathrm{D}$} & EMC-93 & 5104 & \cite{Ashman:1992kv} & 9/10 \\
    & EMC-93 (chariot) & 5105 & \cite{Ashman:1992kv} & 9/9 \\
    & EMC-88 & 5106 & \cite{Ashman:1988bf} & 9/9 \\
    \hline
    $\mathrm{Kr}/\mathrm{D}$ & Hermes & 5158 & \cite{Airapetian:2002fx} & 12/167 \\
    \hline
    $\mathrm{Ag}/\mathrm{D}$ & SLAC-E139 & 5135 & \cite{Gomez:1993ri} & 2/7 \\
    \hline
    $\mathrm{Sn}/\mathrm{D}$ & EMC-88 & 5108 & \cite{Ashman:1988bf} & 8/8 \\
    \hline
    $\mathrm{Xe}/\mathrm{D}$ & FNAL-E665-92 & 5127 & \cite{Adams:1992nf} & 2/10 \\
    \hline
    $\mathrm{Au}/\mathrm{D}$ & SLAC-E139 & 5137 & \cite{Gomez:1993ri} & 3/18 \\
    \hline
    $\mathrm{Pb}/\mathrm{D}$ & FNAL-E665-95 & 5129 & \cite{Adams:1995is} & 3/11 \\
    \hline 
    \hline 
    \textbf{Total} &  &  &  & \hspace{-0.15cm}\textbf{213/913}\tabularnewline
    \hline 
\end{tabular}
  \caption{NC DIS data as ratios to deuterium, $F_2^A/F_2^D$ and $\sigma_{\text{DIS}}^{A}/\sigma_{\text{DIS}}^{D}$.}
  \label{tab:NCDIS2}
\end{table}

\begin{figure*}[!htbp]
\centering
\includegraphics[width=\textwidth]{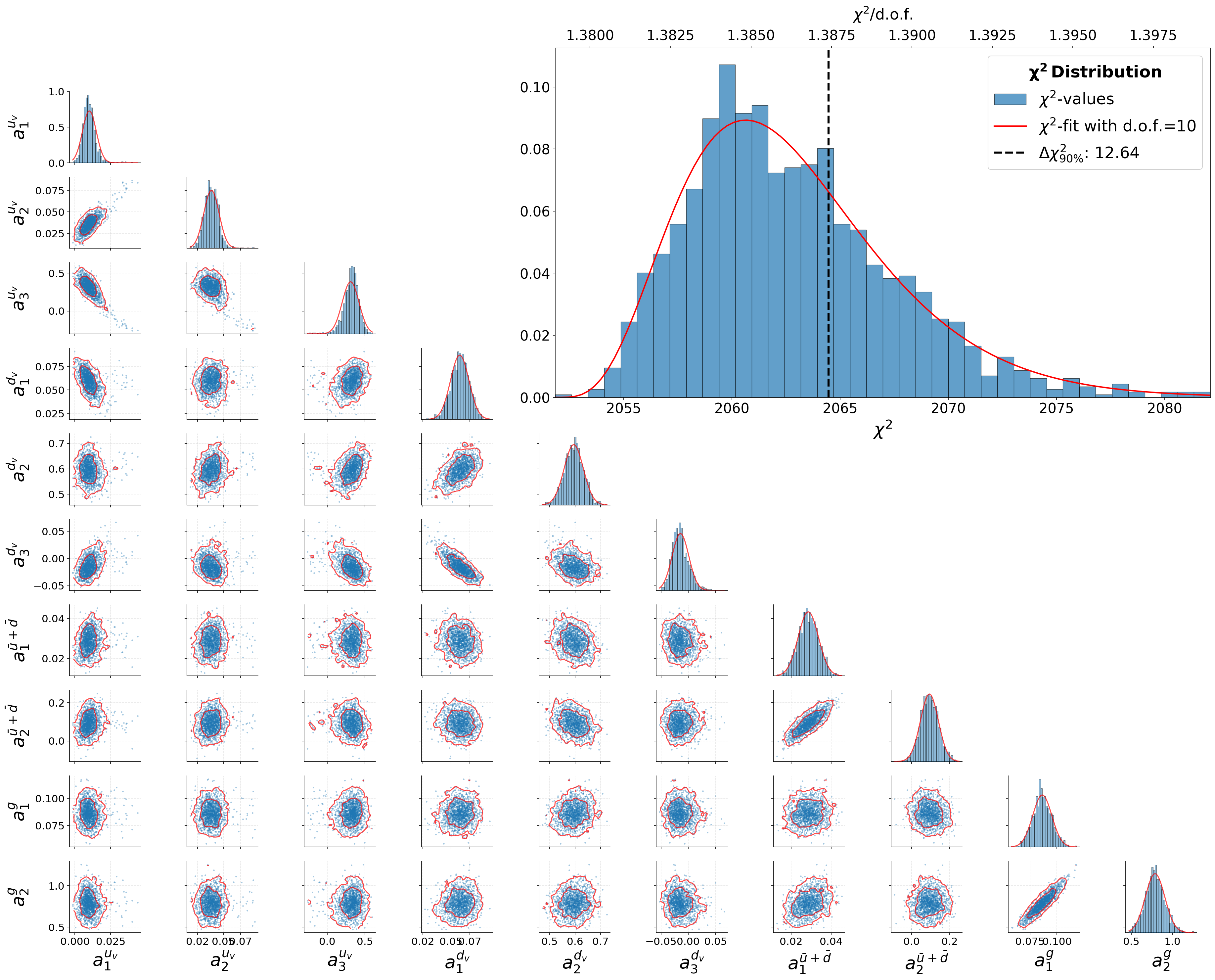}
\caption{Pairwise plot of the posterior distribution of the \multinuc{}, analogous to Fig. \ref{fig:pairwise}}
\label{fig:Nupairwise} 
\end{figure*}

\begin{figure*}[!htbp]
\centering
\includegraphics[width=\textwidth]{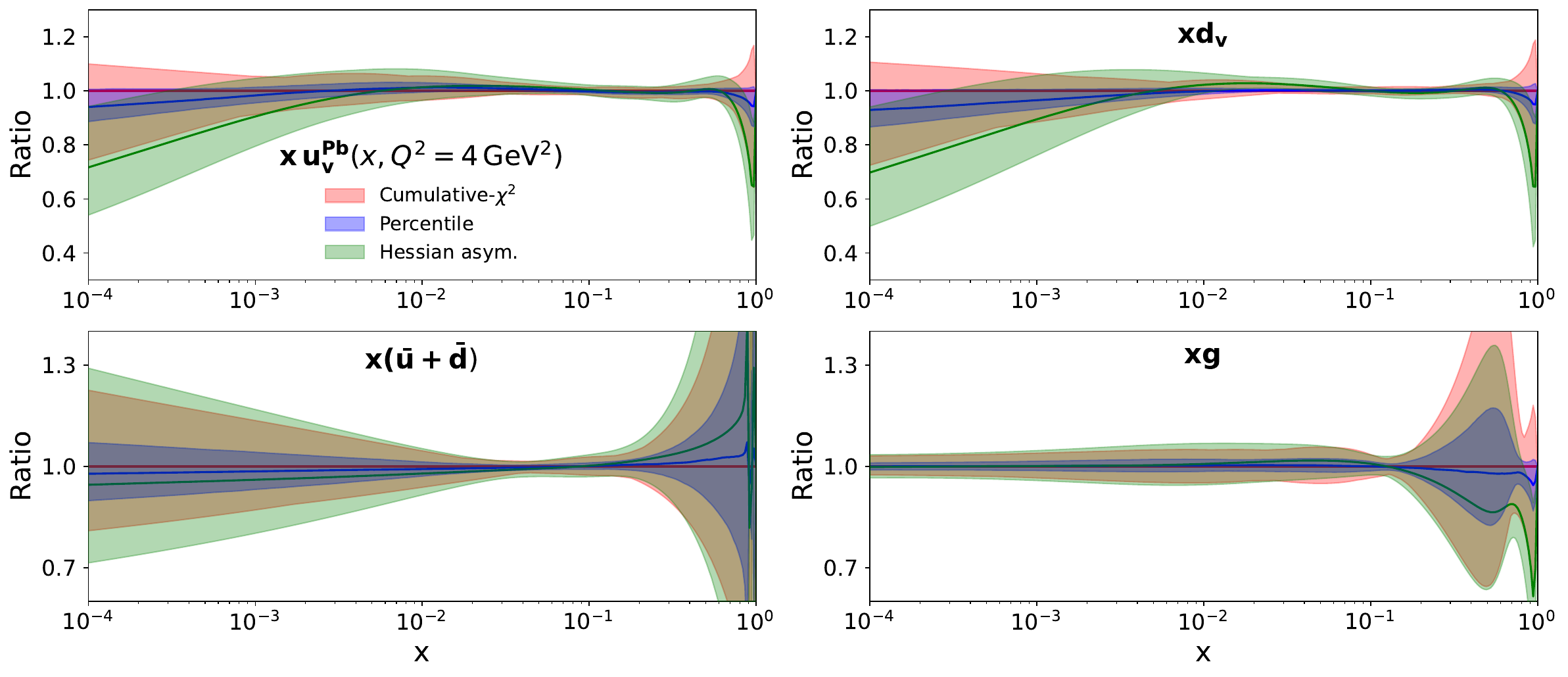}
\caption{Comparison of lead PDFs error estimations for $u_v$, $d_v$, $\bar{u}+\bar{d}$, and $g$ distributions for the \multinuc\ MCMC and Hessian analyses. We present the ratio to the best fit from the \cumchi{} method.}
\label{fig:PDFsMultiNuc} 
\end{figure*}

We also include here some additional results related to the \multinuc\ analysis. First of all, in \cref{fig:Nupairwise}, we show the pairwise plot for the final sample. It is analogous to \cref{fig:pairwise} and demonstrates correlations between parameters in this case. Then, in Fig.~\ref{fig:PDFsMultiNuc} we display a comparison of the obtained Pb PDFs comparing different methods for computing PDF uncertainties. This is a plot analogous to Fig.~\ref{fig:PDFcompPb-MCMChess} where the corresponding comparison for \Pbonly{} analysis was presented but in this case we show only the ratio panels. We observe here similar behavior as in Fig.~\ref{fig:PDFcompPb-MCMChess}. Namely, we see that the percentile method provides much smaller error bands than the \cumchi{} method. We also see that in the case of gluon and light sea quarks, the Hessian uncertainty is quite similar to the one obtained from MCMC using the \cumchi{} method. This happens because marginal distributions for the corresponding parameters are close to Gaussian. Again for the valence PDFs, for which the parameter distributions are non-Gaussian, we observe a substantial deviation between the MCMC and Hessian error estimates, see \cref{fig:pairwise}. Interestingly, in the case of the \multinuc{} analysis, the \cumchi{} uncertainty estimates provide larger error bands than the estimates from the Hessian, which is opposite to what we observed in the case of \Pbonly{} analyses.

%
\section{MCMC Performance Diagnostics}
\label{appen:B}
\begin{figure*}[!htb]
  \centering{}
    \includegraphics[width=0.49\textwidth]{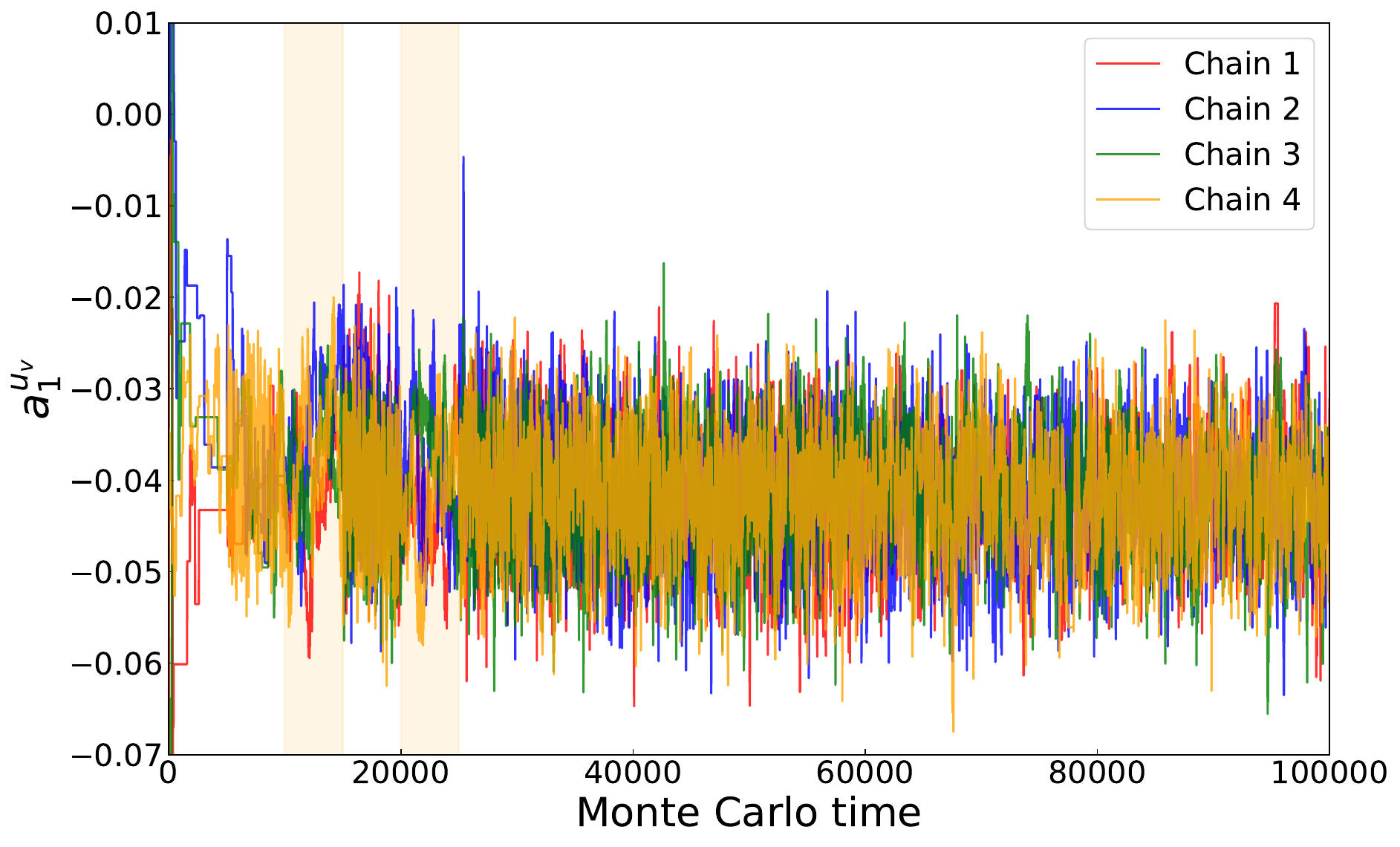}
    \includegraphics[width=0.49\textwidth]{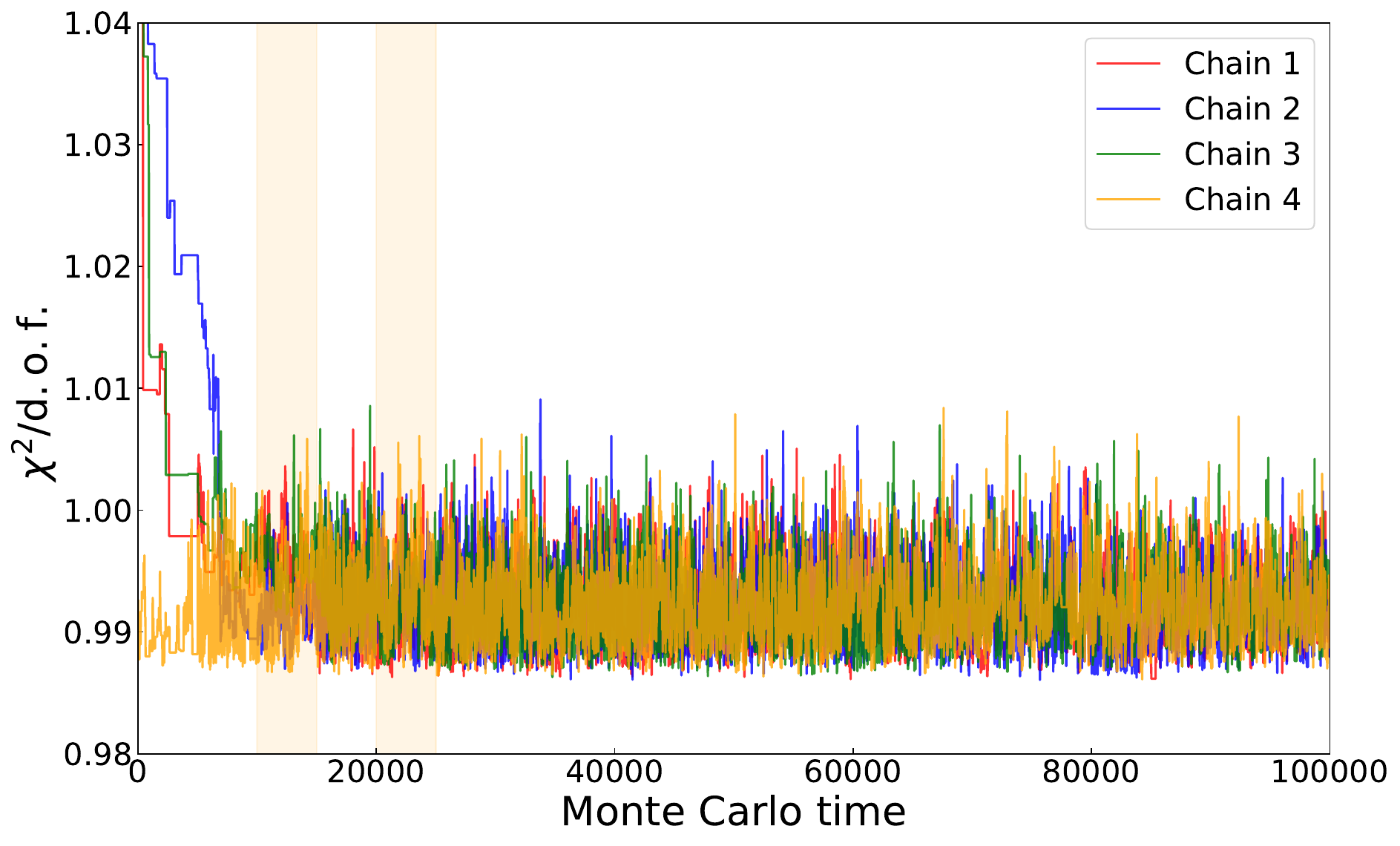}
  \caption{Left: Evolution of the $a_1^{u_v}$ parameter over Monte Carlo time for four independent Markov chains initiated from different starting points. Right: Corresponding $\chi^2/\text{d.o.f.}$ evolution for these chains, confirming consistent convergence and effective sampling from the target distribution.}
  \label{fig:app.3}
\end{figure*}

\begin{figure*}[!htbp]
    \centering{}
    \includegraphics[width=0.49\textwidth]{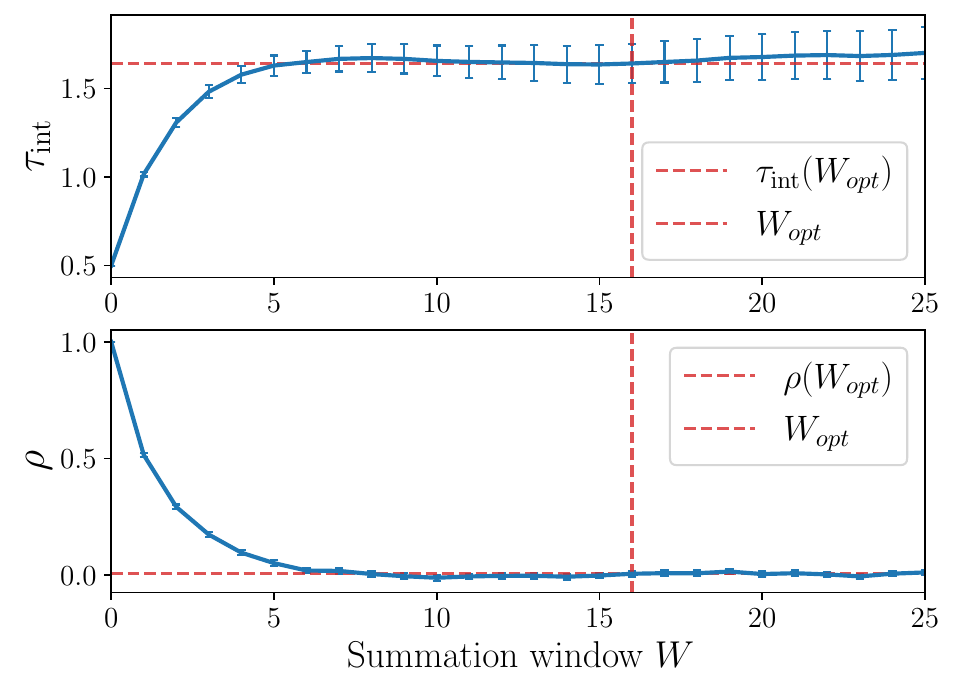}
    \includegraphics[width=0.49\textwidth]{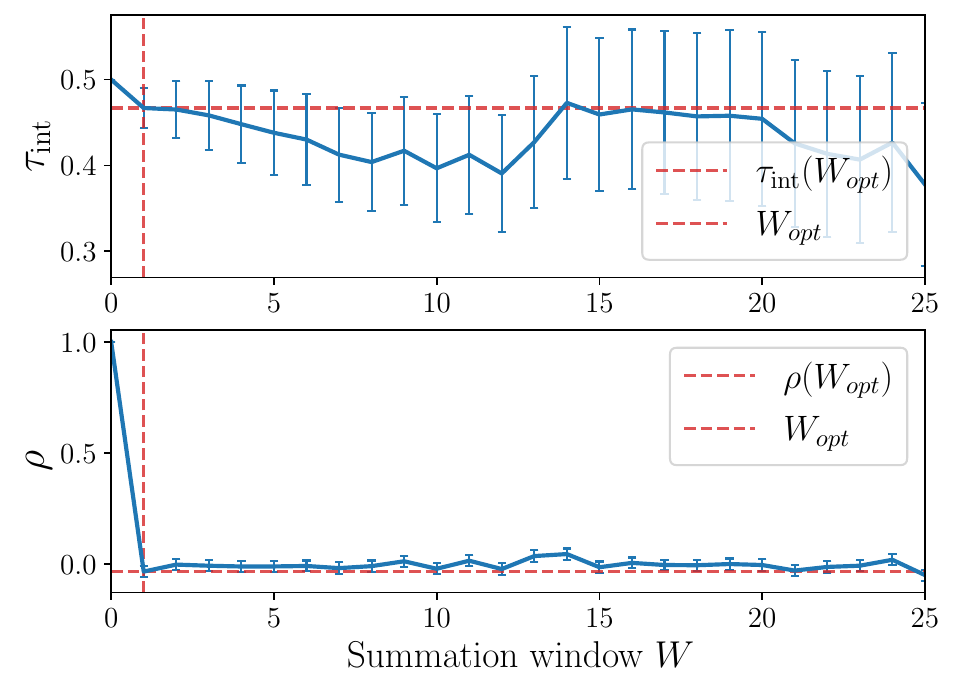}
    \caption{ACF and $\tau_{\mathrm{int}}$ as a function of the summation window $W$ for the \Pbonly{} chain thinned by rate 50 (left) and 400 (right). The dashed red vertical line indicates the optimal summation window, while the horizontal dashed red line marks the corresponding $\tau_\text{int}$ or $\rho$ value at that window, both determined by the $\Gamma$-method.}
    \label{fig:app.4}
\end{figure*}

A representative Markov chain from the \Pbonly{} analysis, containing about 720\,000 samples, is examined to illustrate the sampling behavior. The full evolution of the chain is displayed in the trace plot in Fig.~\ref{fig:timeseries} in Sec.~\ref{sec:res}, where the initial 80\,000 steps (the red-shaded region) are identified and discarded as the thermalization (burn-in) phase. Following this period, all ten parameters exhibit stable fluctuations around constant means, confirming that the sampler has successfully reached equilibrium and effectively explores the posterior distribution.
Figure~\ref{fig:app.3} compares four independent chains initialized from different random starting points. The evolution of the $a_1^{u_v}$ parameter and the corresponding $\chi^2$ values reveal consistent convergence toward the same stationary region, confirming the reproducibility and robustness of the sampling. Finally, Fig.~\ref{fig:app.4} shows the autocorrelation function, $\rho$, and the integrated autocorrelation time, $\tau_\text{int}$, as functions of the summation window $W$ for two thinning rates, $\eta$ = 50 and 400. The results confirm that a thinning rate of $\eta \simeq 400$ yields nearly uncorrelated samples, supporting the choice adopted in the main analysis.

%
\section{Details about Secondary Minima}
\label{appen:C}
\begin{figure*}[t]
\centering
\includegraphics[width=\textwidth]{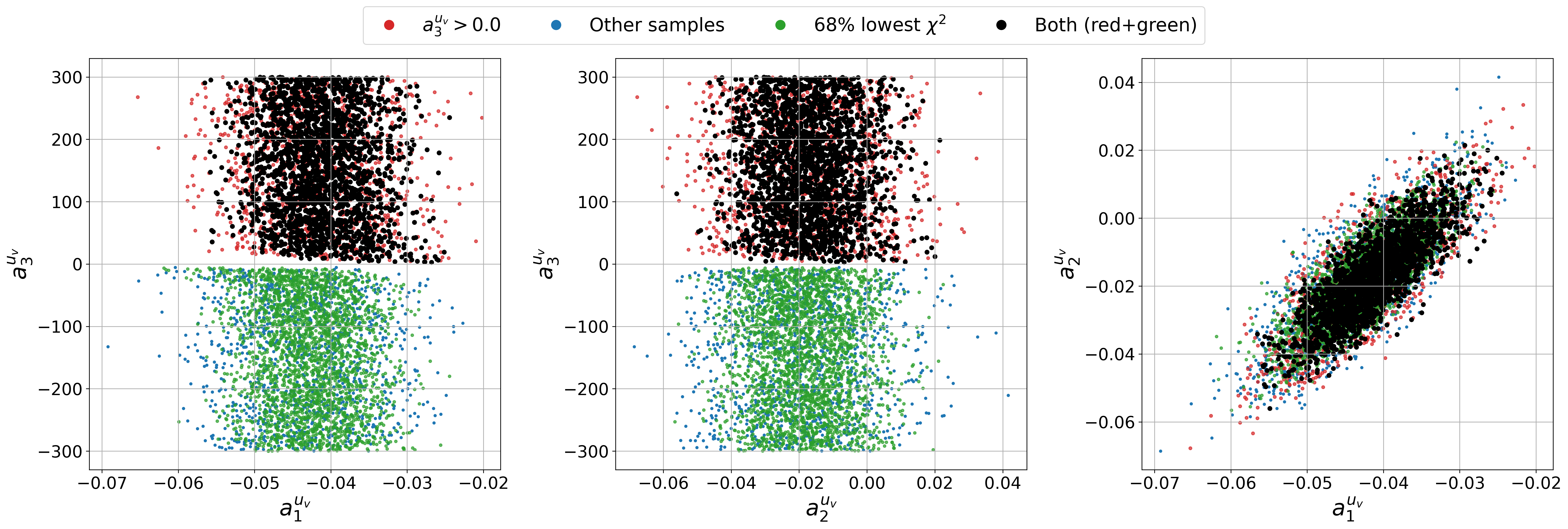}
\caption{Same as Fig.~\ref{fig:dv68CLvs90CL} but for $u_v$ parameters. Here 68\% CL is used.}
\label{fig:altMINuvpairwise}
\end{figure*}

\begin{figure*}[t]
\centering
\includegraphics[width=0.48\textwidth]{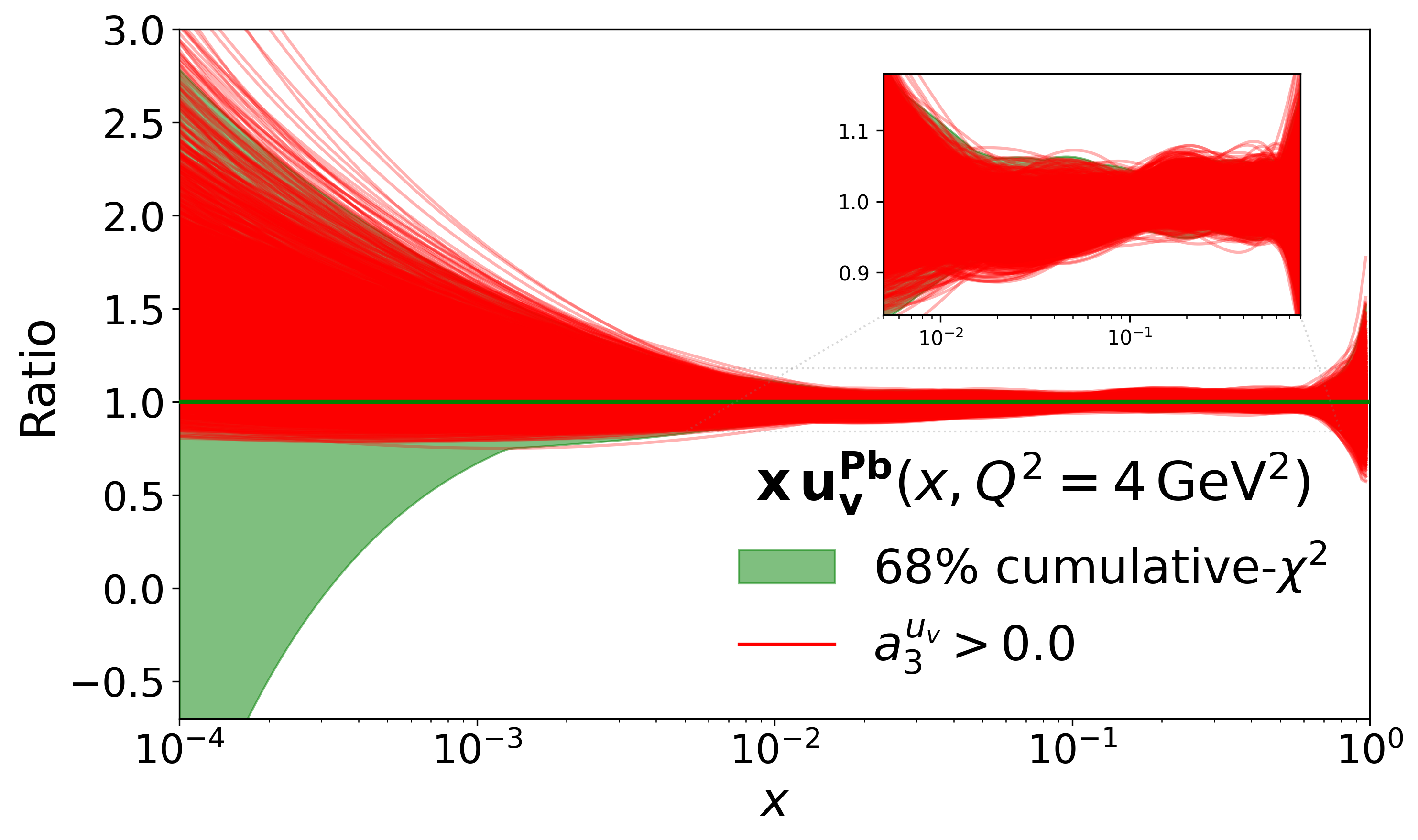}
\includegraphics[width=0.48\textwidth]{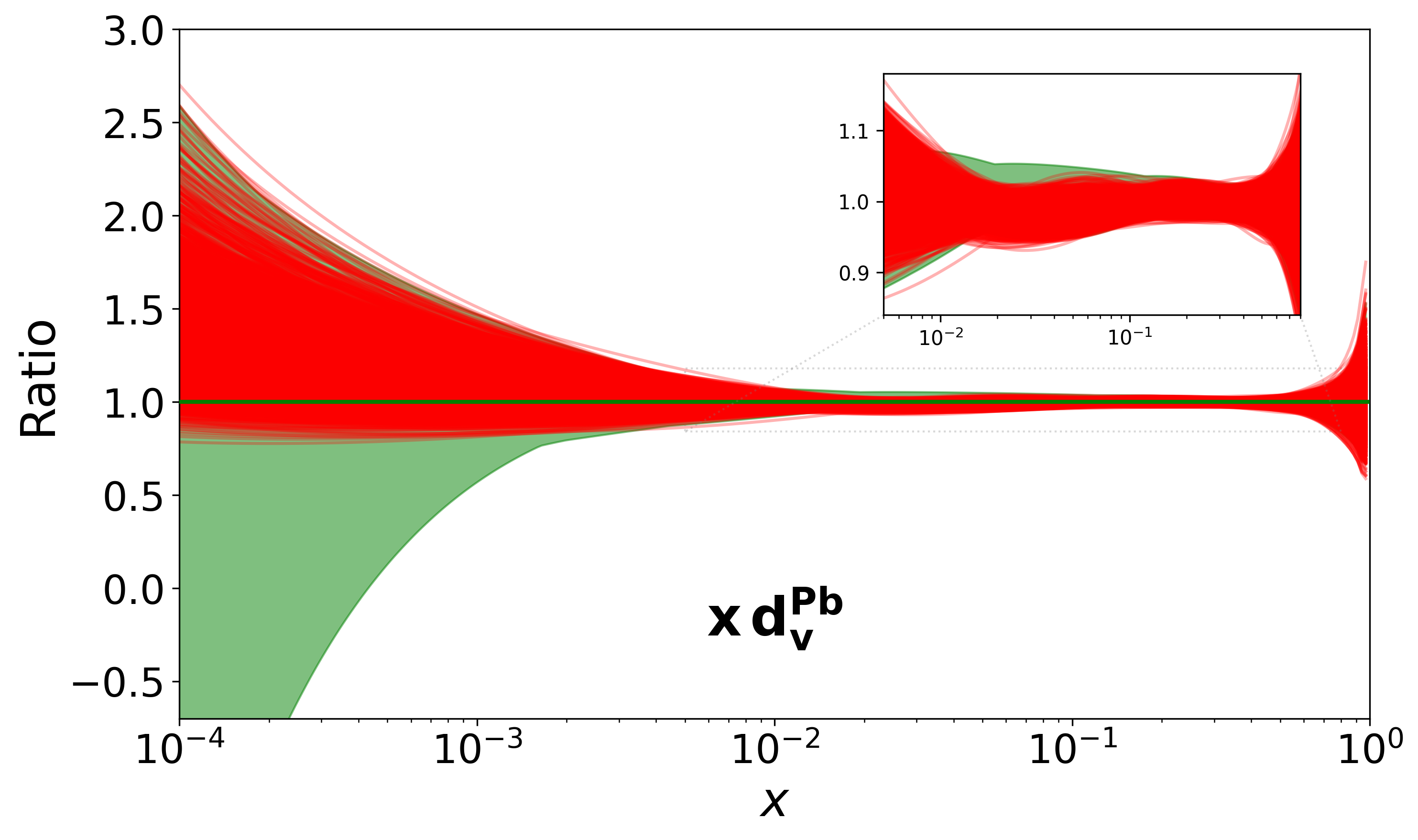}
\includegraphics[width=0.48\textwidth]{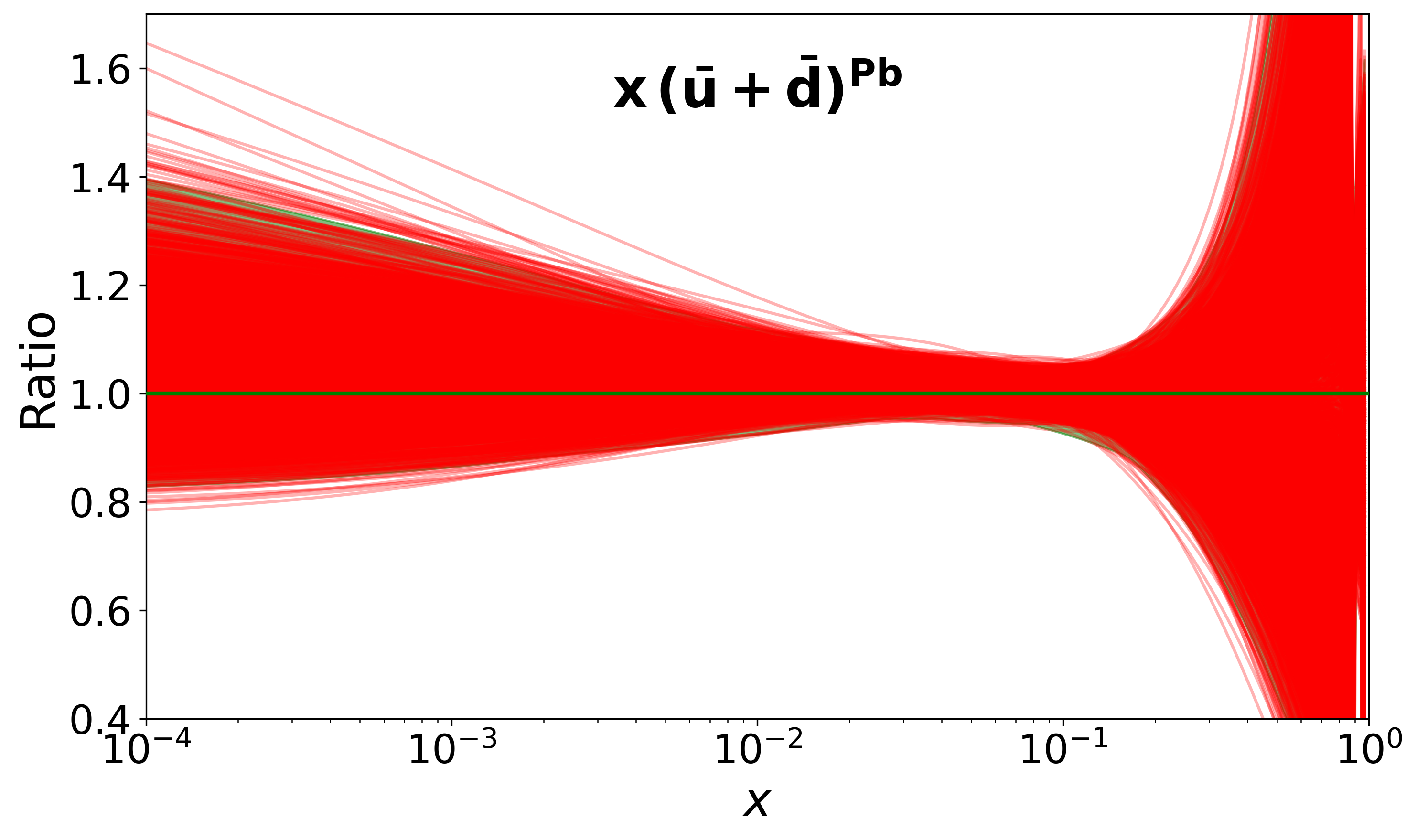}
\includegraphics[width=0.48\textwidth]{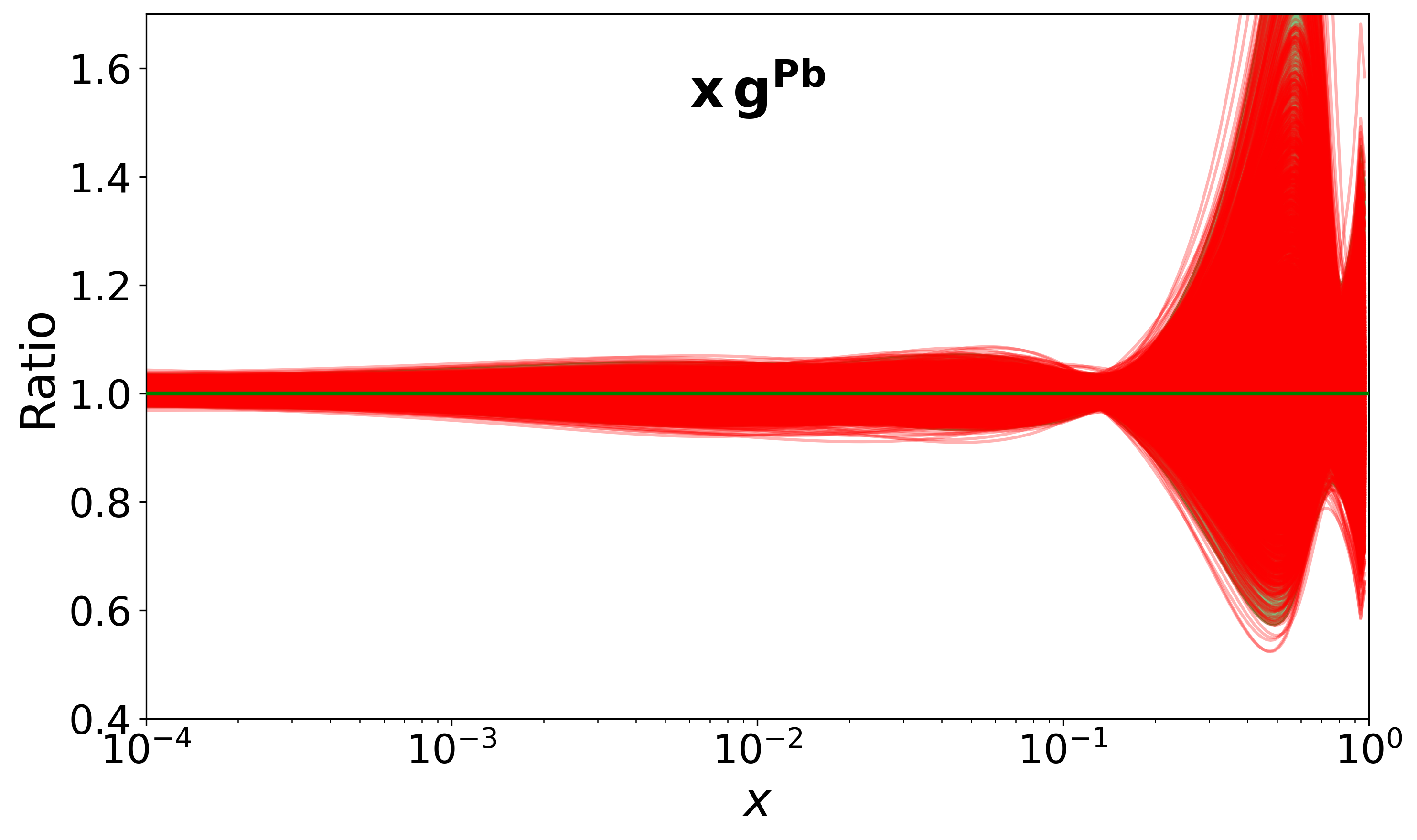}
\caption{Same as Fig.~\ref{fig:PDFsaltMinPb} but for minimum related to $a_3^{u_v}$ parameter. Replicas plotted in red correspond to $a_3^{u_v}>0$ (red dots in Fig.~\ref{fig:altMINuvpairwise}).}
\label{fig:altMINuvPDF}
\end{figure*}

We provide here additional details about the secondary minimum in the \Pbonly{} analysis related to the $a_3^{u_v}$ parameter discussed in Sec.~\ref{sec:Multiple_minima}. In Fig.~\ref{fig:altMINuvpairwise} we display pairwise correlations for the $u_v$ parameters. In contrast to what we saw for the $d_v$ parameters shown in Fig.~\ref{fig:dv68CLvs90CL}, for $u_v$ there is a clear separation between the positive and negative values of $a_3^{u_v}$, with an excluded region of $a_3^{u_v}\sim0$ (see Fig.~\ref{fig:chi2Scan}). However, we do not observe a clear correlation between its positive/negative values and the $a_1^{u_v}$ and $a_2^{u_v}$ parameters.
In Fig.~\ref{fig:altMINuvPDF} in green we plot 68\% CL PDFs (corresponding to green and black points in Fig.~\ref{fig:altMINuvpairwise}) and overlay them with the red PDF replicas corresponding to the PDFs for which $a_3^{u_v}>0$ (red dots in Fig.~\ref{fig:altMINuvpairwise}). We can see that the split in $a_3^{u_v}$ parameter affects only the low-$x$ valence PDFs. The high-$x$ is mostly unaffected and we do not see any change in the shape of the valence distributions in this region. The fact that both $u_v$ and $d_v$ PDFs are affected by the split in the $a_3^{u_v}$ parameter is caused by the fact that the parameters we are using are defined for the bound proton PDFs (see Eq.~\eqref{eq.2}) but the plots show the full lead PDFs (Eq.~\eqref{eq.1}). Hence, there is a mixing between $u_v$ and $d_v$ distributions. The value of the $a_3^{u_v}$ parameter does not have any impact on the sea-quark and gluon distributions.

Finally, for the \multinuc{} analysis, we also observe structures resembling secondary minima. The behavior is similar to what we discussed in case of the \Pbonly{} analysis but less pronounced, \textit{i.e.,} showing that an increase in constraints moves the parameter distributions closer to Gaussian.

\section*{Acknowledgments}
The authors would like to thank our nCTEQ collaborators, in particular Jan Wissmann and Ingo Schienbein for many useful comments and discussions. 
The work of T.J., M.K. and K.K. has been supported by the BMBF under contract 05P24PMA.
A.K. and N.D. acknowledge the support of the National Science Centre Poland under the Sonata Bis grant No. 2019/34/E/ST2/00186. N.D. acknowledges financial support provided by the NAWA-STER Program project no. PPI/STE/2020/1/00020/U/00001. We also acknowledge the computational resources provided by PLGrid in early stages of the project under the grant number PLG/2022/016037.
The work of P.R. was supported by the U.S. Department of Energy under Grant No. DE-SC0010129, and by the Office of Science, the Office of Nuclear Physics, within the framework of the Saturated Glue (SURGE) Topical Theory Collaboration. P.R. thanks the Jefferson Lab for their hospitality. This work was supported by the US Department of Energy Contract No.~DE-AC05-06OR23177, under which Jefferson Science Associates, LLC operates Jefferson Lab.
\clearpage
\bibliographystyle{utphys}
\bibliography{refs}
\end{document}